\documentclass[twocolumn,rmp,aps,reprint, psfig,showpacs,amsmath,longbibliography]{revtex4-1}

\usepackage{graphicx}
\usepackage{epstopdf} 
\usepackage{pdfsync}
\usepackage{color}
\usepackage[dvipsnames]{xcolor}
\usepackage{amsmath, amssymb}
\usepackage[normalem]{ulem}

\newcounter{mycount}

\newcommand{\be}[1]{ \begin{eqnarray} \mbox{$\label{#1}$} }

\newcommand{\ee}{\end{eqnarray}}
\newcommand{\eeq}{\end{equation}}

\newcommand{\pref}[1]{(\ref{#1})}

\newcommand\ie {{\it i.e. }}
\newcommand\eg {{\it e.g. }}

\newcommand\cf {{\it cf.  }}

\newcommand\half{\frac 1 2 }

\newcommand\ket [1] {|#1 \rangle }
\newcommand\bra [1] {\langle #1 |}
\newcommand{\bracket}[2]   {  \left<#1 |  #2\right>}
\newcommand{\av}[1]{\langle #1\rangle}

\newcommand{\bkappa}{\boldsymbol{\kappa}}

\newcommand\jasf[3]{(z_{#1} - z_{#2})^{#3}}

\newcommand{\grupp[3]} {\{{#1}_i \}_{#2}^{#3 } }
\newcommand{\sgrupp[1]} {\{{#1}_i \} }

\begin{document}
\title{Quantum Hall Physics -- 
hierarchies and CFT techniques  }

\author{T. H. Hansson$^{1,2}$}
\author{M. Hermanns$^3$}
\author{S. H.  Simon$^4$}
\author{S. F. Viefers$^5$}

\affiliation{
$^1$Department of Physics Stockholm University,
AlbaNova University Center, 106 91 Stockholm, Sweden}
\affiliation{$^2$Nordita,
KTH Royal Institute of Technology and Stockholm University,
Roslagstullsbacken 23, SE-106 91 Stockholm, Sweden}
\affiliation {$^3$Institute for Theoretical Physics, University of Cologne, \\ 50937 Cologne, \\ Germany }
\affiliation {$^4$ Rudolf Peierls Centre for Theoretical Physics, University of Oxford, 1 Keble Road,  \\ Oxford OX1 3NP, UK}
\affiliation {$^5$ Department of Physics, University of Oslo, Box 1048 Blindern, \\ 0316 Oslo, \\ Norway}

\date{\today}

\begin{abstract} 

The fractional quantum Hall effect, being one of the most studied phenomena in condensed matter physics during the past thirty years, has generated many  groundbreaking new ideas and concepts.   Very early on it was realized that the zoo of emerging states of matter would need to be understood in a systematic manner.  The first attempts to do this, by Haldane and Halperin, set an agenda for further work which has continued to this day. Since that time the idea of hierarchies of quasiparticles condensing to form new states has been a pillar of our understanding of fractional quantum Hall physics.    
In the thirty years that have passed since then, a number of new directions of thought have advanced our understanding of fractional quantum Hall states, and have extended it in new and unexpected ways.   
Among these directions is the extensive use of topological quantum field theories and conformal field theories, the application of the ideas of  composite bosons and fermions,  and the study of nonabelian quantum Hall liquids. This article aims to present a comprehensive overview of this field, including the most recent developments. 

\end{abstract}

\maketitle
\tableofcontents

\section{Introduction}    

\subsection{Manifesto}
The  aim of this article is to give a systematic account of several of the main approaches to  quantum Hall physics. 
In particular we have aimed at a comprehensive review of  conformal field theory techniques and of the various approaches to quantum Hall hierachies. We describe the original ideas of Haldane and Halperin, the composite fermion approach, and the various field theory descriptions based on composite bosons or fermions. We also, for the first time, give a comprehensive review of the conformal field theory (CFT) approach to both abelian and nonabelian hierarchy states.

We have tried to make this paper self-contained.   We carefully introduce the main ideas and results relating to hierarchy states, and also
the CFT approach.   For brevity, the main text assumes a basic knowledge of quantum Hall physics.  For those needing a review of basics, a brief introduction is provided in appendix \ref{app:QHE}. We similarly assume some familiarity with a few concepts and techniques of CFT \cite{di1997conformal,belavin1984infinite}, so for the non-expert we provide a very brief introduction to these ideas in appendix \ref{app:CFT}. There are naturally several  comprehensive texts  that  cover part of the material in this review, and we would particularly mention \cite{jain2007composite,ezawa2008quantum,fradkin2013field}. 
 
\subsection{Brief historical introduction}
\label{ssec:history}
It is hard to overemphasize the importance of the discoveries of the integer \cite{klitzingdiscovery} and fractional \cite{tsuidiscovery} quantum Hall effects.  Experimentally, the integer quantization of the Hall conductance, while being crucial for modern metrology\footnote{The quantum Hall effect defines the standard of electrical resistance. The uncertainty of metrological quantum Hall measurements has been determined to be less than 1 part in $10^{10}$ despite the fact that the samples have substantial disorder. This is like measuring  the distance from Stockholm to San Francisco to within one millimeter!}\cite{weisklitzing},  is perhaps even more important for its theoretical implications.   The precise quantization of the integer quantized Hall conductance was initially understood as a reflection of gauge invariance \cite{laughlingauge,halperinedge} or as a measurement of the quantized electron charge.  However, soon thereafter it was realized that the quantized conductance could also be understood as a robust topological invariant \cite{TKNN,niu1985quantized},
which led to
 our modern understanding of topological insulators and superconductors \cite{TopInsulators}. 
Perhaps even more revolutionary from the theoretical point of view was the discovery of the  
fractional quantum Hall effect by \textcite{tsuidiscovery}, and the resulting realization that quantum liquids made of electrons could support excitations with fractional charge \cite{laughlintheory} and fractional statistics \cite{halperin1984statistics,arovas1984fractional}.  This led to the concept of topological order, which is central to the modern classification of phases of matter \cite{wenbook}. 

Topological order is a way to characterize phases of matter that cannot be distinguished by the pattern of spontaneous symmetry breaking, and the associated expectation values of local order parameter fields. 
A topologically ordered state of matter has an energy gap to bulk excitations, some of which carry fractional quantum numbers. In two spatial dimensions, the fractionalized particles are anyons, obeying fractional statistics that can be abelian or nonabelian.   The notion of nonabelian fractional statistics\footnote{The first theoretical works on nonabelian statistics of quasiparticles were by \textcite{Bais,witten,Gabbiani,Fredenhagen,moore1991nonabelions}, where the work of \textcite{moore1991nonabelions} first suggested that such statistics may occur in quantum Hall systems.} opened up the possibility of using nonabelian quasiparticles as a resource in quantum information processing \cite{nayakreview,kitaev2003}. 
When defined on a topologically nontrivial closed manifold, a topologically ordered state exhibits a  characteristic ground state degeneracy. 
On a manifold with boundaries, there are -- at least in the cases we will be concerned with -- gapless edge modes.\footnote{
This list of properties provides a working definition of topological order, but there are also formal definitions in terms of modular tensor chategories \cite{wangbook,kitaev2003,bondersonthesis}} 
The robust connection between the bulk topological order, and the existence and properties of  edge states is made most clearly through the connections forged by conformal field theory \cite{moore1991nonabelions}.

\begin{figure}
\begin{center}
\hspace*{0 pt}\includegraphics[width=\columnwidth]{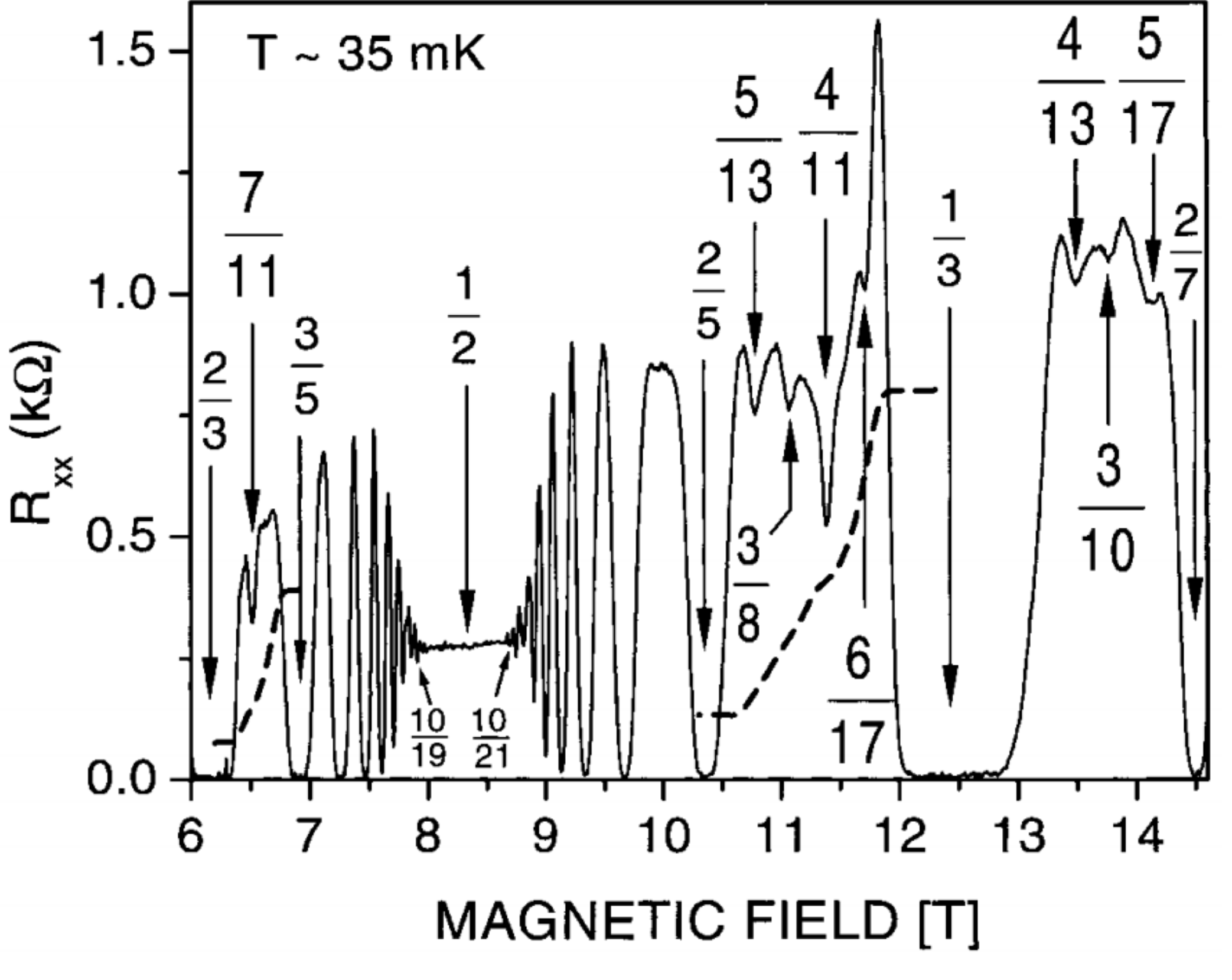}
\end{center}
\caption{Very high quality quantum Hall data from \textcite{pan2003fractional} showing many plateaus in the Lowest Landau level.}\label{LLL}
\end{figure}

\begin{figure}
	\begin{center}
		\includegraphics[width=\columnwidth]{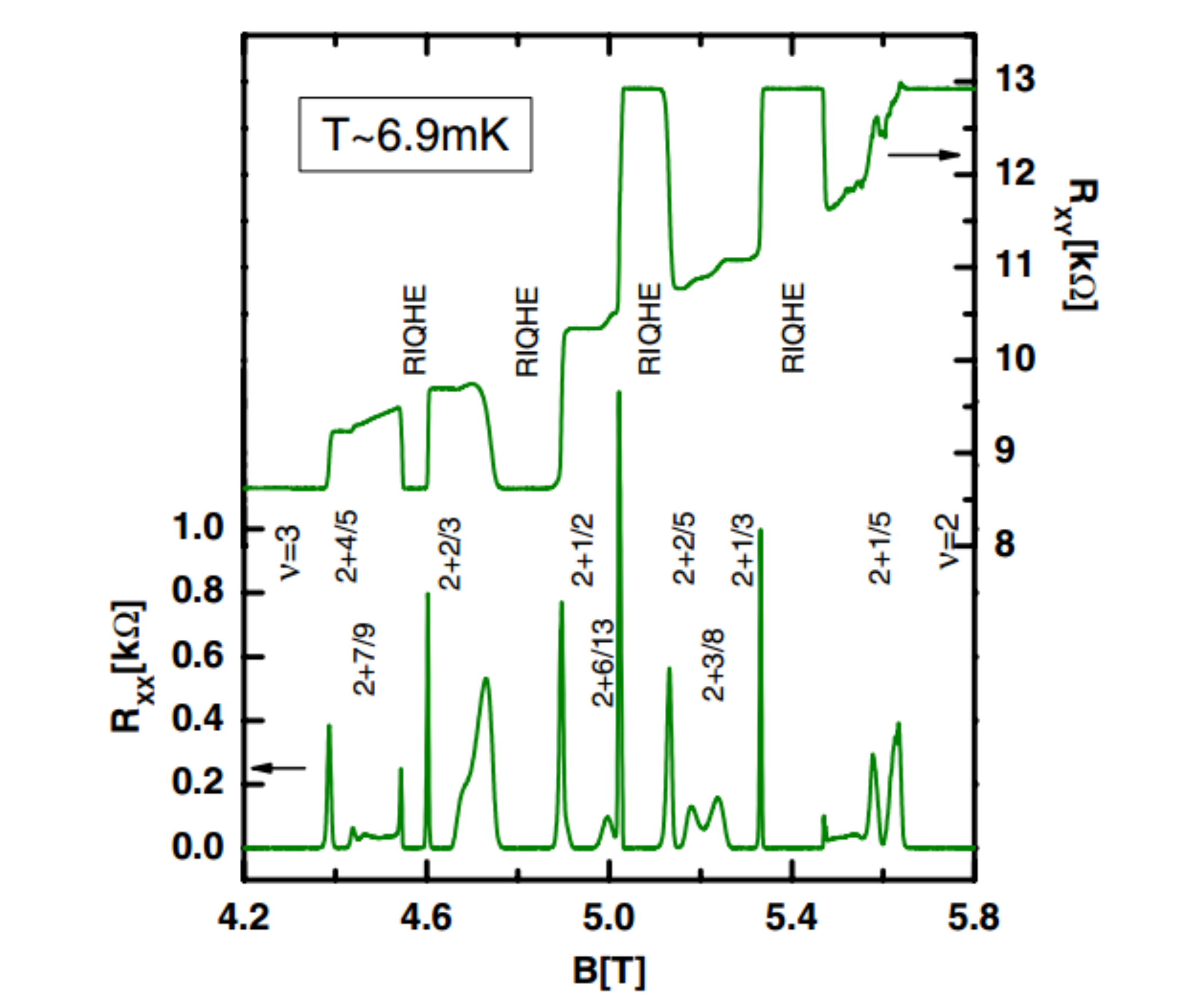}
	\end{center}
\caption{Very high quality quantum Hall data from \textcite{kumar2010nonconventional} showing plateaus in the second Landau level.   Note that many of the plateaus, labeled RIQHE (meaning Re-entrant Integer Quantum Hall Effect) are not fractional quantum Hall states but are believed to be some sort of charge density wave.
			}
\label{2ndLL}
\end{figure}

The first observed fractional quantum Hall liquid \cite{tsuidiscovery} was the state at filling fraction\footnote{The filling fraction $\nu$ is the number of complete Landau levels filled, see Appendix \ref{app:QHE}.}  $\nu=1/3$ which was soon understood by Laughlin's seminal theory \cite{laughlintheory}.  This theory predicted states of the form $\nu = 1/m$ with $m$ odd.  A few additional states were experimentally observed that could be understood as simple generalizations of this, including particle-hole conjugates of these simple Laughlin states and analogous fractions in partially filled higher Landau levels.   However, very soon thereafter, new states were found experimentally which did not fit this framework \cite{stormer2}. In particular, many fractions were observed in the  lowest Landau level (LLL), generally of the form $\nu = n/(2mn \pm 1)$ which are now known as ``Jain Fractions''  \cite{jain2007composite}.   
Additional fractions, not fitting this Jain form, were later discovered  in the LLL \cite{pan2003fractional} as well as in partially filled higher  Landau levels \cite{panLL2}. 
See Figures~\ref{LLL} and \ref{2ndLL} for high quality experimental data in the lowest and second (or `first excited')  Landau levels.

Clearly it was necessary to find an organizing principle in order to understand the plethora of experimentally observed states. The first idea of this kind was due to \textcite{haldane1983fractional} and  \textcite{halperin1983theory,halperin1984statistics},  who suggested that the states in the LLL are hierarchically ordered. Moving away from the centre of a quantum Hall plateau by changing the filling fraction (by changing magnetic field or electron density) amounts to creating quasielectrons or quasiholes. These are then assumed to condense in a Laughlin-like state, thus forming a daughter state. In this way one can obtain any fraction with odd integer denominator. 
A further major step was taken by Jain \cite{jain1989composite,jain1990theory,
jain2007composite,heinonenbook} who 
constructed  trial wave functions for the fractional quantum  Hall effect at  $\nu = q/(2qp \pm 1)$ as being  the integer quantum Hall effect  of ``composite fermions". 
The latter can roughly be thought of as electrons with  $2p$ quanta of vorticity (or ``flux") attached,  which fill  $q$  effective Landau levels.
These hierarchies (and other generalizations of the same ideas) are a main focus of this article. 

The theoretical understanding of the fractional quantum Hall effect has progressed on several fronts.    The first approach, starting with the seminal work of Laughlin, and later that of Jain,  has come from the analysis of cleverly guessed, and  numerically highly accurate,  trial wave functions.   Later, conformal field theory (CFT) was used to engineer model wave functions with interesting properties \cite{moore1991nonabelions}, which allowed the first construction of nonabelian quantum Hall states such as the Moore-Read \cite{moore1991nonabelions}, Read-Rezayi \cite{read1999beyond}, and nonabelian spin singlet \cite{ardonne1999new} states.   A second front of attack was the development of effective field theories.  These are of two different types. The first amounts to a rewriting of the original microscopic field theory of electrons moving in a strong magnetic field, as a field theory of composite bosons in zero field \cite{zhang1989effective}, or composite fermions in a weaker magnetic field \cite{lopez1991fractional}.  
In principle these theories  describe the microscopic physics of 
abelian quantum Hall states, but they can only be solved using mean-field methods.   The second type of field theories are topological quantum field theories (TQFT)
 based on Chern-Simons gauge fields. These theories can be formulated for both abelian and nonabelian states, but, as the name indicates,  only encode topological information,  such as filling fractions and charge and statistics of the quasiparticles \cite{nayakreview,wen1995topological}. 

Connections between TQFTs and CFTs have been made extensively throughout the literature, starting with the pioneering work of Witten \cite{witten}. These same connections are extremely powerful in the quantum Hall context. The CFT approach \cite{moore1991nonabelions} gives a description of the dynamical (1+1 dimensional) theory of the quantum Hall edge \cite{KaneFisher,WenEdge} and at the same time gives an explicit wave function for the (2+0 dimensional) quantum Hall bulk as a correlator of certain operators in the CFT.  The primary operators of the CFT then define the fields of the corresponding TQFT, and hence define the full statistics of quasiparticles present as excitations above the ground state. 

Perhaps surprisingly, it has been hard to fit the prominently observed abelian hierarchy states in the LLL into the CFT framework. The effective Chern-Simons description of  these states is well developed, defining the universal properties of the edge CFT as well, but most work on trial wave functions has been in terms of composite fermions, with no obvious connection to a hierarchy formed by successive condensations of quasiparticles. This situation has  changed over the past decade with the realization that CFT techniques can be used to construct representative wave functions for any state in both abelian and nonabelian hierarchies. Also, using these methods, the composite fermion wave functions can  be written as condensates of quasiparticles of  a parent state, and thus belong to the Haldane-Halperin hierarchy as well. A main objective of this article is to review these advances. 


\subsection{More detailed aims and organization of this article}

In most of this review we will focus our attention on the simplest cases where there is a single, partially filled, Landau level with only one species of fermion (or boson).   It should be noted, however, that more complicated situations may be considered where each particle carries an additional nontrivial quantum number. This has been explored in many experiments  -- for example, for a spin-unpolarized Landau level one must keep track of both spin species \cite{du1995fractional}.  Similarly in bilayer quantum Hall effect \cite{EisensteinBilayer}, the layer index may play the role of  a pseudospin.  The situation may be more complicated still in multi-valley semiconductors such as graphene \cite{dean2011multicomponent} where one may have to keep track of both spin and valley indices.

This paper is organized as follows. In section \ref{sect:hierarchy} we introduce the concept of quantum Hall hierarchies and then give a short review of the various theoretical approaches that have been employed to describe them. In section \ref{sec:experiment} we discuss the current experimental status and  relation the proposed theoretical approaches, and in section \ref{sect:CSCFT} we review some important  theoretical results with bearing on the connection between Chern-Simons theory, conformal field theory and quantum Hall  physics. The next two sections describe how to obtain explicit  wave functions for abelian and nonabelian hierarchies respectively. In particular we explain the need for ``quasi-local" CFT operators for describing quasiparticles, and give a fairly detailed discussion of various approaches to nonabelian hierarchies.  In section \ref{othergeom} we generalize the construction of wave functions to the sphere and the torus, and explain why this is of interest. We conclude with a very brief summary and outlook.  

Several appendices are included for those needing a bit more background.  In Appendix \ref{app:QHE} some basic facts about quantum Hall physics are elaborated.  Appendix \ref{app:CFT} gives basic facts about conformal field theory.   Appendix \ref{sec:commutator} gives some details of derivations for one-dimensional edge theories.


\subsection{Notation and conventions} 
For  electron coordinates we shall use $\vec r = (x,y)$, $z = x+iy$ and $\bar z = x -iy$, and similarly $\vec\eta$, $\eta$ and $\bar\eta$ for quasiparticle coordinates. 
We sometimes use the abbreviation $\grupp [z] a b $ for the set $(z_a,z_{a+1} \dots z_b)$ {\it etc.}, and just $\sgrupp [z]$ when the label $i$ runs from 1 to $N$.

The term ``quasiparticle" will refer to either a quasielectron or a quasihole.  
Up to a gaussian factor, electronic wave functions will be holomorphic functions of the $z_i$'s only and depend parametrically on the quasihole 
coordinates $\eta_i$ and the  quasielectron coordinates  $\bar\eta_i$ (see Appendix \ref{app:QHE}). Note however, that at an intermediate step, the hierarchy wave functions will be constructed as to  have anti-holomorphic components which will either be integrated out or  projected onto the LLL. For sake of brevity, we will often omit the gaussian factors. 

Late Greek letters, $\mu,\ \nu,\ \sigma \dots$ will be used for Lorentz indices, and early Greek letters  $\alpha,\ \beta,\ \gamma \dots$, to label  different CS gauge fields. In most of the paper we shall use a radial gauge $\vec A = (B/2) (-y, x)$, where a single electron wave function in the LLL is $\sim f(z) \exp[- |z|^2/(4\ell^2)]$, and often we will put the magnetic length $\ell^2 = \hbar c /eB = 1$. The area element on the plane is denoted by $d^2r = dxdy = d^2 z$. We  will always put $c=1$,  often $\hbar = 1$, and always assume zero temperature.


\section {Quantum Hall hierarchies}   \label{sect:hierarchy}

\subsection{Laughlin states -- plasma analogy and quasiparticles} \label{sub:plasma1}

In his  original paper  \textcite{laughlintheory} argued  that  his wave function describes an incompressible liquid, and that the quasihole excitations are fractionally charged.   Soon thereafter it was shown that the quasiholes are anyons, \ie particles obeying fractional quantum statistics \cite{leinaas1977theory, wilczek1982quantum}.  All these insights were based on using the so-called plasma analogy\footnote{It is the common nomenclature to call it an `analogy', even though it is in fact a precise mapping. }, which we here briefly review using  a modified version of the original argument given by  \textcite{arovas1984fractional,halperin1984statistics}.  More details are presented in section \ref{subsub:mon} below.

The Laughlin wave function \cite{laughlintheory} at filling fraction $\nu=1/m$ with two quasiholes at positions $\eta_1$ and $\eta_2$ is given by
\be{laughlinholes}
\Psi(\eta_1,\eta_2 ; z_1\dots z_N)& =& N(\eta_1,\eta_2 ) \prod_{i=1}^N  (\eta_1 - z_i) (\eta_2 - z_i),  \nonumber  \\
&\times& \prod_{i<j} \jasf i j m e^{-\sum_i  |z_i|^2/(4\ell^2)} 
\ee
where $N(\eta_1,\eta_2)$ is a normalization constant. 
 By writing
\be{laughnorm}
\label{eq:Ngauge2}
N(\eta_1,\eta_2 ) = \tilde N |\eta_1 - \eta_2|^{\frac 1 m} e^{-\frac 1 {4 m \ell^2} (|\eta_1|^2 + |\eta_2|^2)}  \, ,
\ee
$|\Psi|^2$ can be related to the Boltzmann  factor  $e^{-\beta U}$ of a two-dimensional Coulomb plasma, meaning that two point charges, $q_1$ and $q_2$ repel each other by the potential $q_1 q_2 \ln |\vec r_1-\vec r_2|^2$. The corresponding partition function is
$$
 Z(\eta_1,\eta_2)  =  \prod_{i=1}^N \int d^2 z_i\, |\Psi|^2  =      |\tilde N|^2 \prod_{i=1}^N \int d^2 z_i\, e^{-\beta U}  \, .
$$
The inverse temperature of this analogue plasma is $\beta=1/m$, the charge of the $z$ particles is  $m$, the charge of the $\eta$ particles is unity, and there is a quadratic confining background potential $m |r^2|/(4\ell^2)$ corresponding to a constant background charge density  $2\pi \rho_{pl}= eB$.  
The crucial observation by \textcite{laughlintheory} was that  the classical plasma is in a screening phase, at least for $m < 70$.

From this follows a number of  important results.   First, in the absence of the quasiholes,  a uniform density of $z$-charges would (by Gauss' law) precisely screen the constant background charge. Thus, the wave function describes a  system with uniform charge density, and since the plasmon in this system is gapped, the long range density fluctuations are negligible and the quantum liquid is incompressible.  
Second, it is easy to see that the electrical charge associated with $\eta$ must be exactly $-1/m$ of the charge of an electron --- and is spread over a distance given by the screening length of the plasma (on the order of $\ell$). 
 An important point is that these fractional charges are sharp in the sense that properly defined, the variance of their charge is essentially zero \cite{kivelson1982fractional} as discussed by  \textcite{kjoensberg1999charge}.
Third, the partition function $Z$ is independent of the positions of the charges,  and thus $|\tilde N|^2$ is a constant independent of $\eta_1$ and $\eta_2$ as long as they are far separated compared to the magnetic length \cite{halperin1984statistics}. From the properly normalized wave function, one can verify that these particles obey fractional exchange statistics, as is explained below.

A heuristic argument for the exchange phase was first given by \textcite{halperin1984statistics}, arguing that removing the absolute value of the relative factors of Eq.~\eqref{eq:Ngauge2} amounts to a singular gauge transformation that leaves the electronic wave function `unchanged', but allows us to interpret  Eq.~\eqref{laughlinholes} as the proper wave function for anyonic particles in the LLL with exchange phase $\pm \pi/m$. 
Alternatively, one can compute the Berry phase factor  $e^{i \gamma}$, or ``holonomy'', associated with moving one quasiparticle at $\eta_1$ adiabatically around another at $\eta_2$.  
The general expression for the Berry phase is,
\begin{equation}
\label{eq:berry1}
i \gamma = \oint d\tau \left\langle \Psi(\eta_1(\tau), \eta_2) \left| \frac{d}{d \tau} \right| \Psi(\eta_1(\tau),\eta_2)\right\rangle \, ,
\end{equation}
where $\tau$ is a parameter that takes the particle $\eta_1$ once around $\eta_2$ along some closed loop.    
Such an integral was first considered by \textcite{arovas1984fractional}, who argued that the presence of a quasiparticle within the loop gives an additional $\pm 2 \pi/m$ contribution to the Berry phase. Thus the Berry phase can be written as 
\begin{equation}
\label{eq:gammares}
\gamma =   \frac{2\pi}{m} \, \frac{\Phi}{\phi_0} + 2 \frac  \pi m \, ,
\end{equation}
where $\phi_0 = h/e = 2\pi/e$ is the unit quantum of flux, and $\Phi$ is the flux enclosed  by the  loop. Thus the first term gives the usual Aharonov-Bohm phase for a particle of charge $e^\star = e/m$ moving in a magnetic field, while the second term is twice the fractional exchange statistics phase for anyons with statistical angle $\theta = \pi/m$.  For more details on this calculation we refer the reader to \textcite{stonebook}. We will present an alternative derivation of the Berry phase in section \ref{qp&mon}, which is based on the particular properties of the wave functions when written in terms of CFT correlators.

A more detailed derivation of this result was done by \textcite{kjonsberg1997anyon}, who used the plasma analogy to map the expressions to the two-anyon system. 
This analytically derived result has been confirmed to be true numerically \cite{kjoensberg1999numerical,zaletel2012exact}, so long as the quasiholes remain sufficiently far (on the order of $\ell$) from each other.  
While there is no plasma analogy for handling quasielectrons (rather than quasiholes), it is often assumed that putting the  factor  $\prod_{i=1}^N (2 \partial_{z_i} - \bar \eta)$, proposed by \textcite{laughlintheory}, in front of the Laughlin wave function would amount to placing a quasielectron at position $\vec\eta$. This description is not unproblematic; 
we will discuss quasielectrons in  detail in section \ref{ssec:qe}.

Laughlin's 1981 argument for the quantized conductance in case of the integer quantum Hall effect was based on adiabatic flux insertion in an annulus (Corbino) geometry where the edges are held at a fixed voltage difference $V$. Each inserted flux quantum can be shown to effectively move one unit of charge from one edge to the other giving an energy shift $\Delta E = eV$. On the other hand the Hall current is given by $I =\Delta E/\phi_0$,  so $I = (e^2/h) V = \sigma_0 V$\cite{laughlingauge}.  
Having established the presence of charge $e/m$ quasiparticles, \textcite{laughlintheory} used the same argument to show that for the fractional effect the Hall conductance is $\sigma_0/m$.

\begin{figure}[htbp]
\begin{center}
\includegraphics[width=\columnwidth]{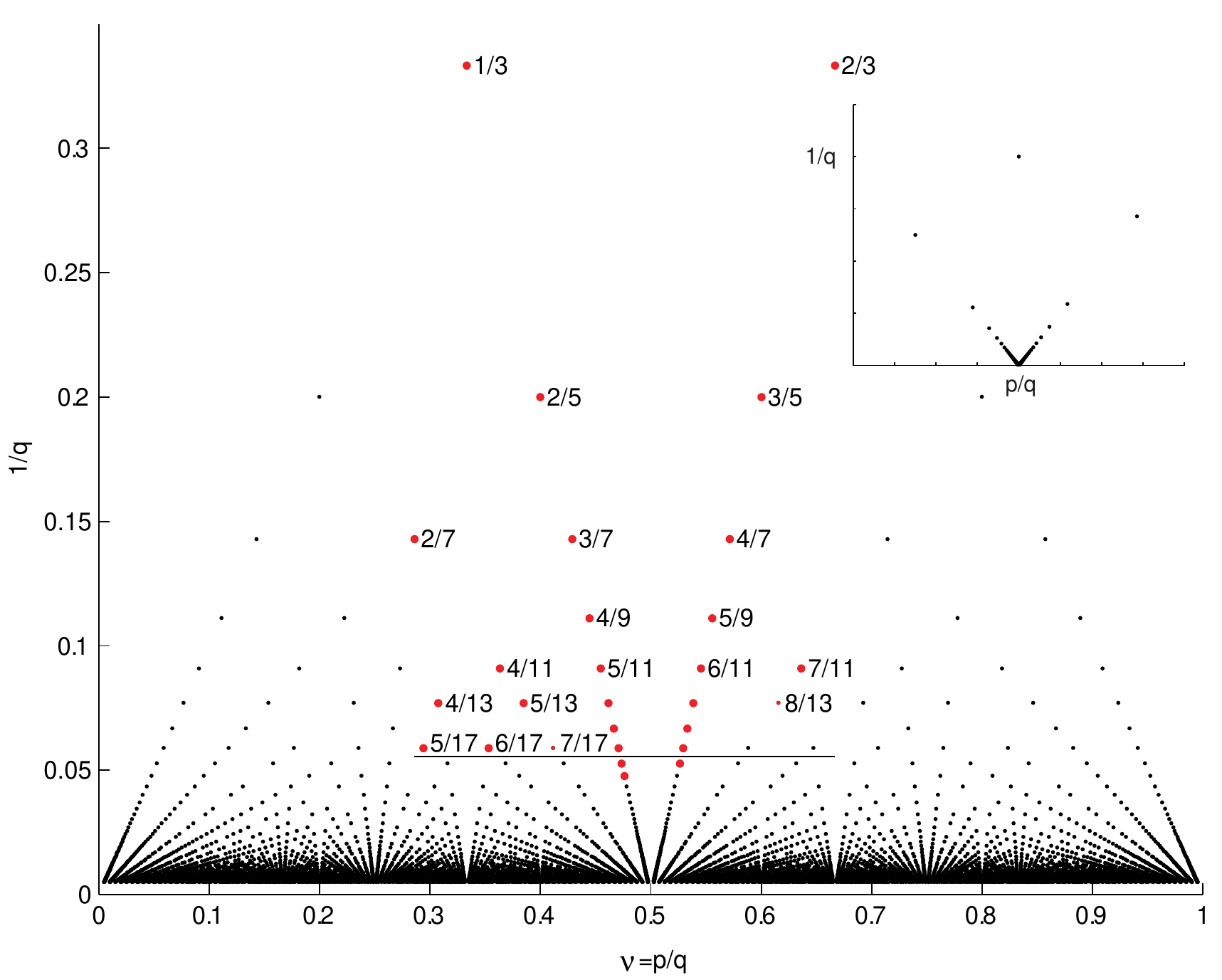}
\end{center}
\caption{Observed states and fractal structure.
Quantum Hall  hierarchy states in the lowest Landau 
and their relative stability. For each rational, $\nu=p/q<1, \, q$ odd, 
there is a unique  state and its stability increases monotonically with $1/q$. 
Red dots denote observed states in the region $2/7\le \nu  \le 2/3$ \cite{pan2003fractional}. The horizontal line marks the extent in $\nu$ 
of the experiment and is a line of constant gap. This line provides a good (though not perfect) approximate boundary for the observed states.   
The inset shows the  structure of hierarchy states: At each $\nu=p/q$, $q$ odd, there is a state with gap $\sim 1/q$ and 
quasiparticles with charge $\pm e/q$. When these condense, two sequences of states approaching $p/q$ with decreasing gap are obtained.
Figure taken from \textcite{bergholtz2007microscopic}.}
\label{figure1}
\end{figure}

The above argument can in fact be used in ``reverse" to establish that a  quasihole created by adiabatic  insertion of a unit flux in {\it any} incompressible quantum Hall state with filling fraction $\nu$ has a fractional charge $e^\star = \nu e$ \cite{karlhede1992quantum}. The argument, which does not invoke the plasma analogy, goes as follows: 
Imagine inserting a thin (radius $\ll \ell$) solenoid, and then adiabatically turning on a flux $\Phi (t)$. This will, by Faraday's law, induce an azimuthal electric field $E_\alpha = (\partial_t \Phi(t))/2\pi R$, where $R$ is the distance from the solenoid. Because of the quantum Hall response $\sigma_H= \nu e^2/h$
this results in a radial current density $j_r = \sigma_H E_\alpha$. So during a process which introduces one quantum of flux, $\phi_0 = h/e$, the charge transported through the circle at $R$ is $e^\star = \int dt\,  2\pi R j_r (R) = \sigma_H   \phi_0 = \nu e $.    One can now invoke the Byers-Yang theorem which states that if a system consists entirely of particles with charge $e$, there is no physical consequence of having an integer number of elementary fluxes $h/e$ added through a puncture in the plane \cite{byers1961theoretical}. This is precisely the configuration we end up in after inserting the flux, so without changing any physical properties of the system, it can be gauged away  leaving  a local charge $Q =  \nu e$ excitation. We do not know of any  similarly clear and general argument for the statistics of the Laughlin quasiholes. 

\subsection{The Haldane-Halperin idea}  \label{ssec:hierarchy}
If only a few quasiparticles are present, they will be pinned to impurities and remain inert, so that the system will have the same conduction properties as if the quasiparticles were not present (indeed, this is the origin of the finite width of the quantum Hall plateau). If many quasiparticles are present, and are unpinned, they will form a system of itinerant  charged particles in a magnetic field, which, at an appropriate filling fraction, is likely 
to condense into a Laughlin-like state. This is the heuristic picture behind the 
 original proposal of a hierarchy   due to  \textcite{haldane1983fractional} and   \textcite{halperin1983theory,halperin1984statistics}. Given this idea, one can show that any filling fraction with an odd denominator can be obtained by successive condensations of quasiparticles into Laughlin-like states (see Sections \ref{ss:full} and \ref{sec:cfthier}).
 Since the charge of the elementary quasiparticle in a state $\nu = p/q$ is $\pm e/q$, the Coulomb gap is expected to be a decreasing function of $q$, and thus one expects states to be increasingly fragile, and thus harder to observe, as $q$ increases. This tendency is clearly seen in experiments as illustrated in Fig. \ref{figure1}.

 
\subsubsection {General form of an abelian hierarchy wave function} \label{sec:genform}

 Following Halperin, we shall refer to the wave functions for the quasiparticles,  $\Phi(\vec \eta_1 \dots \vec \eta_M ) $,  as a {\it pseudo wave function}, and write a general hierarchy state at level $n+1$ as 
\be{hiwf}
\Psi_{n+1} ( \vec r_1\dots \vec r_{N}) &=&   \int  d^2\vec \eta_1 \dots   \int  d^2\vec \eta_M    \, \Phi_n^\star(\vec \eta_1 \dots \vec \eta_M )  \nonumber\\
&\times&  \Psi_n (\vec \eta_1 \dots \vec \eta_M ; \vec r_1 \dots \vec r_{N})  \, ,
\ee
 where $\Psi_n$ is a state with $M$ identical quasiparticle excitations in a parent state at level $n$. 
Recall that throughout this paper we shall use $\vec r_i$ or $z_i$ for the position of the $i^{th}$ electron, and $\vec \eta_i$ for the $i^{th}$ quasiparticle.

The reader should be aware that the hierarchical constructions of Haldane and Halperin are slightly different. For instance, Halperin obtains the filling fraction of $3/7$ by condensing quasielectrons of the $2/5$ state into a Laughlin 1/2 state, while Haldane condenses quasiparticles of the $1/3$ state into a $2/3$ quantum Hall state (the bosonic version of 2/5). Also their arguments on how to determine the possible filling fractions differ. Despite these differences, the two constructions yield the same filling fractions with identical topological properties.
In this manuscript, we will mostly follow Halperin's construction; in particular, the pseudo wave function in Eq.~\eqref{hiwf} will always be of Laughlin form. 
Below we exemplify Haldane's construction by giving a heuristic picture based on \cite{haldane1983fractional} on how to determine the filling fraction and quasiparticle properties of a hierarchical daughter state.

Assume that $M$ particles at positions $\xi_i$ have condensed to form a parent quantum Hall state.  (Note that these particles can be either electrons, or  the quasielectrons or quasiholes of some grandparent quantum Hall state).   A quasihole in this state   at position $\vec\eta$ amounts to having a factor  $  \prod_{i=1}^M (\xi_i - \eta) $   and for a quasielectron we might take  $\prod_{i=1}^M (2 \partial_{\xi_i} - \bar \eta) $.  
 Effectively this means that each quasihole at position $\eta $ ``sees" each of the particles
$\xi_i$  as if it were a single quantum of flux.\footnote{
What we precisely mean here is that the wave function for the $\eta$  particle
is an analytic polynomial whose overall degree is $M$ --- this is exactly of the same
form as we would have for a unit charge particle in a magnetic flux of $M\phi_0$.
It is tempting to imagine that the $\eta$ particles ``see" a flux quantum at the position of each of the
$\xi$ particles.   This is not correct, as a real flux quantum would just imply a phase winding ({\it i.e.}, it is
a singular gauge transformation).     It is more correct to think of the $\xi$ particles as having captured
the vortices of the wave functions for the $\eta$ particles.  At a ``mean field" level, one can heuristically
think of the $\xi$ particles as providing a uniform magnetic field that the $\eta$ particles experience, see 
sections \ref{sub:compbosons} and \ref{sect:CF}. }
We can now form multi-quasiparticle states simply by taking a product of
such factors, and if we use the normalization \eqref{laughnorm} the wave
function is symmetric in the $\eta_i$ (or the $\bar\eta_i$ in case of quasielectrons), so the quasiparticles are bosons. In order for
the $\eta$ particles to condense they must see enough total flux to form a quantum Hall state.
Since the quasiholes  are bosons in this description, we must have an  effective filling fraction $\nu^{\rm eff}=1/p$ with $p$ even, meaning that there must be $p$ times as many particles in the parent layer of the hierarchy as in the daughter.

Let us exemplify the hierarchical construction by  starting from $\nu=1/3$, with $N$ electrons in  $3 N$ flux, and then adding an extra $N/2$  flux. This creates $N/2$ (bosonic) quasiholes, each experiencing one flux quantum per each of the original electrons, hence $N$ fluxes.  These $N/2$  quasiholes in  $N$ flux can condense in  a $\nu=1/2$ Laughlin liquid of bosons.  We conclude that this first level daughter state would have  $N$ electrons in   $3 N + N/2$ flux,  or a filling fraction  $\nu=1/(3 + 1/2) = 2/7$.

To determine the charge of the fundamental quasihole in this state we follow \textcite{HaldaneGirvin}, and imagine removing one electron at fixed magnetic field.  This creates three quasiholes of the original electron Laughlin state, leaving us with $N/2 + 3$ bosons in $N-1$ flux. 
 We must now figure out how many defects (quasiparticles) we obtain when we have $N/2+3$ bosons in $N-1$ flux at filling fraction $\nu=1/2$.  Since there should be twice the number of flux quanta as bosons in the ``ground'' state of such a system, this would be $2(N/2+3) = N+6$ flux.  With $N-1$ flux, we are clearly $7$ flux quanta short, and therefore there are seven  quasiparticles present.   Going back to our original system where we removed a charge of $-e$, we conclude that the elementary  quasihole charge is $e/7$.  
Note that there are no rigorous arguments for deducing the fractional statistics of such quasiholes. 
However, Wen's $K$-matrix expression in section \ref{ssec:effectiveCS} predicts a statistical phase $-3\pi /7$ for the elementary quasihole $e/7$. Moreover, it has been shown that predictions based on clustering arguments as invoked by \textcite{su1986statistics} are generally consistent with those from the $K$-matrix formalism \cite{fulsebakke}.

\textcite{read1990excitation} has suggested an alternative way to extract the topological content of hierarchy wave functions written as in Eq.~\eqref{hiwf}. The basic assumption is that two such wave functions are orthogonal whenever the quasiparticle positions differ much more than a magnetic length. Using this orthogonality postulate the expectation values of operators, obtained by integrating over the electron coordinates, simplify to integrals that can be reinterpreted in terms of a multi-component plasma, and arguments similar to the ones given above for the Laughlin case will apply. We will discuss the status of the plasma analogy for hierarchy states further in section \ref{hiplasma}.  

The hierarchy argument can be continued to  higher levels to  give  filling fractions of the continued fraction form \cite{haldane1983fractional,halperin1984statistics}, 
\begin{align}
\label{eq:fillingfraction}
\nu = \cfrac{1}{m - \cfrac{1}{\sigma_1 p_1 - \cfrac{1}{\sigma_2 p_2 -   \cfrac{1}  {\cfrac{\ddots}{~~~~~ - \frac{1}{\sigma_{n-1}p_{n-1}}}}}}}   
\end{align}
where $1/m$ with $m$ odd is the filling fraction of the $1^{st}$ level state, $1/p_\alpha$ with $p_\alpha$ even is the filling fraction of the $\alpha^{th}$ level daughter bosonic condensate of quasiparticles and  $\sigma_j=\pm 1$ indicates whether this condensate is formed by quasiholes ($-$) or quasielectrons ($+$).   Note that with these conventions the Laughlin states are at level $n=1$, and a general level $n$ state is formed by $n-1$ condensates determined by $p_\alpha, \ \ \alpha = 1, 2 \dots n-1$.

An often aired criticism of the hierarchy scheme is that it is based on a quasiparticle picture, which is only valid when the separation between quasiparticles is large compared to their radius. However, condensation into a daughter state occurs precisely when the quasiparticles are overlapping, and hence where a quasiparticle description loses its integrity. Thus the very notion of condensation is questionable \cite{jain2007composite}. We believe that this point of view is based on demanding more than what can be expected from a `picture'; the real content of the hierarchy (in the Halperin sense) is embodied in Eq.~\eqref{hiwf}, as well as in the general arguments about the relative stability and the properties of their excitations, see e.g. \cite{halperin1984statistics}. The description in terms of condensation is only a (very suggestive!) picture. 
	It is  worth mentioning here that also ordinary electrons in the LLL  have a finite size ($\sim \ell^2$) and that they  condense in a Laughlin state precisely when the distance between them becomes comparable with this length scale. Of course, electrons are in a sense ``real'' particles, but it is far from clear that this distinction should be very important.\footnote{This is commented upon in 
section \ref{nonlocalop}.}


\subsubsection {Difficulties, and early numerical work }     \label{sub:difficulties}

Although it is in principle straightforward to generate hierarchy wave functions using Eq.~\eqref{hiwf}, in practice this formula turns out to be very difficult to work with. As a concrete example, consider the very simplest case of a quasihole condensate in a $\nu = 1/3$ Laughlin state. 
The corresponding quasihole wave function is the obvious generalization of \pref{laughlinholes} with the normalization factor $\prod_{\alpha < \beta}^M  |\eta_\alpha - \eta_\beta|^{1/3} $, which assures that the  normalization, up to a phase, is independent of the positions  $\eta_i$, as long as these are far separated. 
The pseudo wave function is  taken as
$
  \Phi= \left[ \prod_{\alpha<\beta}^N (\eta_\alpha - \eta_\beta)^2  \right] e^{-\frac{1}{3} \sum_\alpha |\eta_\alpha|^2/(4\ell^2)}
$, 
which is a  Laughlin state treating the quasiholes at positions $\eta_\alpha$ as bosons \cite{halperin1984statistics,haldane1983fractional}, and the factor of 1/3 in the gaussian exponent accounts for the  quasiholes having charge 1/3, and hence a longer magnetic length than the electrons.   
Choosing $M = N/2$ and using the above expressions in  the wave function Eq.~\eqref{hiwf} gives a quantum Hall wave function at $\nu = 2/7$\footnote{Note that at this density, there is no reason to believe that this wave function is correctly normalized. 
As already mentioned in section \ref{sec:genform}, the real meaning of ``condensation'' is embodied in Eq. \eqref{hiwf}, and here we note that it must also come with a definite prescription for how to construct the pseudo wave function for the condensing particles.}:
\be{eq:hier2/7}
		   \Psi &=& \int \prod_{\alpha=1}^M  d^2 \eta_\alpha e^{-\frac{1}{3} \sum_\alpha |\eta_\alpha|^2/(4\ell^2) }  \prod_{\alpha < \beta}^M  (\bar\eta_\alpha -\bar \eta_\beta)^2  |\eta_\alpha - \eta_\beta|^{1/3} \nonumber \\
		   &\times&  \left[ \prod_{\alpha=1}^M \prod_{i=1}^N   \rule[0pt]{0pt}{20pt} (\eta_\alpha - z_i) \prod_{i<j}^N (z_i - z_j)^3  \right] e^{-\sum_i  |z_i|^2/4\ell^2}\, .
   \ee
Even for this simplest case of a hierarchy wave function, there is unfortunately no known way to evaluate the  integrals  analytically, and numerical methods are feasible only for a small number of particles. 
Such calculations were done early on by \textcite{greiter1994microscopic}, who compared hierarchy states\footnote{
 In order to evaluate the expressions \eqref{eq:hier2/7}, Greiter omitted the factor  $| \eta - \eta |^{1/q}$, which amounts to   a  change in the short distance behavior thought to be physically unimportant, \cf section \ref{quasilocal}. } 
 with results of exact diagonalization; mainly at level 2 (\ie first level daughter) for up to 8 particles, but also level 3 for up to 6 particles.
The overlaps for these small systems are very good,  but it is clearly questionable to draw conclusions from studying a level 3 hierarchy state of only 6 electrons.  

The Laughlin wave function is the exact ground state wave function for a certain ultrashort-ranged electron-electron interaction \cite{haldane1983fractional,trugman,Pokrovsky} which makes it very amenable to certain analytic approaches.  Similarly, some of the more exotic wave functions such as the Moore-Read wave function \cite{moore1991nonabelions} are also the exact ground states of simple interactions \cite{ReadRezayi1996,greiterwenwilczek}.   Unfortunately, there is no known electron-electron interaction for which any of the hierachy wave functions (those discussed here or below) are exact ground states, and this substantially complicates the  detailed analysis of their properties.

If we relax the condition that the wave function should reside in the LLL, there are potentials for which the \emph{unprojected} Jain state\footnote{Jain's composite fermion states will be introduced in more detail in section \ref{sect:CF}.} 
at $\nu=2/5$  is the ground state \cite{jain1990scaling}. 
This might have been a satisfactory situation if components in higher Landau levels had been relatively small, but in reality this is not the case.\footnote{Already for relatively small systems, such as $6$ particles in the (bosonic) $\nu=2/3$ Jain state on the torus, only about 32\% of the unprojected state reside within the LLL, and this percentage  decreases rapidly with increasing system size. Similar numbers for the LLL weight have been reported for the spherical geometry, both for bosonic and fermionic Jain states (N. Regnault, private communication).} 
Another argument for restricting the discussion to within a single Landau level is that the large $B$ (or small $m_e$) limit provides a theoretically well-defined limit that is believed to capture the essence of quantum Hall physics.  So it seems like a good strategy to first try to solve that problem, and only later include effects of higher Landau levels  (See the discussion at the end of section \ref{sub:connection}).

\subsubsection{{Why wave functions, and which?}}
At this point it is fair to ask why we are at all interested in constructing explicit many body wave functions. None of the model wave functions considered here are the exact ground states of any experimentally relevant Hamiltonian, and the overlap with the realistic ground state will always be zero in the thermodynamic limit. In addition, one may ask whether model wave functions can tell us anything beyond what we can deduce from the  effective theories that are discussed in the following sections. 

First we should make clear that the wave functions we are considering in this review are ``representative wave functions" in the sense of Moore and Read.\footnote{As far as we know this concept was first introduced in \cite{moore1991nonabelions}.} 
This means that even though they might not be eigenstates of any physical Hamiltonian, they have the topological  characteristics of some  distinct phase of matter. These are typically the charge and statistics of quasiparticles, as well as properties of edge states and the ground state degeneracy on topologically nontrivial manifolds. It is these topological features we are interested in, rather than microscopic details of the actual wave functions. 
This, however, still begs the question of relevance --- why are we not satisfied with an effective field theory description, since they are after all constructed precisely in order to capture the topological characteristics  of a phase. There are at least three answers to this. 

The first, which was already stressed in section \ref{sub:plasma1}, is that in many cases the construction of wave functions has preceded the effective field theories. This was true for both the Laughlin states, the Jain states, and the states in the Read-Rezayi series. In fact, for the latter there is to this day no generally accepted low energy field theory. Thus, constructing explicit model wave functions has been an important step in understanding many topological phases, and this might well continue to be true in the future.

Secondly, one should be aware that  the effective field theories can rarely be \emph{derived} from the microscopic physics. In fact, even with rather low standards of rigor, this can only be done for the Laughlin \cite{PhysRevLett.66.1220} and  the Jain states \cite{lopezfradkinwavefunction1992}. 
Thus it is not at all guaranteed that given an effective field theory, there will be \emph{any} microscopic wave function with the same topological properties. In this case the construction of trial wave functions provides a sanity check of the predictions based on the effective field theory, provided one can determine the topological properties, either by analytical or numerical methods\footnote{Note that some  properties, such as the edge theory or the topological ground state degeneracy, cannot be probed using the sphere geometry, so constructing model wave functions on different geometries is an important aspect in numerics.}. The latter has often proven hard, especially when dealing with composite fermions. Here, the CFT methods detailed in later sections have proven useful, as their direct connection to TQFTs allows us to conjecture their topological properties. In this sense, rewriting the composite fermion wave functions using CFT is more than just an academic exercise, but allows us more direct access to the important properties of the representative wave functions.

Thirdly, as will be discussed many times later,  explicit wave functions for a small number of particles can be compared with wave functions obtained by numerically solving the Schr\"odinger equation for a realistic interaction. 
Often good agreements (typically as measured by overlaps of wave functions) are taken as a proof of the validity of the model wave function, but here we should add a note of caution. First, it is clear that \emph{any} conceivable model wave function will have zero overlap with a  numerically obtained one in the thermodynamic limit.\footnote{An amusing exception to the rule that overlaps decrease with increasing system size is the composite fermion trial wave function for the so-called single vortex state of bosons in the LLL with contact interaction. Its overlap with the exact ground state {\it increases} with the number of particles and approaches 1 in the limit $N \rightarrow \infty$ \cite{korslund}.}
Moreover, for a too small number of particles, the relevant Hilberts space, taking symmetries into account, might not be very large, so that any model wave function that is constructed as to keep the particles well separated, typically by including Jastrow factors, will have a good chance to have a large overlap. Thus it is to some extent a matter of judgement and experience to decide upon the success of trial wave function by only studying overlaps for a small, but not too small, number of particles. The Laughlin state is often considered as something of a gold standard for overlaps.  For example, for $N=9$ electrons at $\nu=1/3$ on a spherical geometry (see section  \ref{sect:sphere}), the dimension of the symmetry-reduced Hilbert space is 84 and the squared overlap of the trial state with the exact coulomb ground state is 0.988 \cite{Fano}.   The Jain wavefunctions are similarly (if not more!) impressive.  For example for $\nu=2/5$ with $N=10$ electrons on a sphere, the symmetry-reduced Hilbert space has dimension 52, and the squared overlap with the exact Coulomb ground state is .9956 \cite{Toke2}.    A model wave function that gives overlaps this large, for comparably large Hilbert space is usually considered quite good.   However, overlaps must be used with caution.  For example, the Gaffnian wavefunction mentioned in section \ref{ssec:otherhierarchies} (which, as discussed there is gapless and therefore not a valid quantum Hall state) also has very high overlaps:  for the same $\nu=2/5$ system the squared overlap is .954.
Because of this difficulty in interpreting overlaps, other methods for determining the validity of  wavefunctions have become important. A particularly useful one is the entanglement spectrum \cite{LiandHaldane}, which is discussed in section \ref{sect:edgesandentanglement} below.


\subsection{Bosonic Chern-Simons theory and the Quantum Hall phase diagram}  \label{sec:GLCS}
It took some time after the discovery of the Laughlin wave functions to unravel the physical nature of the quantum Hall liquids. In modern terminology, they are topologically ordered states \cite{wenbook}. An important step was taken by \textcite{girvin1987off}, who proposed an order parameter and  showed that the Laughlin states are characterized by  algebraic off-diagonal long range order (ODLRO).  In essence, by removing all phases from the Laughlin wave function using a singular gauge transformation (see section \ref{sub:compbosons} below) the remaining wave function has algebraic long ranged order. 
The same authors also suggested that this order pointed towards a field theory description in terms of composite objects consisting of charge and flux \cite{girvin1987off,GirvinGirvin}. 

This idea was exploited in a later paper by Zhang, Hansson, and Kivelson \cite{zhang1989effective} who proposed the Lagrangian
for the $\nu=1/m$ Laughlin state
\be{GLCS}
{\cal L} &= &\phi^\star (i\partial_0 - a_0 ) \phi - \frac 1 {2m_e} |(\vec p -e\vec A + \vec a )\phi |^2 - V(\rho)  \nonumber \\
&+& \frac 1 {4\pi m} \epsilon^{\mu\nu\sigma} a_\mu \partial_\nu a_\sigma,
\ee 
where $m_e$ is the electron band mass. Here $\phi$ is a complex scalar order parameter field that is minimally coupled to the ``statistical" gauge field $a_\mu$,  $A_\mu$ is the  external electromagnetic field, and $\rho = \phi^\star\phi$ is the particle density. It had already been shown that  coupling a gauge field with a Chern-Simons (CS) term in the action, will effectively change the statistics of the original bosonic or fermionic matter field \cite{wilczek1983linking,pisarski1985topologically,wilczekbook}.  The choice of coefficient in front of the Chern-Simons term in Eq.~\eqref{GLCS} (the final term) corresponds to a change from fermions to bosons when $m$ is odd, {\it i.e.}, the original underlying electrons are fermions, but the field $\phi$ is bosonic.   When the statistical gauge field is treated in a mean-field approximation, where $\vec a = e\vec A $, this Ginzburg-Landau-Chern-Simons theory (GLCS)\footnote{
This name has stuck, although it is an exact rewriting of the underlying fermionic theory, and as such a proper microscopic theory with the extra condition of 
the bosons having hard cores. The usual GL theory for superconductors are more like the hydrodynamic theory discussed in section \ref{wenzee}. }
 thus describes interacting charged bosons without any  magnetic field.\footnote{
  There is a precursor to this picture were the physical magnetic field is ``cancelled'' by a collective effect due to a condensation of bosons. Starting from a Wigner crystal, \textcite{kivelson1987cooperative} use semi-classical methods to study instabilities that occur at the Laughlin filling fraction, due to coherent tunneling events that can be viewed as a condensation.}
In this effective description, the system is  essentially a charged superfluid and  captures many topological properties of the Laughlin states, including the algebraic ODLRO discussed above. Later it was also shown that the Laughlin wave functions can be derived by including fluctuations around the mean-field solution in  a random phase approximation \cite{kane1991general,zhang1992chern}. The GLCS approach was further developed to  include effects of orbital spin, and the response to curvature \cite{abanov2013effective,son2013newton},  see also section \ref{sect:torus}.

Applying the same ideas to the hierarchy states, \textcite{kivelson1992global} used the GLCS approach to generate the full abelian quantum Hall hierarchy, and to study the resulting phase diagram and associated phase transitions.  An important tool in this analysis was the concept of ``corresponding states" \cite{jain1990scaling}, which relates transport properties of quantum Hall liquids at different filling fractions.
 \textcite{kivelson1992global} also proposed the existence of another phase of a two-dimensional electron gas in a magnetic field, which they named a ``Hall insulator".  In this phase $\sigma_{xy} \sim \sigma_{xx}^2 \rightarrow 0$   as $T\rightarrow 0$, so $\rho_{xy}$ remains constant at very low temperatures. Early experiments on heterojunctions giving evidence for this phase were made by \textcite{shahar1995universal}, and recently  it has also been found in disordered indium-oxide films \cite{breznay2016}. 
L\"utken and Ross have analyzed the quantum Hall phase diagram using an effective field theory and certain, rather strong, symmetry assumptions \cite{lutken1992duality,lutken1993delocalization}, see also \cite{fradkin1996modular}.

The bosonic Chern-Simons approach correctly obtained much of the topological data about each of the hierarchy states (including Hall conductivity, quasiparticle charge and statistics), and set the stage for field theoretic evaluation of many additional quantities.    While the construction of the Chern-Simons Lagrangian Eq.~\eqref{GLCS} is in principle exact, the evaluation of any quantity beyond mean-field or random phase approximation level is quite challenging, and the lack of any small parameter to control perturbation theory makes it essentially impossible to confirm the correctness of any such calculation without support from other approaches such as numerics.   Particularly problematic is the issue of obtaining the proper energy scale.   While the bare band mass of the electron $m_e$ is present in the Lagrangian Eq.~\eqref{GLCS}, all low energy physics (which describes physics of a partially filled single Landau level) should be independent of the value of this parameter --- particularly in the limit of large 
Landau level spacing.  In principle, this result should emerge were calculations done exactly, but all approximations so far attempted have failed to achieve this.

 
\subsection{Composite bosons}
\label{sub:compbosons}
\newcommand\jas[3]{(z_#1 - z_#2)^{#3}}

In a first quantized language, the wave function for the charge-flux bosons described by Eq.~\eqref{GLCS} is related to the wave function of the original fermions via\footnote{ In the context of anyon superconductivity, attachment of phase factors was already used by \textcite{laughlin1988superconducting} to describe fractional statistics particles in terms of fermion wavefunctions.}
\be{wfrel}
\Psi_F (\vec r_1, \dots \vec r_N) =   \prod_{i <  j}  \left( \frac {z_j - z_i} {|\bar z_j - \bar z_i|}  \right)^{m}  \Psi_B (\vec r_1, \dots \vec r_N) \, ,\nonumber
 \ee
\ie they differ by a phase factor which can be interpreted as attaching singular (infinitely thin) flux tubes each carrying $m$ flux quanta at the position of the particles. 
 An alternative  order parameter was introduced by \textcite{read1989order},
\be{read}
\phi_R^\dagger (\vec r) & =&\psi^ \dagger(\vec r)\,  U_m(\vec r)  \\
&=&\psi^\dagger (\vec r) \exp\left[ - \frac {|z|^2} {4\ell^2} + m\int d^2 r'\,  \ln(z'-z)\rho(\vec r\,') \right]\, , \nonumber
\ee
who showed that it exhibits true long-range order. Here, $\psi^\dagger (\vec r)$ creates an electron,   $\rho = \psi^\dagger \psi$ is the electron density operator,  and $U_m  $  a vortex of strength $m$.\footnote{
    Note that $\psi^\dagger$ is the full electron creation operator. In Read's original formulation, $\phi_R^\dagger$ is defined in terms of the electron operator projected on the LLL. }
 In a first quantized formalism,  such a vortex, centered at the particle position  $\vec r$, is $U_m( \vec r) = \prod_{ i} (z_i  -z )^m e^{-|z|^2/(4\ell^2)}$.

The  Read operator $\phi_R$ is bosonic, but it is {\it not} related to the original fermion  
by a unitary transformation, and it is therefore more difficult to derive an effective GLCS theory \cite{read1989order,rajaraman1996field}.
The difference, compared to the pure phase transformation leading to Eq.~\eqref{GLCS}, is that  the Read operator builds in 
the characteristic vanishing property, $\lim_{z_i\rightarrow z_j} \Psi_L(z_1 \dots z_N) \sim (z_i - z_j)^m$, and exhibits true (\ie not 
algebraic) ODLRO. The Laughlin state can be constructed in a way similar to a Bose condensate, either as a coherent state where 
the $\phi_R$ has a finite expectation value, or as the state,
\be{read state}
\ket {1/q ; N}_L = \left[ \int d^2 r\, \phi_R^\dagger (\vec r)  \right]^N \ket 0  \, \nonumber
\ee
at fixed number of particles, which directly yields the $\nu = 1/m$ Laughlin wave function for $N$ particles. 


\subsection{Composite fermions } \label{sect:CF}
The idea of composite fermions \cite{jain2007composite,heinonenbook} is most easily understood as the fermionic version of the transformation \eqref{read}, where an even number of vortices  is attached to the original fermions in order to form a composite fermion (CF). The  transformation between the 
fermion wave function $\Psi_F$ and the CF wave functions, $\Psi_{CF}$ is thus, 
\be{cfwfrel}
\Psi_F (\vec r_1, \dots \vec r_N) = \Phi_{0}^{2p}(\vec r_1, \dots \vec r_N) \Psi_{CF} (\vec r_1, \dots \vec r_N),
\ee
where $\Phi_{0}$ is the wave function for a filled LLL.  This formula can be written in an arbitrary geometry, and specializing to  the plane we have 
$\Phi_0  =   \prod_{i <  j}  \left(  {z_j - z_i}   \right)e^{-1/(4\ell^2)\sum_j |z_j|^2}$. Note that $(\Phi_0)^2 = \prod_i U_2(z_i)$
which provides the link to the composite boson transformation \eqref{read}. 
Using $\Phi_{0}^{2p}$ for the ``flux attachment'' instead of the commonly employed Jastrow factor $\prod_{i<j}(z_i-z_j)^{2p}$ in Eq.~\eqref{cfwfrel} has the advantage that it gives  the correct gaussian factors independent of the geometry or gauge choice. If $\Psi_{CF}$ is not a LLL wave function, then neither is $\Psi_F$, and one must project to the LLL to get a valid (fully holomorphic) wave function. 


\subsubsection{ The Jain wave functions}
The composite fermion transformation \eqref{cfwfrel} was used by Jain to establish a powerful correspondence between fractional and integer 
quantum Hall states. 
His basic insight was that Eq.~\eqref{cfwfrel} allows us to think of the $\nu = 1/(2p+1)$ Laughlin state as a filled LLL of composite fermion.
One may then consider other homogeneous CF wave functions in order to form new model wave functions. 
In particular,  filling $q$ Landau levels of composite fermions --- which we call $\Lambda$-levels  following Jain's terminology ---  
 gives a wave functions at $\nu = q/(2pq + 1)$ in the positive Jain series \cite{jain1989composite,jain1990theory},  
\be{jainwf}
\Psi_{\frac{q}{2pq+1}}(\sgrupp[\vec r] ) = {\cal P}_{LLL} \left[ \Phi_{0}^{2p}(\sgrupp[\vec r])   \Phi_q(\sgrupp[\vec r] )  \right] \, ,
\ee
where ${\cal P}_{LLL} $ projects on the LLL.  The filling fraction follows from a simple counting argument: Since the state is expected to be homogeneous (up to edge effects), the filling fraction is given by $\nu = N/L_{\mathrm{max}} = N/(2pN + N/q) + O(1/N)$, where we used that each Landau level contains $N/q$ particles and that all filled Landau levels have the same maximum angular momentum $L$ up to boundary effects, \cf appendix \ref{app:QHE}.  Similarly, the states in the negative Jain series,  $\nu = q/(2pq - 1)$ are obtained by attaching an even number of anti-vortices to the electrons \cite{jain2007composite}.

Quasiparticle excitations amount to putting  composite fermions in an unfilled $\Lambda$-level  (usually the lowest),  or holes in any of the filled $\Lambda$-levels
--- entirely analogous to how we would make the corresponding low energy excitations for integer quantum Hall states. 
This simple mapping  between fractional quantum Hall states and systems of free, or at most weakly interacting, fermions, gives a very fruitful intuitive picture that extends far beyond the above example. Another example is paired states of composite fermions, which will be discussed later in the context of nonabelian quantum Hall states.   
The ground state wave functions \pref{jainwf}, as well as those  for its quasiparticle excitations, have an excellent  overlap with the states found by exact diagonalization  of realistic Hamiltonians, thus providing a strong {\it a posteriori} justification for the CF ansatz. 
There is a large and successful body of work based on 
the composite fermion wave functions which is reviewed by  \textcite{jain2007composite}  and in  the book by  \textcite{heinonenbook}. 


\subsubsection{The Lopez-Fradkin and Halperin-Lee-Read effective field theories} 

Just as the GLCS theory \eqref{GLCS} can be considered as a field theory for bosonic flux-charge composites, there  is 
also a  field theory for fermionic flux-charge composites that was proposed and analyzed by \textcite{lopez1991fractional}.
The only difference  in the Lagrangian from Eq.~\eqref{GLCS} is that the integer in the coefficient in front of the CS term is now  even, $m=2p$, which means that the composite object $\phi$ is still a fermion.  The mean-field solution does, however, differ from the one in the GLCS theory, in that the statistical mean-field only partially cancels the external magnetic field.
This  leaves the composite objects moving in a weaker effective field.
For special filling fractions, these composite objects completely fill a number of Landau levels and form a gapped state.  The effective mean-field seen by the composite fermions is
$$
 B^{\rm eff} = B^{\rm ext} - 2p \phi_0 \rho,
$$
where $\rho$ is the density of electrons, $\phi_0$ the flux quantum,  and $B^{\rm ext}$ is the externally applied field.   Thus, the effective filling fraction is 
$$
 \nu^{\rm eff} =\frac{\rho \phi_0}{B^{\rm eff}} =  \frac{\rho \phi_0}{B^{\rm ext} - 2p \phi_0 \rho} = \frac{\nu}{1-2p\nu},
$$
which is an integer $ q$ exactly when $\nu=q/(2pq + 1)$.    At mean-field level, the fractional quantum Hall gap is  given by the effective cyclotron energy $e B^{\rm eff}/m_e = \phi_0 \rho /(2 q m_e)$.  
As with the Chern-Simons boson theory, the (exact) derivation of the Chern-Simons fermion Lagrangian results in the bare electron band mass $m_e$ appearing  in the kinetic term (and this sets the size of the mean-field gap).    However, since the low energy physics occurs within a single Landau level, $m_e$ should not be an important parameter of the problem (and should vanish altogether in the limit of large Landau level spacing).   As with the bosonic case, this problem occurs since the Chern-Simons theory can only be evaluated using rather crude methods such as mean-field or random phase approximation.  Presumably a more precise (or exact) calculation would result in the dependence on $m_e$ vanishing.   As with the bosonic case, there is, however, serious difficulty in achieving any more precise calculation, since there is no 
natural small parameter in the theory.

The composite fermion field theory approach was further developed by \textcite{HLR} who considered the case of $\nu=1/2$ (or any $1/m$ with $m$ even) where the composite fermions move in zero effective field (i.e., $\nu^{\rm eff}=\infty$) and thus can form a Fermi liquid. In the Fermi liquid context, a  natural solution to the problem of energy scales results from viewing the strongly interacting Fermi liquid via  the phenomenological Landau Fermi liquid theory where an energy functional is proposed to describe the deformation modes of the Fermi surface.   In this framework the electron mass can be renormalized to an effective mass $m^*$ set by the interaction strength of the electrons so as to correctly give a fractional quantum Hall gap (the composite fermion cyclotron frequency $e B^{eff} / (m^* c)$) on the interaction scale  \cite{HLR,simonhalperin}.  As is usual in Landau theory, the mass renormalization is linked to a certain Fermi surface deformation parameter so as to preserve Galilean invariance of the system \cite{PinesNozieres}.

A feature of  the composite fermion field theory, as well as the composite fermion wavefunctions, is that they do not manifestly obey particle-hole conjugation.  In the limit of infinite Landau level spacing, the lowest Landau level is perfectly particle-hole symmetric\footnote{In the presence of particle-hole symmetric disorder one can also show that $\sigma_{xy}$ should be exactly $e^2/2h$ when the Landau level is half filled \cite{Krotov}).}, so, for example, the energy gap for $\nu=1/3$ should be identical to the energy gap at $\nu=2/3$.    In the wavefunction approach this symmetry is very close to true, but not exactly true \cite{MollerSimon}.  In the field theory approach, since the Lagrangian \eqref{GLCS} is in principle exact, the particle-hole symmetry should emerge if calculations are performed exactly.   However, any practical calculation involves approximations, and no known approximation preserves the particle-hole symmetry (although RPA methods  come very close \cite{simonhalperin}),  similar to the wavefunction approach.

The composite fermion field theory gave rise to other field theory approaches to composite fermions which emphasized the dipolar nature \cite{Read94} of the composite object \cite{shankarmurthy,readLLLtheory,dhleedipole,sterndipole,haldanepasquier}. 
There is a recent surge in interest in constructing a particle-hole symmetric description of the composite fermion 
liquid in terms of neutral Dirac particles \cite{PhysRevX.5.031027,barkeshli2015particle,geraedts2015half,wang2015dual,murthy2015landau}. This approach has the notable advantage that it is manifestly particle-hole symmetric at all orders.

It should be noted that just as in the GLCS theory, in the composite fermion field theory approach, the Laughlin wave function does not appear at the mean-field level. This is because in both cases the composite object is formed by a charge and a point-like statistical flux, so the extra ``correlation hole" $\sim \prod_{j\neq i} (z_j - z_i)^{2p}$ around the electron at position $z_i$, must arise from fluctuations around the mean-field \cite{lopezfradkinwavefunction1992}.

Sometimes the wave functions \eqref{jainwf} are also motivated in terms of ``flux attachment"  --- the cartoon picture being that each electron grabs  $2p$ flux quanta, leaving a weaker effective magnetic field.   
This picture is somewhat misleading since in the approaches by Read and Jain the charge binds to a {\it vortex} that  includes the correlation hole, and directly yields good wave functions with filling fractions following from a counting argument without any appeal to effective  fields.


\subsection{The Wen-Zee effective Chern-Simons theory } \label{wenzee}
\label{ssec:effectiveCS}
\newcommand{\kb}{\mathbf K}
\newcommand{\tb}{\mathbf t}
\newcommand{\lb}{\mathbf l}
\newcommand{\Sb}{\mathbf s}

In the Chern-Simons approaches discussed above one can imagine integrating out the particles of the theory to leave an effective Lagrangian for the longest length scales of the system.  Since the quantum Hall states are gapped, one might naively expect that this leaves no physics at all remaining. This is however not true. First, as discussed in some detail in the next section, in realistic geometries, there are low energy modes propagating along the edges of the sample. Second, even in closed geometries the topological data remains at arbitrarily long length scale and low energy.

Before presenting the general low energy Lagrangian first proposed by \textcite{wen1992classification} we make some general comments about its structure and status.
It is useful to think of  this kind of effective theories in terms of hydrodynamics, implying that the strategy for deriving them is not to keep the original degrees of freedom below some cut off scale, but rather to introduce new ones. In hydrodynamics, the  conserved currents are natural variables, and in two dimensions they are conveniently parametrized as $j^\mu = \epsilon^{\mu\nu\sigma}\partial_\nu\tilde a_\sigma$. In this way, current conservation is built in, at the expense of introducing a gauge symmetry.  
 The hydrodynamical gauge field $\tilde a$ is not to be confused with the statistical gauge field $a$ in the previous sections. 
 In the case of the Laughlin states where we have a GLCS theory,  or  a composite fermion field theory, we can derive the hydrodynamical theory by directly integrating over the microscopic field $\phi$ and the statistical gauge field  $a$. 
 For general QH states this procedure is not possible, and one must revert to the usual philosophy of effectve low-energy theories, and include  in principle all terms allowed by symmetry. In practice one is  usually  keeping only the lowest derivative terms. In the QH context, these are Chern-Simons terms $\sim \epsilon^{\mu\nu\sigma}\tilde a_\mu\partial_\nu\tilde a_\sigma$,  which are linear in derivatives and preserved rotation invariance but break parity and time-reversal symmetry. This term is also Lorentz invariant, which is however  a consequence of gauge symmetry. Higher derivative terms are, in general, not Lorentz invariant. What is special with the Wen-Zee effective action, is that the coefficients in front of the various terms  have a direct topological meaning, and thus can only take discrete values.

In fact, all topological information about an abelian quantum Hall state at the $n^{\rm th}$ level of hierarchy can be coded in the following effective Lagrangian based on $n$ Chern-Simons gauge fields \cite{wen1992classification,wen1995topological},
\be{wenlag}
{\cal L}_{WZ} &=& - \frac 1 {4\pi} K_{\alpha\beta} \epsilon^{\mu\nu\sigma} a^\alpha_\mu \partial_\nu a_\sigma^\beta - 
\frac e {2\pi} t_\alpha  A_\mu \epsilon^{\mu\nu\sigma} \partial_\nu a_\sigma^\alpha \nonumber  \\
& - & \frac{ s_\alpha} {2\pi}  \omega_i \epsilon^{i \nu\sigma} \partial_\nu a_\sigma^\alpha  + a^\alpha_\mu  l^q_\alpha  j_q^\mu   \, ,
\ee
where  we suppressed  the tilde on the field $a$ for ease of notation, where $\mu,\nu,\sigma$ denote space-time indices and  while the index $i$ runs over spatial indices only.    
$A_\mu$ is an external electromagnetic field, and $a_\mu^\alpha$ are the $n$ internal Chern-Simons fields. 
  The indices $\alpha,\beta$ run from $1\ldots n$, where $n$ is the rank of  $K_{\alpha\beta}$ which is a symmetric matrix with odd integers along  the diagonal and with all other entries even integers, and $t_\alpha$ is a vector of $n$ integers defining the electric charge of the currents $\epsilon^{i \nu\sigma} \partial_\nu a_\sigma^\alpha $. 
  The coupling of the $n$ quasiparticle currents $ j_q^\mu $ to the Chern-Simons fields is determined by the integer vector $l^q_\alpha$, and taking $l^q_\alpha = \delta_{q\alpha}$ corresponds to the ``elementary" currents $j^\mu_q$ that couple minimally to the gauge fields. 
In Eq.~\eqref{wenlag} $s_\alpha$  is a vector of spins  associated with the different Chern-Simons fields   and $\omega_i$ a spin connection, such that the corresponding field strength, ${\mathcal R} = \epsilon^{ij}\partial_i\omega_j$, is the gaussian curvature of the surface on which the quantum Hall state lives.\footnote{The Ricci curvature $R$ is twice the gaussian curvature, $R=2\mathcal R$.}
We shall return to this in Sect. \ref{sect:sphere} where we study quantum Hall states on a sphere.  On a flat surface, however, $\omega$ can be taken to be zero and the spin does not enter. 

The full topological information can be extracted from the Lagrangian \eqref{wenlag}.  For example, using an obvious vector notation, the filling fraction is given by \cite{wen1992classification},
 \be{wenres1}
 \nu &=& \tb^T \kb^{-1} \tb 
\ee
and the ground state degeneracy on a manifold with genus $g$  is given by 
\begin{align}\label{eq:gsdegeneracy}
(\det\, {\bf K})^{g}.
\end{align}   
Another important topological quantum number is the  shift $S$ defined by 
\begin{align}\label{eq:shift}
N_e=\nu(N_\Phi+S),
\end{align} 
where  $N_\Phi$ denotes the total number of flux quanta (on the sphere) or the highest occupied angular momentum orbital (on the disk), respectively. 
Its value is determined by the topological data as
\begin{align}\label{eq:shiftK}
S=2\bar s =\frac{2}{\nu}\,\tb^T \kb^{-1} \Sb.
\end{align}
where $\bar s$ is known as the average orbital spin.
The electric charge of the quasiparticle of the  $q^{\rm th}$ type is,
 \be{wenres2}
 Q_q &=&   -e\tb^T \kb^{-1} \lb_{q} \, ,
\ee
and clockwise braiding of quasiparticle type $q$ around one of type $r$ gives a phase of 
\be{wenres3} 
 \theta_{q,r} &=&  2 \pi \lb_{q}^T \kb^{-1} \lb_{r},    
\ee
whereas clockwise exchange of two quasiparticles of type $q$ gives
\be{wenres4} 
 \theta_{q} &=&  \pi \lb_{q}^T \kb^{-1} \lb_{q}.    
\ee

A given topological order does not uniquely define the necessary data $(\kb,\tb,\Sb)$, in the Lagrangian \eqref{wenlag}.   In fact, one is free to make the transformation \cite{read1990excitation,wen1992classification} 
\begin{align}\label{trans1}
  \kb &\rightarrow {\bf W}^T \kb {\bf W}  &  \Sb &\rightarrow {\bf W}^T \Sb     \\
  \tb &\rightarrow {\bf W}^T \tb &   \lb_{q} &\rightarrow {\bf W}^T \lb_{q} \nonumber \, ,  
\end{align}
where $\bf W$ is an integer matrix with unit determinant, without changing the values obtained from Eqs.~\eqref{wenres1}, \eqref{eq:gsdegeneracy} and  \eqref{eq:shiftK}-\eqref{wenres4}.  There are two very natural bases to work in.  The first, stemming from the Haldane-Halperin hierarchy and thus  called hierarchical basis,  has $t_1 =1$ and $t_\alpha=0$ for $\alpha > 1$.    The second basis is the symmetric, or composite-fermion basis, which has  $t_\alpha =t = 1$ for all $\alpha$. 
The  transformation that maps the symmetric to the hierarchical  basis, \ie $\mathbf K^h=\mathbf W^T \mathbf K^s \mathbf W$, is given by $W_{ij}=\delta_{i,j}-\delta_{i+1,j}$.

Wen has argued that for a state at level $n$ in the hierarchy, there must be $n$ linearly independent combinations with charge  $e$, or in other word, $n$ independent electron currents \cite{wen1995topological}. 
These are described by the $n$ vectors $\bf l_{e}^\alpha$.   In the symmetric basis these  take the simple form
$
l_{e, r}^\alpha = K_{r\alpha}
$
where $\alpha$ labels the vector and $r$ its component, resulting in $Q_e = -e t = -e$ as expected. 
This will be important when we later use the topological data $(\kb,\tb, \Sb)$  to construct explicit electronic wave functions. 
Finally we give the explicit form, in the symmetric basis, for the $K$-matrix and the spin vector that  describe 
the full hierarchy obtained by successive condensations of quasiparticles: 
\be{chirK}
K_{\alpha \beta} &=&  \left\{ \begin{array}{rl} 
m+ \Sigma_{\alpha}& \mbox{ if } \alpha=\beta \nonumber\\
-\sigma_\alpha+m+\Sigma_{\alpha}& \mbox{ if } \alpha<\beta \nonumber\\
\end{array}\right.\nonumber\\
\Sigma_\alpha&=&\sum_{j=1}^{\alpha-1}(p_j-2)\sigma_j\nonumber\\
s_{\alpha} &=& \half K_{\alpha\alpha} + \alpha - 1   \label{chirs},
\ee
with $m$  odd and $K_{\alpha \beta}=K_{\beta\alpha}$, if $\alpha>\beta$. The even integers $p_j$ specify the density of the quasiparticle condensates, 
and $\sigma_j=+1(-1)$ for condensing quasielectrons (quasiholes) (compare Eq.~\eqref{eq:fillingfraction}).


\subsection{Physics at the quantum Hall edge}  \label{sub:edge}

On a closed surface (such as  a sphere or torus -- see section \ref{othergeom} below), the above Lagrangian \eqref{wenlag} is completely self-consistent.   However, for a system with edges, such as a disc, the action turns out to not even be gauge invariant, as we will discuss further in section \ref{sub:csedge} below.   The reason for this failure is that quantum Hall systems always have non-trivial edge states -- which must be included in the theory.   We will thus briefly turn our attention now to the physics of quantum Hall edge states -- and in particular edge states of the hierarchy.  There are a number of  papers which  give more detail about quantum Hall edge physics \cite{Chang,KaneFisher,WenEdge}.  Here we  give a very brief review just to indicate how these ideas fit into the overall framework of hierarchies. 


\subsubsection{Introduction to edge physics}\label{ssec:edge}

The existence of edge currents in a quantum Hall system is virtually guaranteed by the presence of crossed magnetic field $\bf B$ and electric field $\bf E$ (the electric field being caused by the confining potential keeping the electrons inside the sample).   Classically one  expects a drift velocity ${\mathbf v} = {\bf E} \times {\bf B}/|{\bf B}|^2$.  Thus a perturbation in density will propagate along the edge, in only one direction, at this  velocity --- thus giving a chiral hydrodynamics.    Another semi-classical picture of these edge states is that of electrons  executing so-called ``skipping orbits" that bounce off the edge and travel in one direction.   What is less obvious is how to  quantize this sort of motion.     An argument by \textcite{WenEdge} gives a proper quantum mechanical treatment for simple Laughlin states, which we shall now review. 

Let us imagine a quantum Hall system filling the lower half plane, with an edge running in the $x$ direction at position $y=0$.   Given that the quantum Hall system is incompressible, any perturbation of the edge must come from a physical displacement  of the edge by some distance $h(x)$ in the $y$ direction so that the excess density along the edge is $\rho(x) = \bar n h(x)$  where $\bar n = \nu/(2 \pi \ell^2)$ is the electron density of the bulk quantized Hall state.    The energy of the displacement should be given simply in terms of the electrostatic potential $V(x,y) = E y$ on the edge\footnote{With a sufficiently abrupt edge, this approach breaks down and very different edge physics can result \cite{Fern}.}.  We expect \cite{WenEdge}
\be{edgeh}
  H &=& \int dx \int_0^{h(x)} dy \,  e V(x,y) \bar n    = \frac{E e \bar n}{2}  \int dx \,  h(x)^2  \nonumber  \\
   &=&  \frac{\pi v}{\nu} \int dx \, \rho(x)^2 = \frac{2\pi v}{ \nu} \sum_{k>0} \rho_k \rho_{-k}
\ee
with the Fourier transform convention $\rho_k = \frac{1}{\sqrt{L}} \int dx e^{-i k x} \rho(x)$. Now, given that any density perturbation should move along the edge at the drift velocity, we expect dynamical equations $(\partial_t - v \partial_x) \rho(x) = 0$ or  equivalently  \begin{equation} (\partial_t - i v k) \rho_k = 0.  \label{eq:dynameq}
\end{equation}
From the Hamilton equations $\dot q = \partial H/\partial p$ and $\dot p = -\partial H/\partial q$, it follows that 
 the momentum conjugate to the coordinate $q=\rho_k$ for $k>0$, must be taken as $p=2 \pi \rho_{-k} /(i k \nu)$, 
in order for the dynamical equation \eqref{eq:dynameq} to be consistent with the Hamiltonian \eqref{edgeh}.   Canonical quantization is now straightforward and one obtains the so-called $U(1)$ Kac-Moody algebra,
\begin{align}
 [\rho_k, \rho_{k'} ] = \frac{\nu k'}{2 \pi} \delta_{k+k'}  \label{eq:KacMoody} \, .
\end{align}
This well known, but nevertheless quite subtle result, is derived and commented upon in Appendix \ref{sec:commutator}. 

Next we introduce a bose field $\varphi(x)$ related to the density by
\begin{equation}\rho(x) = 
\frac{1}{2 \pi} \partial_x \varphi(x) \, ,
\label{eq:rhophi}
\end{equation} 
where the field $\varphi$ has equal-time commutation relations 
\begin{equation}
 [\varphi(x), \varphi(x')] = i \nu \pi \, {\rm sign} (x - x').
\label{eq:phicommute}
\end{equation}
For the normalization convention, see appendix \ref{app:chiralboson}.
Using these commutation relations together with the relation \eqref{eq:rhophi} for  $\rho(x)$,  and Fourier transforming we recover the Kac-Moody commutators, \eqref{eq:KacMoody}.

From Eq.~\eqref{eq:phicommute} and \eqref{eq:rhophi} it easy  to see that $\partial_x \varphi/(2 \pi \nu)$ is canonically conjugate to $\varphi(x)$  (their commutator is proportional to a delta function).     Using this information with the Hamiltonian given above, \textcite{WenEdge} wrote the Lagrangian for the edge as
\begin{equation}
{\cal L}_{\rm edge} = \frac{1}{4 \pi \nu}  (\partial_x \varphi) (\partial_t \varphi) - \frac{v}{4 \pi \nu} (  \partial_x \varphi)(\partial_x \varphi), \label{eq:edgeLagrangian1}
\end{equation}
which is that of a one dimensional chiral boson.

One can write a creation operator for a  bump of charge $\alpha$ at position $x$ as
\begin{equation} 
\label{eq:vertexonedge}
  V_\alpha(x) = \, :e^{i \alpha \varphi(x)}: \, ,
\end{equation}
with colons denoting normal ordering.  It is easy to check that this inserts charge $\alpha \nu$ by noting that 
$$
  [\rho(x'),V_\alpha(x)] = \nu \alpha V_\alpha(x)  \delta(x - x') \,  .
$$
Crucially, we can create a  charge $e$ using the operator $\Psi_e = V_{1/\nu}$.   Note, however, that this operator has proper fermionic commutations, $\Psi_e(x) \Psi_e(x') = -\Psi_e(x') \Psi_e(x)$ only when $\nu$ is one over an odd integer, corresponding to half integer conformal spin  (see Appendix \ref{app:CFT}.).   This implies that only these simple Laughlin states can be described by a single edge mode.  


\subsubsection{Chern-Simons approach and Wen-Zee edge} 
\label{sub:csedge}
 Another way to understand the edge physics is via the Chern-Simons approach.  In the case of a  Laughlin state, the Wen-Zee effective Chern-Simons Lagrangian \eqref{wenlag} is very simple: 
\be{cssimple}
 {\cal L} = -\frac{1}{4 \pi \nu}  \epsilon^{\mu \nu \sigma} a_\mu \partial_\nu a_\sigma - \frac e {2\pi}   \epsilon^{\mu \nu \sigma} a_\mu \partial_\nu A_\sigma \, ,
\ee
where we have left out the coupling to external currents, and possible curvature of the two dimensional surface.   

On a closed manifold, it follows by integration by parts that the corresponding action is gauge invariant.   However, as mentioned above, the action is not gauge invariant in the presence of a boundary.   Let  again  our fluid  occupy $y<0$, and our edge  run along the $y=0$ axis.   Now, we cannot integrate by parts in the $y$ direction without getting a boundary term. What happens is that the would-be gauge  mode at the boundary becomes a physical degree of freedom, which is shown as follows:
First notice that  $a_0$ is not dynamical, but  a Lagrange multiplier implementing the constraint $ \epsilon^{ij}\partial_i a_j    + \nu e \epsilon^{ij}\partial_i A_j  = b + \nu e B= 0$, which is solved by $a_i = -\nu e A_i +  \partial_i \varphi$. 
Substituting this back in the action \eqref{cssimple} and picking the $a_0=0$ gauge, we get after partial integration and some 
rearrangements:
\begin{align}\label{edgekin}
S &= \int dtd^2x\,{\cal L}_{0} \nonumber\\
&= \frac{1}{4 \pi \nu} \int dtdx\, (\partial_x \varphi - \nu e A_x) (\partial_t \varphi - \nu e A_0)   \nonumber \\
 &+ \frac{\nu e^2}{4 \pi }  \int dt d^2x\,   \epsilon^{\mu \nu \sigma} A_\mu \partial_\nu A_\sigma   - \frac e {4\pi} \int dt dx\, \varphi E_x \, ,
\end{align}
where $\varphi(x,t) \equiv \varphi(x,0,t)$ is a scalar field with support  along the boundary $y=0$ and $E_x = \partial_t A_x - \partial_x A_0$. Written in this form it is clear that the only dynamical degrees of freedom reside on the edge, and from the ``symplectic''  term  $\sim \partial_x\varphi \partial_t\varphi$ we  immediately get the canonical commutation relation \cite{faddeev1988hamiltonian}
	\be{cancomrel}
	[\varphi(x), \partial_{x'} \varphi (x')] = {2\pi i  \nu} \delta(x-x') \, ,
	\ee
	which  again means that $\partial_x \varphi/(2 \pi \nu)$ is canonically conjugate to $\varphi(x)$ as was inferred from \eqref{eq:KacMoody}. 
For the action \eqref{edgekin} to be invariant under gauge transformations of this background field, $A_\mu \rightarrow A_\mu + \partial_\mu \xi/e$,   the scalar field  $\varphi$ must transform like $\varphi\rightarrow\varphi +\nu  \xi$ which identifies it as a phase field. Also note that  the edge action on the first line is separately invariant under the gauge transformation, while the ``anomaly'' term $\sim E_x$ on the second line exactly cancels the gauge variation of the Chern-Simons term for the background field \cite{hansson2000edge}.

Neglecting the background field,  the Lagrangian \eqref{edgekin} matches the above Eq.~\eqref{eq:edgeLagrangian1} except that it has zero edge velocity!  This is perhaps not surprising since the velocity is a non-universal parameter of the edge --- depending on the details of the edge confinement.  In order to obtain precisely the edge Lagrangian Eq.~\eqref{eq:edgeLagrangian1}, one can choose a different gauge fixing $a_0 - v a_x = 0$, as discussed by \textcite{WenEdge}, or alternatively one can simply add the potential term
$ -( v /{4\pi\nu})(\partial_x \varphi - \nu e A_x)^2$  by hand.

Calculating the edge charge and current by varying the action \eqref{edgekin} with respect to $A_0$ and $A_x$ we get $\rho_{edge} = (e/2\pi)(\partial_x\varphi - \nu e A_x)$ and $j_{edge} = 0$. Using the equation of motion for $\varphi$ we get  $\dot\rho_{edge} = -(\nu e^2/2\pi)E_x$, while a direct variation of the bulk action with respect to $A_y$ gives the  outgoing bulk current perpendicular to the edge, and we find $J_y = \dot\rho_{edge}$.
This ``anomaly cancellation'' which was originally pointed out in the quantum Hall context  by \textcite{stone1991edge}, based on earlier work by \textcite{callan1985anomalies} is a general phenomenon. 
Charge non-conservation in a CS theory, which physically corresponds to a current perpendicular to the boundary, is exactly compensated by the charge not being conserved in the edge theory since the corresponding  current is anomalous. Here we considered the charge current, but we shall later also discuss the energy-momentum current related to heat transport.

It is quite easy to generalize to arbitrary hierarchy states.  One starts with the first term of the Lagrangian \eqref{wenlag} and similarly enforces gauge invariance to obtain an edge Lagrangian generalizing Eq.~\eqref{edgekin} to hierarchy states,
$$
 {\cal L}_0 = \frac{1}{4 \pi} K_{\alpha \beta}  (\partial_x \varphi^\alpha)  (\partial_t \varphi^\beta),
$$
where here $\varphi^\alpha$ are a set of bose fields with commutations  
$$
 [\varphi^\alpha(x), \varphi^\beta(x')] = i  \pi K^{-1}_{\alpha \beta} \, {\rm sign}(x - x').
$$
As in the case of Eq.~\eqref{eq:edgeLagrangian1} there is also a non-universal interaction term $\cal H$, which depends on details of the edge structure as well as the strength of the interaction between the different edge modes.  We generally write the full Lagrangian as  ${\cal L}_{\rm edge} = {\cal L}_0 - {\cal H}$ with the Hamiltonian 
$$
 {\cal H} = \frac{1}{4 \pi} V_{\alpha \beta} (\partial_x \varphi^\alpha) (\partial_x \varphi^\beta) 
$$
where $V$ is a positive definite matrix. 

Given the coupling of the gauge field $a$ in Eq.~\eqref{wenlag} to the external electromagnetic field $A$, it follows that the charge density associated with an edge excitation must be 
$$
  \rho(x) = \frac{1}{2\pi} t_\alpha (\partial_x \varphi^\alpha) \, ,
$$
and a quasiparticle described by a vector $\bf l$ with entries $l_\alpha$ as in Eq.~\eqref{wenres2} can be created using the operator
$$
 \Psi^\dagger_l(x) =  :e^{i\sum_\alpha l_\alpha \varphi^\alpha(x) }:.
$$

The emerging picture  of the edge of a hierarchy state is one involving multiple chiral bose modes --- one for each level of the hierarchy.   However, one should be aware that modes corresponding to any negative eigenvalue of the $K$ matrix will be reverse moving.    
One should also remember that there may be processes which can remove an electron from one of the edge modes and put it back into a different edge mode.   
In composite fermion language this would correspond to removing an electron from one $\Lambda$ level edge and placing it in another. 


\subsubsection{The conformal edge}
Whether one is considering a single edge mode, or a system with multiple edge modes, one obtains a gapless (massless) one dimensional system --- which is necessarily described by a conformal field theory (CFT).  
There are, however, several added features.  To begin with, multiple edge modes will generically have different velocities, whereas in a true CFT there should be only one velocity of light.   More  generally the edge can be a perturbed CFT where the perturbations may be irrelevant (vanishing at low energy), marginal, or relevant.   In some cases, a relevant perturbation can drive the edge to another gapless fixed point \cite{Haldane95,ReadEdgeTheory,cano2014bulk}.   Nonetheless, even when we consider more complex quantum Hall states (such as nonabelian states), the edges will always be described by (possibly perturbed) CFTs. 

The connection to CFT turns out to be far more significant than it might appear so far.   As we will see in section  \ref{bulkedgecorr} below, a so-called bulk-edge correspondence allows us to describe the bulk of the quantum Hall state with the same CFT that we use for a minimal description of the edge.  


\subsubsection{Edge CFT and entanglement spectrum}  \label{sect:edgesandentanglement}
For quantum Hall wave functions that are the exact zero-energy ground states of a model Hamiltonian, one can explicitly derive the edge excitations and show that the state counting (as a function of angular momentum) is the same as in the associated CFT \cite{ReadRezayi1996,ardonne2005filling,Read2006wavefunctions}.  For other quantum Hall states, one has to resort to numerical simulations. 
While the study of quantum Hall edges potentially gives a direct way to study the spectrum of the 1+1 dimensional CFT associated with a quantum Hall state, it can also be complicated by the fact that these spectra depend on non-universal properties of the edges (such as the slope of the potential at the edge) and can also be further complicated by so-called ``reconstruction'' where edges become non-monotonic in density \cite{MacDonaldEdgeReconstruction,JohnsonEdgeReconstruction,chamonwen1994}. 

Recently it has been realized that one can learn about the \emph{excitations} by looking at the entanglement properties of the representative \emph{ground state} wave function.  The entanglement entropy, independently introduced by \textcite{levinwen,KitaevPreskill}, is currently the most commonly used entanglement measure.\footnote{For a mathematical perspective on topological entanglement entropy in Chern-Simons theories and quantum Hall states, see \cite{Dong} and \cite{Hikami}.}
It contains, in principle, all the information about the $S$-matrix, and thus the quasiparticle excitations \cite{zhang2012quasiparticle},  and gives direct access to the total quantum dimension of the underlying field theory. The numerical evaluation is, however, very difficult for FQH systems \cite{haque2007entanglement}, and only very recently it was possible to obtain reliable results \cite{zaletel2013topological}.  
 A numerically more accessible quantity is the `entanglement spectrum' \cite{LiandHaldane}, which allows one to examine the CFT spectrum of a virtual edge and has quickly become an indispensible tool of the community.

We consider a  system bi-partitoned into two subsystems $A$ and $B$ (we can choose $A$ and $B$ to be the top and bottom halfs of a sphere divided along the equator, although many other geometries are possible).   The ground state wave function $|\Psi\rangle$ is contained in the tensor product of the Hilbert spaces of the $A$ subsystem and the $B$ subsystem.  We make a Schmidt decomposition of the state as
\begin{equation}
\label{eq:schmidt}
|\Psi\rangle = \sum_n  e^{-\xi_n/2} |\psi_A^n\rangle \otimes |\psi_B^n\rangle \, ,
\end{equation}
 where  $|\psi_A^n\rangle$ and $|\psi_B^n\rangle$ are states forming orthonormal sets 
completely contained in the $A$ and $B$  subsystems respectively.   The $\xi_n$ are known as ``entanglement energies'', but one should realize that they are properties of the ground state  decomposition and have nothing to do with the physical energies of the quantum Hall state.   The rather remarkable conjecture by  \textcite{LiandHaldane} (see also  \cite{KitaevPreskill}) is that the spectrum of entanglement energies should match that of the 1+1 dimensional CFT defining the quantum Hall state.

In order to examine these spectra more closely, one typically divides the system into $A$ and $B$ so as to conserve several quantum numbers.  For example, if we divide the sphere into an upper and lower part along a latitude, the division conserves  the total $L_z$ angular momentum, and,  assuming a rotationally invariant interaction, it will conserve the number of particles in each part. In this case one can plot the entanglement energies as a function of these conserved quantities, for example $\xi(N,L_z)$.    One can choose to bipartition the system in several different ways: an orbital cut, where certain single electron orbitals are placed in subsystem $A$ and others in subsytem $B$  \cite{LiandHaldane}; a real-space cut, where particles in one physical region are placed in subsystem $A$ and others in subsystem $B$ \cite{Dubail1,SterdyniakRealSpace,RodriguezRealSpace}; or a particle cut, where certain particles are placed in subsystem $A$ and others in subsystem $B$  \cite{SterdyniakParticle,Zozulya}. 

For topologically ordered states, both  the orbital and the real-space cut can be used to extract important information about the edge spectrum \emph{viz.} the type of CFT that describes the edge spectrum and the compactification radius of the charge field, which is directly related to the filling fraction \cite{Dubail1,hermanns2011haldane}. 
Both cuts should also give the same topological entanglement entropy \cite{Zozulya,Dubail2,zaletel2013topological}, even though the finite-size corrections in numerical simulations can be quite substantial for the currently accessible system sizes \cite{SterdyniakRealSpace}. 
There are, however,  differences which reflect that the orbital cut strictly speaking does not mimic a spatial edge, as can be easily seen in the case of the IQHE.\footnote{Being a single slater determinant, this is trivially  a short-range entangled states, and an orbital cut gives an entanglement spectrum with a single state. 
A real-space cut does reveal that the edge supports a gapless fermion, \ie a $c=1$ CFT with unit charge radius.}
From practical point of view, the orbital cut is the favoured tool since it is much easier to evaluate numerically.

The particle cut provides information about bulk excitations. More precisely, for trial wave functions that are zero-energy ground states of model Hamiltonians, the particle entanglement spectrum reproduces the quasihole state counting: the number of non-infinite eigenvalues equals the number of zero-energy states for the appropriate number of particles/flux. For other trial states, where no model Hamiltonian exists, the particle entanglement spectrum may instead be used to `define' the  quasihole state counting. This `definition' is consistent both with  the counting obtained using the composite fermion picture, as long as one considers holes in \emph{any} of the filled $\Lambda$ levels,  and with that obtained using CFT techniques, as long as all quasihole operators (see section \ref{sect:expwfs}) are taken into account. 

Even though entanglement spectra of strongly correlated states are very complicated and rigorous proofs are rare, there have been some advancements recently. 
Several works have formalized the connection between the  CFT spectrum and entanglement spectrum \cite{QiKatsura,Swingle,Dubail1,Dubail2} at various levels of rigor.
In addition, one can derive a bulk-edge correspondence in the entanglement spectra, analogously to the well-established bulk-edge correspondence in the physical system \cite{chandran2011bulk}.

The ``simple'' quantum Hall states, \ie those that can  be written as correlators of primary fields of a CFT, have correspondingly simple entanglement spectra. The spectra for hierarchy states are somewhat more complicated, and have been studied  to a much lesser extent.  At least for $\nu=2/5$ and $3/7$ the counting of low entanglement energy modes matches that of 2 or 3 chiral bosons respectively \cite{RegnaultEntanglementJain,RodriguezEntanglementCF}. It has also been suggested that these states can be described fairly simply in terms of composite fermions \cite{DavenportEntanglementCF} although this has only been explicitly tested for the fractional quantum Hall effect of bosons at $\nu=2/3$  (which is presumably similar to fermions at $\nu=2/5$).


\subsection{Effective response action} \label{sc:effresp} 
Since the Wen-Zee action \eqref{wenlag} is quadratic in the topological gauge fields $a_\mu^\alpha$, these can be integrated to give an effective action that describes the response to external fields. Neglecting the quasiparticle currents,\footnote{
These can be included, but will give a non-local action, which is needed in order to encode the statistical phases related to the braiding of the quasiparticles.}
 a (nontrivial) calculation and some rearrangement gives \cite{gromov2015framing}, 
\be{resplag}
S_{eff} &=& \frac \nu {4\pi} \int d^3x\, \epsilon^{\mu\nu\sigma}[ (e A + \bar s \omega)_\mu \partial_\nu   (e A + \bar s \omega)_\sigma \nonumber  \\
&+&  \beta  \omega_\mu\partial_\nu\omega_\sigma ] - \frac c {48\pi} \int d^3x\, \epsilon^{\mu\nu\sigma} \omega_\mu\partial_\nu\omega_\sigma .
\ee
Note that the first term is what we naively get from the Gaussian integration. The filling fraction $\nu$ and the average orbital spin $\bar s$ are given by  \eqref{wenres1} and \eqref{eq:shiftK} respectively, and the \emph{orbital spin variance}, $\beta$, is  $\beta = \nu(\overline{s^2} - \bar s^2)$  with  $\nu\overline{s^2} =  \Sb^T \kb^{-1} \Sb$.
For  a $\nu = 1/q$ Laughlin state $\bar s=q/2$ and $\beta  = 0$, since in a one-component state all electrons have the same orbital spin so the variance vanishes. For general abelian hierarchy states, as \eg the Jain states, $\beta \neq 0$. 

In the last term, the chiral, or  topological, central charge $c$ equals the signature of the $K$-matrix which is the difference between the number of positive and negative eigenvalues \cite{read2000paired,KaneFisherThermal}
$$\kappa_H = c \frac {\pi k_B^2 T} 6 \, .$$

Although the first term in \eqref{resplag} goes back to the original work of \textcite{wen1992classification}, the history of the second term is more involved. 
It was  introduced in the condensed matter context  by \textcite{volovik1990gravitational} in  his work on liquid ${}^3\rm{He}$ films, and later \textcite{read2000paired} argued that it was needed to cancel the anomaly in the energy-momentum current  conservation on the boundary, in analogy with the anomaly cancellation in the charge current described in section \ref{sub:csedge}. Later it was found by an explicit calculation using free fermions (\ie integer QH effect) by \textcite{abanov2014electromagnetic}, and subsequently by \textcite{gromov2015framing,PhysRevLett.114.149902} by carefully performing the integral over topological gauge fields in \eqref{wenlag}. The subtlety is that although the action is independent of the metric, performing the integral requires a gauge-fixing which will depend on the metric. As shown by \textcite{witten}, this generates an extra term, which essentially is the the second one in \eqref{resplag}. In \textcite{bar1991perturbative} this is referred to as the ``framing anomaly''. In a parallell development, \textcite{bradlyn2015low} determined the general form the effective low-energy  action obtained by integrating out the matter fields.
Later they calculated the second  term in \eqref{resplag} directly from particular CFT wave functions \cite{bradlyn2015topo}. 
They also argued, based on  anomaly cancellation (for this see also \cite{stone2012gravitational}) that the orbital spin variance, $\beta$ should vanish in a general one-component (non-abelian) quantum Hall state.
Related work on the gravitational response of Hall fluids was done by \textcite{hoyos2012hall,son2013newton}, \textcite{can2014fractional,can2015geometry} and  \textcite{cho2014geometry}.


\subsection{Nonabelian quantum Hall states and nonabelian  hierarchies }
The defining feature of a nonabelian quantum Hall state is that the quasiparticles are described by multi-component wave functions and obey nonabelian fractional statistics due to the appearance of nonabelian Berry phases \cite{moore1991nonabelions, wilczek1984appearance} during braiding. The first concrete proposal was the Moore-Read Pfaffian state discussed in detail in section \ref{sub:MRstate}.  Several more filling fractions --- all of them in the second Landau level (see Fig.~\ref{2ndLL})--- have been conjectured to be nonabelian. It is natural to wonder, whether there is a single governing principle that can explain all the observed states as well as their properties and relative stability, similar to the Haldane-Halperin hierarchy in the LLL. Another obvious question is whether it is possible to form hierarchy states by condensing nonabelian quasiparticles. Several proposals in these directions are discussed in Sect. \ref{sect:nahier}. 


\subsection{Other approaches}\label{ssec:otherhierarchies}

There has been no shortage of attempts to construct new representative wave functions for the quantum Hall effect.   Ideally we would like these to be based on some physical principles, and we would like them to be gapped ground states of some (even slightly realistic) Hamiltonians.    This combination of demands makes successful wave function building extremely difficult. Establishing that a ground state is gapped is particularly difficult, and is only ever really confirmed by extensive numerics, requiring large system size extrapolation, which is always open to at least some degree of doubt.   Even the conventional hierarchies discussed above are not beyond question.   For example, a detailed analysis of the conventional hierarchy has been given in a series of papers reviewed by \textcite{Quinn}. 
Further it has been proposed that certain experimentally observed fractional quantum Hall states may not be of the hierarchy/composite-fermion form \cite{mukherjee2014enigmatic,mukherjee2014enigmaticErr} --- although it is not 
at all clear if the proposed competing states would be viable gapped states in the thermodynamic limit.

One very straightforward way to build new wave functions is to perform one of several well-established transformations on a known wave function.   One such transformation is to build the same wave function in a higher Landau level, which should be a simple alteration --- indeed, given a Hamiltonian with a corresponding ground state in one Landau level, it is easy to produce another Hamiltonian that will give precisely the same ground state within another Landau level \cite{HaldaneGirvin,SimonRezayiCooperPseudo}.   Another transformation is particle-hole conjugation, which can potentially produce new topological orders \cite{Joli} even for a half-filled Landau-level \cite{AntiPfaffian1,AntiPfaffian2}.  A third transformation is flux attachment, which has seen great success in the composite-fermion \cite{jain2007composite} and composite-boson approaches.   Finally, one can consider simply multiplying together two known wave functions, which can be interpreted as a parton construction \cite{wen1999projective}, see also section \ref{ssec:parton}. 
Combinations of these techniques have been used to generate a number of novel filling fractions outside of the conventional hierarchy such as $\nu=3/8$  \cite{Mukherjee,Toke,hutasoit} and an alternative candidate for $4/11$ that is distinct from the hierarchical one \cite{mukherjee2014enigmatic}. 

There have also been approaches proposed that precisely reproduce known conventional hierarchies from different physical intuition.  In two interesting papers using a formalism entirely in the LLL, Haxton and co-workers  \cite{Haxton1,Haxton2} provide a quite different way to understand  why the hierarchy/composite-fermion states are energetically favorable. 
Another work by \textcite{JainKamillaLLL} physically motivates the fully chiral subset of the CF hierarchy without appeal to higher Landau levels. This is also the case for the CFT approach \cite{bergholtz2008quantum,hansson2009quantum,suorsa2011quasihole,suorsa2011general}, which is discussed in detail in section \ref{ss:chiral}.
 
Aside from the above  methods, there have been attempts to unify fractional quantum Hall states, and to construct new states, by focusing on clustering patterns.   There are several closely related ways of approaching this which we will review below. 
Generalizing the clustering patterns of the Laughlin and Moore-Read state lead to the Read-Rezayi series of quantum Hall states 
\cite{read1999beyond}, which can also be defined as correlators of primary fields of CFTs.   
In addition to the Read-Rezayi series, several other wave functions can be described in this clustering language including the Haffnian wave function \cite{dmitrygreenthesis} describing electrons at $\nu=1/3$ and the Gaffnian wave function \cite{YoshiokaGaffnian,Gaffnian} describing electrons at $\nu=2/5$.  Unfortunately, despite having good overlaps with exact diagonalizations for small systems sizes, the Gaffnian and Haffnian wave functions are defective in certain ways. 

The Haffnian turns out to be a correlator of a non-rational conformal field theory which invalidates it as a candidate for a sensible quantum Hall state as discussed in detail by  \textcite{read2009nonabelian}.   
Some of its resulting defects are that it has an infinite number of particle types and correspondingly an infinite ground state degeneracy on the torus in the thermodynamic limit -- both of which strongly suggests that it is gapless in this limit \cite{read2009nonabelian}.   Indeed, arguments have been made that this wave function represents a particular critical point \cite{dmitrygreenthesis}.

Similarly, the Gaffnian is a correlator of a non-unitary conformal field theory \cite{Gaffnian}, and \textcite{ReadEdgeTheory,read2009nonabelian} has given a series of arguments  demonstrating that no non-unitary CFT can represent a gapped phase of matter. Recently, numerical studies on the Gaffnian verified that its excitation spectrum is gapless 
\cite{jolicoeur2014absence,papic2014solvable,Weerasinghe2014thin} and that charge is indeed not screened \cite{screeningunpublished,wuestienne,Estienne2015correlation}. This is an unfortunate  feature of the clustering approaches -- focusing on the single component case for simplicity, none of these clustering approaches have convincingly yielded {\it any} new gapped phases of matter, beyond the Read-Rezayi series.

One of the clustering approaches to building new wave functions is to identify polynomials (wave functions) with certain clustering or vanishing properties.  For example, the Laughlin $\nu=1/3$ wave function is uniquely identified by being the lowest degree antisymmetric polynomial in $z_1, \ldots, z_N$ which vanishes as three or more powers when any two coordinates $z_i$ and $z_j$ approach the same point.  This direction of thought has resulted in the detailed study of Jack polynomials \cite{BernevigHaldane,GreiterBulletin}, which are a family of special polynomials having well defined vanishing properties, where each polynomial can be identified with a particular CFT \cite{Feigin,BernevigGurarieSimon,Estienne}.  While this Jack approach successfully describes the Read-Rezayi series, and is very powerful for both numerical and analytic computation \cite{BernevigRegnaultAnatomy,BernevigHaldane2}, all of the other CFTs it describes (including the Gaffnian for example) are non-unitary and therefore cannot  describe gapped phases of matter \cite{ReadEdgeTheory,read2009nonabelian}.

A very closely related approach is the so-called ``pattern of zeros'' approach \cite{WenWang1}.   In this case, one tries to fully describe polynomials by the way in which zeros emerge as various numbers of particles approach each other.   Again this works well for the Read-Rezayi series, and can also define wave functions such as the Gaffnian and Haffnian (which, as discussed above are not acceptable quantum Hall states).  However, the approach runs into trouble in many other cases since the pattern of zeros is not generically sufficient to uniquely define a polynomial \cite{SimonS3,Jackson,SimonRezayiCooperPseudo} and must be supplemented by further information \cite{WenWang2} to fully define a wave function, which ends up being almost the same as simply defining a full CFT.

Another closely related approach is to examine quantum Hall states in the limit of a thin torus or cylinder geometry \cite{BergholtzThinTorus1,BergholtzThinTorus2,SeidelThinTorus}.  
In this ``TT-limit"\footnote{Here TT can stand either for thin torus, or Tou and Thouless, who first considered this limit \cite{tao1983fractional}.} the problem of finding the ground state and low energy excitations of strongly interacting quantum particles is reduced to finding the lowest energy configurations of  classical, static charges, see e.g. \cite{thintoruslong} for a detailed discussion on this. 
In the TT-limit,  the quantum Hall ground states become simple charge density waves\footnote{ Note that there is a close relation to the charge density waves occuring in one-dimensional coupled electron-phonon systems as studied by  \textcite{SSH} and later works \cite{SuSchrieffer81, JackiewSchrieffer81, GoldstoneWilczek81}. These systems also harbor fractionally charged excitations on the domain walls between different ground states.} where the pattern of occupied and unoccupied orbitals maps to the pattern of zeros, or vanishing properties in the above described clustering approaches.  Indeed, each Jack polynomial can be directly mapped to a TT-pattern of occupancies, known as its ``root state".   Furthermore the domain walls in these charge density wave states map precisely to the operators of the corresponding CFT \cite{ardonne2009domain} and give a very simple way to understand many of  the complicated quasiparticle properties of these quantum Hall wave functions.   Nonetheless, with the exception of the  abelian hierarchy states and the Read-Rezayi series, the CFTs are all nonunitary. 

 There is an  interesting alternative way of looking at the TT-limit that is also explained by \textcite{thintoruslong}. Rather than thinking of making a torus very thin, one can equivalently keep the geometry fixed, but change the interaction to become very anisotropic. Viewed in this way, the TT-limit amounts to a RG flow,  towards (an admittedly unphysical) fixed point Hamiltonian.  Conversely, moving
from the TT-limit to a realistic Hamiltonian, the topological properties, that are manifest in the TT-limit, will persist as long as the excitation gap does not close.

Yet another clustering approach is to define model Hamiltonians that enforce a particular clustering behavior \cite{SimonRezayiCooperPseudo,SimonRezayiCooperHamiltonian}. Quite similarly to the approaches discussed above one can design a Hamiltonian to forbid clusters of $p$ particles having relative angular momentum less than some number $m$.   Again this works well for the Read-Rezayi series but beyond this, appears to generate gapless states.  

Given that there are many CFTs that are unitary (in fact an infinite number!), it seems that one should be able to define a Hamiltonian whose ground state is the correlator of some {\it rational, unitary} CFT that would evade the arguments given by  \textcite{ReadEdgeTheory}.   
Indeed, in several cases  a Hamiltonian has been constructed so that its ground state is uniquely given by the correlator of a  unitary CFT  which is different from those defining the Read-Rezayi series \cite{SimonS3,Jackson}.  Unfortunately, unitarity of a CFT still does not guarantee that the system is gapped.  Indeed, in no case considered so far, except the Read-Rezayi series, has anyone written a single-component quantum Hall wave function which is a simple correlator of primary fields of a CFT and which appears to be gapped. Nonetheless, there have been many varieties of hierarchies attempted from the field theoretical or CFT viewpoints \cite{Zee,Cappelli,Frohlich,Frohlich2,Flohr,Estienne1}.
It remains to be seen if any of these will actually result in gapped ground states of some (even slightly realistic) Hamiltonian.



\section{Relation to Experiment}
\label{sec:experiment}

While much of the theoretical study of hierarchies (and fractional quantum Hall effect in general) focuses on the construction of trial wavefunctions, at the end of the day one is inevitably interested in explaining some physical experiment (whether it be real or hypothetical).    As mentioned above (end of section \ref{ssec:hierarchy}) there are special short range interactions for which the Laughlin states are exact ground states \cite{trugman,haldane1983fractional,Pokrovsky}, and similarly for the Moore-Read state \cite{moore1991nonabelions,greiterwenwilczek} as well as the Read-Rezayi series \cite{read1999beyond}.   Yet there is no known special interaction  which yields the ground state for any other spin polarized gapped quantum Hall state (including all of the hierarchy or Jain states except the Laughlin states).  While such short range interactions are perhaps natural in a cold-atom experimental context \cite{CooperReview,SVreview,CooperDalibard,Lukin} or  lattice-based quantum Hall realizations\footnote{Fractional quantum Hall effect of particles on a lattice, also known as ``Fractional Chern insulators",  are a huge subfield.   We will not give any discussion of these interesting proposals, but we refer the reader to the review by \textcite{BergholtzReview}.}
, such interesting systems remain proposals and have not yet been realized experimentally\footnote{An extremely interesting preprint \cite{Gemelke} suggests that one can get small numbers of cold atoms into the Laughlin regime.}.  For real physical systems where the fractional quantum Hall effect has been realized, interactions are much more complicated, and it is usually impossible to make any exact statements.   

\subsection{Connection of Theory to Real Experiments}
\label{sub:connection}

If we focus on real experiments, then the only systems ever to show the FQHE are two-dimensional electron gases in semiconductors.   While the vast majority of fractional quantum experiments have been performed in some variety of GaAlAs heterostructures, a few other semiconductor systems have also allowed observation of FQH physics, including Si MOSFETS \cite{FURNEAUX1986154}, SiGe \cite{SiGeFQHE,PhysRevB.46.7935,PhysRevLett.93.156805}, ZnMnO \cite{OxideFQHE}, and particularly graphene \cite{GrapheneFQHE1,GrapheneFQHE2,dean2011multicomponent}. 

In any given case of experimental interest, the  Hamiltonian should reflect the particular electron-electron interaction relevant to the physical system in question.  While one might naively guess that the physical interaction between electrons is simply the Coulomb interaction, the situation is actually quite a bit more complicated. Wavefunctions of electrons in two-dimensional electron systems typically have a substantial width in the direction transverse to the two dimensional plane.  As a result, one typically models the Coulomb interaction as being softened at short distances depending on the shape of the transverse wavefunction \cite{SternHoward,ZhangDasSarma,GMP1986}.   There may also be some amount of screening due to nearby metallic layers, such as electrostatic gates, which must be taken into account.   When all of these things are done, one obtains an electron-electron interaction within a single Landau level which can be studied to try to predict what FQH states will be seen in a given experiment, and what their  properties will be.   

As mentioned in App.  \ref{app:QHE}, independent of which  Landau level is being studied, it is always possible to map the problem to an effective interaction within the LLL \cite{HaldaneGirvin}, which is often done for simplicity.     We emphasize, however, that the effective interaction resulting from the physical problem of a partially filled $n^{th}$ Landau level is extremely dependent on the value of $n$.  For typical GaAs samples, in the LLL ($\nu<2$ since there are two spin species), the effective interaction is fairly strong at short range, which tends to favor the realization of the Jain series of quantum Hall states \cite{Pan2008}, see Fig.~\ref{LLL}.  On the other hand, in high Landau levels, the effective interaction is longer ranged, as well as  oscillatory.  This type of interaction favors charge density wave ground states, such as stripes and bubbles \cite{FoglerStripes,MoessnerStripes}.  This is in agreement with experiment: no quantum Hall state has been observed\footnote{Possible exceptions are filling fractions 4+1/5 and 4+4/5 which look like FQHE states at intermediate temperature \cite{FourandoneFifth}.} for Landau levels greater than the first excited level (i.e. $\nu> 4$),  whereas anisotropic compressible states (presumed to be stripe or nematic \cite{FradkinKivelsonLiquidCrystal})  are  observed in ultra-clean samples for filling fractions between 9/2 and 15/2  \cite{LillyStripe,DuStripe}.  The intermediate case of the partially filled second Landau level ($2 < \nu < 4$ for typical GaAs samples) is potentially the most interesting having both fractional quantum Hall physics, as well as some charge density wave physics \cite{RIQHE}.  It is in this regime that the potentially non-Abelian quantum Hall states are experimentally observed at $\nu=5/2$ and $\nu=12/5$ \cite{WillettFiveHalfs,Pan1999,kumar2010nonconventional,Pan2008}, see also Figure~\ref{2ndLL}.  
The reason that this regime is special stems from the form of the intermediate range of the Coulomb interaction projected to the second Landau level, which promotes clustering of electrons, which in turn appears to favor non-Abelian quantum Hall states.  (See also the discussion at the start of section \ref{sect:nahier}).

Theoretically once an interaction is determined for a physical system of interest there are several common methods of analysis.  The simplest is to choose several trial wavefunctions and compare their interaction energies (i.e., their ground state energies within the Landau level) using Metropolis Monte-Carlo integration \cite{chakrabortybook}.  Although this method is quite crude (since one is simply guessing the ground state wavefunction) it has the advantage that such Monte-Carlo calculations can be performed for extremely large systems, with numbers of electrons possibly over a hundred. 

A second method of analysis (and indeed, the gold-standard of fractional quantum Hall theory for a generation) is to perform exact diagonalization of a system for a small number of interacting electrons (typically on a closed surface such as a sphere \cite{HaldaneRezayiSphere} or  torus \cite{Haldane1985periodic} to avoid edge effects).   Once one obtains the ground state, one usually attempts to identify which phase of matter it belongs to by examining its properties, such as its shift (see Eq.~\eqref{eq:shift}) or its excitation spectrum.  More recently the entanglement spectrum (see section \ref{sect:edgesandentanglement})  has been used as a fingerprint for phases of matter. 

A hybrid approach, known as composite fermion diagonalization, has been pursued by Jain \cite{jain2007composite}.   In this work one chooses a small basis of trial wavefunctions corresponding to states that would be low-energy in composite fermion language.   Using Monte Carlo, one can calculate matrix elements in this basis and then diagonalize the resulting matrix.  This approach has the advantage of allowing study of larger systems than exact diagonalization, but presumably introduces a bias towards composite fermion states. 

Very recently another set of extremely powerful tools have been added to the toolbox of numerical techniques.   Using CFT ideas, it is now possible to express certain quantum Hall states, such as the Laughlin, Moore-Read, and Read-Rezayi states in matrix product representation \cite{zaletel2012exact,EstienneMPS1,EstienneMPS2}.  This highly efficient encoding of the wavefunciton enables numerical calculations on very large systems (infinitely long cylindrical geometry with circumference of 30 magnetic lengths) for these particular wavefunctions obtaining quantities such as  quasiparticle braiding statistics, correlation length, pair-correlation function, and so forth.   Very closely related to this technique is the density matrix renormalization group (DMRG) method \cite{DMRG},  which is effectively a variational matrix product approach, which can be used as a replacement for exact diagonalization but for much larger systems, similar to that for matrix product calculations \cite{zaletel2013topological,Shibata,ScottThird,iDMRGlayers,MongTwelveFifths,Kovrizhin,KarlhedeDMRG}.

For simplicity one typically views  the fractional quantum Hall effect as existing within a single partially filled Landau level.  However, in real semiconductor systems, the Landau level mixing parameter, the ratio of the interaction energy $E_{coulomb} = e^2/(\epsilon \ell)$, to the cyclotron energy $\hbar \omega_c = e B/(m^* c)$ is often of order unity, meaning that virtual transitions between Landau levels can potentially be important.    While often it is assumed that such transitions only have a small effect on the physics, in cases where two possible states of matter are very close in energy, the small effects of Landau level mixing can be crucial in determining which one is the ground state.   One case where Landau level mixing terms are obviously crucial is in determining\footnote{Landau level mixing also appears quite important in determining the nature of the experimentally observed $\nu =12/5$ fractional quantum Hall state and the apparent charge density wave at $\nu=13/5$ \cite{MongTwelveFifths,PakrouskiTwelveFifths}.} whether the quantized Hall state experimentally observed at $\nu=5/2$ is the Moore-Read Pfaffian \cite{moore1991nonabelions}, or its particle-hole conjugate, known as the anti-Pfaffian \cite{AntiPfaffian1,AntiPfaffian2} --- the two states being energetically equivalent for any two-body interaction without Landau level mixing.   There are two controlled methods of handling Landau level mixing theoretically, the first being to integrate out inter-Landau-level  transitions \cite{PetersonNayak,Sodemann,SimonRezayiMixing2,HaldaneRezayiLLMixing,QuinnLLMixing} at leading order in the Landau level mixing pararameter ($E_{coulomb}/\hbar \omega_c$) to obtain a modified interaction within a single Landau level\footnote{Beyond leading order one obtains retarded terms, which make it  impossible to represent the resulting interaction terms as simply an equal-time interaction within a single Landau level.}.  A second approach is to include multiple Landau levels within an exact diagonalization or DMRG calculation, and attempt an extrapolation to a system which includes all the Landau levels, as well as the usual extrapolation to large system size \cite{RezayiSimonLLMixing1,iDMRGlayers,MongTwelveFifths}.  In principle the two approaches should agree with each other at least for small values of $E_{coulomb}/(\hbar \omega_c)$.  
However, there are cases where they do not agree (see in particular the discussion regarding $\nu=5/2$ below), and the cause of the conflict remains undetermined.\footnote{This conflict appears to have recently been resolved by Ed Rezayi (to be published), with the conclusion being that the Antipfaffian is favored at low values of $E_{coulomb}/(\hbar \omega_c)$.\label{f:Rezayi}}

\subsection{Status of Real Experiments}

Perhaps the most surprising thing about fractional quantum Hall effect is that it is so definitively observed in experiment.  (Indeed, we should not forget that it was discovered in experiment \cite{tsuidiscovery} before it was explained theoretically!\footnote{It is interesting to debate whether anyone would have believed quantum Hall effect (integer much less fractional) theory if it had been provided before the experimental observation.})  In this section we will discuss what is known from some key experiments.    It is obviously impossible for us to discuss all of the beautiful experiments that have been performed on FQH systems so we have instead focused on those that are most interesting in the context of this review.

\subsubsection{The Nature of Plateaus}\label{sss:fractions}

The observation of any fractional quantum Hall plateau always raises the question of what the properties of that plateau are, that is,  what phase of matter are we observing?   Focusing on high mobiltity GaAs heterostructures within the LLL ($\nu<2$) over 60 FHQ plateaus have been observed \cite{Pan2008,pan2003fractional}.  Of these, all but a very few are of the Jain form $\nu=q/(2pq\pm1)$ or are closely related (particle-hole conjugate for example) and are assumed to be topologically equivalent to simple Jain or Halperin-Haldane hierarchy states.   A very few fractions observed within the LLL do not fit this obvious Jain form, including 4/11, 7/11, 4/13, 5/13, 5/17, 6/17 and 4/19.  None of these are fully formed fractional quantum Hall plateaus, but the evidence for 4/11 (and possibly 5/13) in particular appears fairly good (with higher denominators getting successively weaker as suggested by Fig.~\ref{figure1}).   These anomalous states could potentially all be explained within the hierarchy construction.   Nonetheless there have been a number of other proposals for the nature of these states which are discussed in \cite{mukherjee2014enigmatic}. The most recent study \cite{mukherjee2014enigmatic,mukherjee2014enigmaticErr} has been interpreted as suggesting quantum Hall states arising from an exotic mechanism proposed by \textcite{YiWojsQuinn}\footnote{
This work stems from the proposal that a different kind of quantum Hall state for $\nu=1/3$ can be produced by having an effective hollow-core-like interaction, generating a rotationally invariant ground state on the sphere for $\nu=1/3$ at a shift $S=7$ instead of the usual Laughlin value $S=3$.   
Using the corresponding $\nu=4/3$ wave function as $\Psi_{CF}$ in Eq.~\eqref{cfwfrel} (with $p=1$) then results in a trial wavefunction for $\nu=4/11$ and similarly using $5/3$ gives a trial state for $\nu=5/13$. 
A toy model of this type of wavefunction can be built at $\nu=1/5$ by using an interaction that forbids pairs with  relative angular momentum 3 but allows pairs with relative angular momentum 1.  This model interaction generates a rotationally invariant ground state at shift $S=9$ rather than the Laughlin value of $S=5$.  Such a ground state is unfortunately gapless suggesting that all such states in this family might be also.   Examining the data in \cite{mukherjee2014enigmatic} it looks very likely that the $5/13$ state considered there is indeed gapless, but it is less clear for $4/11$. }, although this conclusion is certainly not generally agreed upon. 

In the second Landau level ($2 < \nu < 4$), where roughly a dozen FQH plateaus have been observed, the situation is potentially even more interesting.  There have been many proposals that many of these states are non-conventional in one way or another.
   
{\bf 5/2:}  The most prominent of the states in this range is the 5/2 state (and its sister the 7/2 state).  Since the seminal work of \textcite{Morf}, it has been increasingly clear (and now fairly solidly established numerically) that this state is either in the Moore-Read Pfaffian phase \cite{moore1991nonabelions} or in its particle-hole conjugate phase, known as the anti-Pfaffian \cite{AntiPfaffian1,AntiPfaffian2}.   As mentioned above, in the absence of Landau level mixing, the two possibilities are equal in energy.   Deciding which is realized  has been difficult.  While \textcite{RezayiSimonLLMixing1,iDMRGlayers} have found the anti-Pfaffian is favored, the work of \textcite{PakrouskiFiveHalfs} has found that the Pfaffian is favored.  It is not yet clear why these do not agree (see however footnote \ref{f:Rezayi}).   The only experiment that attempts to directly distinguish the two possibilities \cite{Radu} was found to be more consistent with the anti-Pfaffian.\footnote{\label{KastnerFootnote} Repeats of this experiment  \cite{Kastner,BaerPointContact} found that the 331 state was favored --- which is very hard to understand, being completely contradictory to a large body of numerical work, and potentially casts doubt on the interpretation of the experiment.} 
The experimental observation {by \textcite{DolevNeutral} of upstream neutral edge currents (flowing opposite the direction of charge transport) seems more natural for the anti-Pfaffian, but could also occur for the Pfaffian if there were edge-reconstruction.     
Finally we mention that recent interferometry experiments  (see \cite{WillettFiveHalfInterferometer} for example) have been interpreted in a simple picture in terms of the anti-Pfaffian, but not the Pfaffian \cite{Keyserlingk}.   
This conclusion should not be taken as definitive given the controvery over interpretation of these experiments (see Sec. \ref{ssec:statistics} below).

{\bf 7/3, 8/3:}  The overlap of the Coulomb ground state with the conventional Laughlin wavefunction is found to be quite low for 7/3 and 8/3 \cite{JainSevenThird}.   It was later suggested by \textcite{PetersonEightThird} that (when Landau Level mixing is included) the 8/3 state might be the $Z_4$ Read-Rezayi phase, and this has been supported using exact diagonalization of up to 12 electrons.  However, more recent DMRG calculations for much larger systems, including Landau level mixing terms, conclude that the 8/3 state is of Laughlin type \cite{iDMRGlayers}.   The experiment of \textcite{BaerPointContact} concludes that both 7/3 and 8/3 are of Laughlin type (although there is some question of the reliabiltity of this type of experiment, see footnote \ref{KastnerFootnote} above).

{\bf 12/5:}  The observation of a 12/5 plateau \cite{Pan1999} is now beyond any doubt \cite{kumar2010nonconventional}.   Due to the small size of the gap (around 80mK in the only experiments that have observed this fraction), it has been impossible to perform many experiments beyond simply the observation of the plateau.   Theoretically, the most recent works by \textcite{MongTwelveFifths,PakrouskiTwelveFifths} confirm that the 12/5 state is (the particle-hole conjugate of) the $Z_3$ Read-Rezayi state, and further that Landau level mixing suppresses the (experimentally unobserved) 13/5 state.

{\bf 2+6/13:} Perhaps the most intriguing fraction yet observed is at filling fraction 2+6/13 \cite{kumar2010nonconventional}.  
While this could potentially fit into a conventional hierarchy, it would be an extremely unusual  creature corresponding to  $q=6$ of the Jain series $q/(2q+1)$, where  $q=3,4,5$ are not observed. 
A very natural explanation of this state would be a first order daughter state of the anti-Pfaffian at 5/2 \cite{levin2009collective}.   
The physics of this hierarchy will be discussed in section \ref{levinhalperin} below.  
The experimentally measured activation gap for this state is exceedingly small at only about 10 mK.

{\bf  2+3/8}	In the LLL there is possibly some (albeit very weak) evidence for a $\nu=3/8$ plateau \cite{pan2003fractional} which has been discussed theoretically by \textcite{Mukherjee,ScarolaThreeEighths}.  (If verified this would be the first even denominator state in the LLL).  
However, in the second Landau level, the evidence for a 2+3/8 state is somewhat stronger, albeit with a very small measured activation gap of roughly 10 mK.    
Recent theoretical work by \textcite{hutasoit} comparing possible trial wavefunctions suggests that this state is of Bonderson-Slingerland type \cite{bonderson2008fractional}, which will be discussed in section \ref{sec:bs} below.

\subsubsection{Quasiparticle Charge Experiments}

One of the particularly interesting features of FQHE physics is the existence of fractionally charged quasiparticles.  
Although in some sense, the fractionally quantized Hall conductance is indirect evidence of fractionally charged quasiparticles, it has nonetheless been a key experimental quest to measure these fractional charges directly.   

The first clear measurements were made via shot noise obtaining charges of $e^* \approx e/3$ at $\nu=1/3$ \cite{dePicciotto1,Saminadayar} compatible with the Laughlin prediction, as well as $e^* \approx e/5$ at $\nu=2/5$  and $ e^\star \approx e/7$ at $\nu=3/7$ (see \textcite{HeiblumReview}).  
These latter charges are compatible with the prediction of the Halperin-Haldane hierarchy. 
More recent shot noise experiments by \textcite{DolevShot} have established charges of $e^* \approx e/4$ at $\nu=5/2$ compatible with the Moore-Read state or the  anti-Pfaffian\footnote{In fact, \textcite{levin09fractional} argued that the additional charge fractionalization of e/2 to e/4 has to occur for \emph{any} valid representative wave function, simply because of the even denominator filling fraction. }, as well as establishing charge $e/3$ at filling $\nu=8/3$ and $5/3$.    
Unfortunately, when the same experiments were performed at extremely low temperature and low voltage other values of charge were sometimes measured, and these results remain  
not fully understood \cite{Dolev2,HeiblumReview}. 

A second approach has been to measure charge motion with scanning tips, or stationary single-electron transistors.  Such experiments similarly give very clear evidence \cite{Martin} of $e^* \approx e/3$  at $\nu=1/3$ and $\nu=2/3$.
The same experiments attempted at $\nu=5/2$ were far more challenging for a number of technical reasons.   Although the expected result of $e^* \approx e/4$ was indeed measured \cite{Venkat}, the result is a bit less clear than for the same experiments at $\nu=1/3$.

\subsubsection{Quasiparticle Statisitics and Quantum Hall Interferometers}\label{ssec:statistics}

Perhaps the single most theoretically exciting quantity to be measured experimentally would be exotic braiding statistics (fractional or non-Abelian).   Unfortunately, despite many attempts over the years there is still no clear demonstration of quasiparticle statistics.    
The main proposals for achieving such a demonstration \cite{ChamonInterferometer} revolve around building Fabry-Perot interferometers out of two-point contacts.\footnote{There have also been impressive experiments with Mach-Zehnder interferometers in the integer quantum Hall regime, but not in the fractional regime.  
See for example, \textcite{Mach}.}
Early experiments reported a number of successes (see for example, \textcite{Camino}) but it was later realized by \textcite{RosenowHalperin} (with later theoretical extensions by \textcite{Neder,Keyserlingk}) that Coulomb charging effects may dominate the physics and as a result the statistical phase may be hidden.  
Further interferometry experiments confirmed the picture of ``Coulomb domination" in some detail \cite{Ofek23032010,Mcclure,Yiming,Choi}. 

Similar interferometry experiments were predicted to give particularly clear evidence of non-Abelian statistics for the $\nu=5/2$ state \cite{BondersonInterferometer,SternInterferometer} given a number of simplifying assumptions (see also the discussion in \cite{RosenowExact,Keyserlingk}).   
A series of detailed experiments were conducted attempting to see this physics (see \cite{WillettFiveHalfInterferometer} and references therein).   
However, the published data actually disagrees with predictions as detailed in \cite{Keyserlingk} and the situation remains controversial.   
Potentially, further experiments will clarify the situation.\footnote{Another experiment to see this statistical physics with an interferometer by examining phase slips claimed to observe both fractional and non-Abelian statistics \cite{AnKang}.  
However, as of yet, this work has not been reproduced.}

\subsubsection{Composite Fermion Experiments}

There have been quite a few experiments demonstrating the detailed physics of composite fermions near the half-filled (often lowest) Landau level.  In particular these have shown that there is indeed a fermionic quasiparticle which moves with a cyclotron radius $R_c^* = \hbar k_F / (e \Delta B)$ where $k_F$ is set by the density of particles and $\Delta B = B - B_{1/2}$ is the deviation of the magnetic field from the half-filled Landau level.   The existence of this length scale is exactly as predicted by \textcite{HLR}.
We refer the reader to \cite{jain2007composite} as well as \cite{heinonenbook} for detailed discussions of these experiments.



\section{Quantum Hall liquids from CFT }  \label{sect:CSCFT}

\subsection{Chern-Simons theory and CFT --- abelian quantum Hall states} \label{CSAbGS}

The presence of fractional statistics particles suggests there should be a Chern-Simons description of the topological properties of the system.  Indeed, this is exactly what is embodied in the Wen-Zee Lagrangian \eqref{wenlag} discussed earlier.  
Once the connection to Chern-Simons  theories is made, one can  exploit the many connections and mathematical results relevant to these well studied theories. 

The Chern-Simons term was first considered in a physics context by \textcite{deser1982topologically}, who showed that when added to a Maxwell, or in the nonabelian case a Yang-Mills, Lagrangian it gives mass to the gauge field without breaking the gauge invariance. Soon afterwards, it was realized that coupling non-relativistic particles minimally to a gauge potential with a Chern-Simons term effectively changes their statistics. 
Starting with the work of \textcite{witten}, it was realized that there is an intimate connection between Chern-Simons theories and CFT.   It is perhaps then not surprising that quantum Hall states can be written in terms of CFT correlators.

As the connection between Chern-Simons theory and CFT was developed, it was noticed that the polynomial part of the Laughlin wave functions  can 
be interpreted as holomorphic conformal blocks \cite{moore1991nonabelions,fubini}.
For example, the  wave function for two quasiholes in  Eq.~\eqref{laughlinholes} can be written as
\be{cftconn}
 \Psi(\eta_1, \eta_2;  \sgrupp [z] ) 
 &=& \av{H(\eta_1) H(\eta_2)  V(z_1)  \dots V(z_N) {\cal O}_{bg} } \nonumber\\
    &=&  (\eta_1 - \eta_2)^{1/m} \prod_{i=1}^N  (\eta_1 - z_i) (\eta_2 - z_i) \nonumber  \\ 
   &\times& \prod_{i<j} \jasf i j m e^{- 1/( 4 m \ell^2) (|\eta_1|^2 + |\eta_2|^2)}  \nonumber \\
   &\times& e^{- 1/(4\ell^2) \sum_i |z_i|^2},    
    \ee
where the average is taken with respect to the action of a massless chiral boson $\phi$, normalized as 
\be{twopoint}  
\langle \phi(z) \phi(z') \rangle = -  \log(z-z') \, .\nonumber
\ee 
 The vertex operators are given by
\be{vo}
V(z) &=& : e^{ i {\sqrt m} \phi (z) } : \nonumber\\
H(\eta) &=& :e^{  i \sqrt{1/m} \phi (\eta) } : \label{hole} \, ,
\ee
 and ${\cal O}_{bg} $ provides a neutralizing background charge that is discussed in more detail below. 
Although closely related,  $\phi$ is distinct from the $\varphi$ used in section \ref{sub:edge}. 
In particular, their normalizations differ.  More details on this chiral boson CFT are found in  appendix \ref{app:CFT}.  
The value of the correlator in Eq.~\eqref{cftconn} is almost precisely that of Eq.~\eqref{laughlinholes} except for the $(\eta_1 - \eta_2)^{1/m}$ factor in Eq.~\eqref{cftconn}, which is holomorphic (although with a branch cut), while Eq.~\eqref{laughlinholes} has a factor $|\eta_1 - \eta_2|^{1/m}$. 
The difference is just a choice of gauge;  the one used in Eq.~\eqref{cftconn}  is often called ``holomorphic'' or ``fractional statistics'' gauge.  
We should keep in mind that the $z$ variables are actual positions of the physical electrons in the system, whereas the $\eta$ variables are just parameters in the wave function --- and we are free to multiply  the wave function with any $\eta$-dependent phase factor.
 
As mentioned in Appendix \ref{app:CFT}, correlators of vertex operators vanish if the fields inside the correlator do not fuse to the identity. This `charge neutrality' is ensured by the insertion of the background charge operator $\mathcal{O}_{bg}$. In principle, there is a freedom of where to put this neutralizing charge.
However, it turns out that if the operator is chosen as  \cite{moore1991nonabelions}
\begin{equation}
\label{eq:background}
  {\cal O}_{bg}  = \exp\left(-i  \rho \sqrt m \int d^2z\,  \phi(z) \right), 
\end{equation}
where $\rho$ is the (constant) fermion density, it provides the  gaussian factors needed to directly interpret the correlators as LLL wave functions. 
Making this operator well-defined is a bit tricky, as discussed in some detail by \textcite{hansson2009quantum}.   On geometries other than the plane,  the gaussian factors  look different, but, {\it mutatis mutandis}, the same recipe will work (see section \ref{sect:sphere} below).

In general, primary fields of the CFT correspond to the Wilson line operators of the corresponding Chern-Simons theory. In the above case, the gauge group is a single $U(1)$  at ``level'' $m/2$.
The operator content of the CFT contains all clusters of multiple elementary quasiholes $H$ and each of these should correspond to a different species of Wilson line.  The Wilson lines generally braid nontrivially with each other, but the electron operator (which topologically amounts to a cluster of $m$ elementary quasiparticles) should actually be topologically trivial. This statement is precisely true if the ``electron'' is a boson, which is true for even $m$.  For odd $m$, braiding is trivial, but exchange of two fermions accumulates a statistical sign.   


\subsubsection{Quasiparticle braiding and monodromies} \label{qp&mon}

\label{subsub:mon}
We now revisit the question of fractional charge and statistics, and use the wave function \eqref{cftconn} to derive the Berry phase in Eq.~\eqref{eq:gammares} for braiding one quasihole at position $\eta_1$ around another at $\eta_2$.
Using the expression \eqref{cftconn} (rather than \eqref{laughlinholes}) for the wave function $\Psi$, and 
the decomposition
$$
\frac{d}{d\tau} = \left( \frac{d \eta_1}{d \tau} \right) \partial_{\eta_1} + \left( \frac{d \bar \eta_1}{d \tau} \right) \partial_{\bar \eta_1} \, ,
$$
it is particularly easy to calculate the Berry phase \cite{gurarienayak,nayakwilczek}. 
In the expression $\bra \Psi \partial_\tau \ket\Psi$  in Eq.~\eqref{eq:berry1} the  $\partial_{\bar \eta_1}$    only acts on the exponent, bringing down a factor of $-\eta_1 / (4 m \ell^2)$.  To handle the $\partial_{\eta_1}$ we integrate by parts so that $\langle \Psi | \partial_{\eta_1} | \Psi \rangle$ is replaced by $(-\partial_{\eta_1} \langle \Psi|) |\Psi \rangle$ noting that $\langle \Psi|\Psi \rangle =1 $ so the total derivative vanishes.   Once in this form, the $\partial_{\eta_1}$ again operates only on the exponent to give $\bar \eta_1 / (4 m \ell^2)$.
 Thus we obtain,
\be{berryresult}
 i \gamma &=& \frac{1}{4 m \ell^2} \oint d \tau   \left[ \bar \eta_1 \frac{d \eta_1}{d \tau}   - \eta_1 \frac{d \bar \eta_1}{d \tau} \right] \nonumber \\
 & =&  \frac{i A}{m \ell^2}   = i  \frac {2\pi} {m} { \Phi} \, ,
\ee
with $A$ being the (signed) area enclosed by the path of $\eta_1$ and $\Phi$ the corresponding enclosed flux in units of the flux quantum.   This is simply the  Aharonov-Bohm phase associated with moving a charge $+e/m$ around the  area $A$. 

At this point, it appears that there is no effect of the other quasihole, $\eta_2$.  Indeed, the position of $\eta_2$ does not enter the Berry phase calculation here.  
However, examining the holomorphic form of the wave function \eqref{cftconn} we see that the wave function is actually multi-valued; mathematically this amounts to having a branch cut in the complex $\eta_2$ plane, or equivalently, defining this variable on a Riemann surface with $m$ sheets.   
The wave function itself accumulates a phase of $e^{2 \pi i/m}$ when $\eta_1$ is moved around $\eta_2$.  
This explicit phase in the wave function is known as the {\it monodromy}.   
Thus the total phase accumulated as one particle is moved around the other is the sum of the Berry phase and the monodromy. The latter gives twice the expected statistical angle $\theta=\pi/m$, precisely the same as that obtained from Eq.~\eqref{eq:gammares}.  
In the current calculation, the fractional statistics is explicit in the phase of the wave function and the Berry phase gives only the 
Aharonov-Bohm phase, whereas if one works in the gauge of Eq.~\eqref{eq:Ngauge2} there is no monodromy and both the statistical phase and the Aharonov-Bohm phase are parts of the Berry phase \cite{kjonsberg1997anyon}.   
Both of these approaches are correct, in that they give the same total phase.   However, the holomorphic gauge appears much simpler for most purposes, and it will generalize very easily to the nonabelian case treated in section \ref{subsub:monhol}.


\subsection{Multicomponent abelian states}  \label{multicomponent}
Although the electrons are spin half particles, the spin degree of freedom can often be neglected in the context of the quantum Hall effect, since it is quenched by the strong magnetic field. This is, however, not always the case \cite{girvin1995multi}, and in general one must consider both components \cite{halperin1983theory,du1995fractional}. Other situations that require a multicomponent, or multilayer, description are bilayers \cite{yoshioka1989fractional,lopez1995fermionic,EisensteinBilayer}, where the layer index is a new quantum number, and semiconductors, like graphene, where there is a valley degeneracy \cite{dean2011multicomponent}. 

All these cases can be described in the $K$-matrix formalism, and the recipe for finding the corresponding CFT wave functions is a straightforward generalization of the Laughlin case: For a state with $n$ components, or layers, the integer valued positive definite $K$-matrix  is symmetric and has rank $n$,  and it can always be factorized  as,
\be{kfact}
{\mathbf K} = {\mathbf Q} {\mathbf Q}^T ,
\ee
where $\mathbf Q$ is a $n\times k$ matrix and $k\geq n$, since the rank of $\mathbf K$ is $n$.\footnote{
An alternative notation is to write  $ K_{\alpha\beta} = {\vec Q}^{(\alpha)} \cdot {\vec Q}^{(\beta)} $ where $ {\vec Q}^{(\alpha)}$ are $n$ different $k$-dimensional ``charge vectors", with elements $Q^{(\alpha)}_\beta = Q_{\alpha\beta}$. 
The vertex operators are then written as $V_\alpha = \, :~\exp( i \vec Q^{(\alpha)}   \cdot \vec \phi  )~: $,    
where  ${\vec\phi} = (\phi^1 \dots \phi^k)$.   }
We can then form $n$ electron operators, using $k$ chiral boson fields:\footnote{While we need at least $n$ chiral bosons to represent the state, in some cases it is advantageous to use more -- for instance when representing the Jain series on the sphere \cite{kvorning2013quantum}.}
\be{multielop}
V_\alpha = \, :e^{iQ_{\alpha\beta} \phi^\beta}:  \, ,
\ee
where the chiral bosons satisfy 
$$
 \langle \phi^\alpha(z) \phi^\beta(z') \rangle = - \delta_{\alpha\beta} \log(z-z') \, .
$$
To neutralize a correlator with $N_\alpha $ particles in the $\alpha^{th}$ layer, we insert a background charge 
\be{multibg}
{\mathcal O}_{bg} = \exp \left( -i \int d^2 r\, \rho(\vec r)  n_\alpha Q_{\alpha\beta} \phi^\beta(z) \right) \, ,
\ee
where $\rho$ is the fermion density (which usually is assumed to be constant), so $\int d^2 r\, \rho(\vec r) = N$. 
The neutrality condition for the field $\phi^\beta$ becomes,
\be{neutrcond}
N_\alpha Q_{\alpha\beta} = n_\alpha N Q_{\alpha\beta}\nonumber
\ee
from which follows $N_\alpha = n_\alpha N$, so $n_\alpha$ is just the relative density of the layers.\footnote{
Note that there is a $U(1)^n$ symmetry, and thus $n$ conserved Abelian charges (the {\it electric} charge is $-e\rho$). 
In many physical situations this charge is not precisely conserved and strictly speaking the wavefunction should be superposition of sectors having different numbers of particles in the different ``layers".   For example, the bilayer 111 state can be more properly written as $\prod_m  ( c^\dagger_{m\uparrow} + c^\dagger_{m\downarrow})| 0 \rangle$.  However, for many purposes it is sufficient to think about a single sector with fixed particle number in each layer. Note that the $K$-matrix of such a state is 1, not $\left(\begin{array}{cc} 1 & 1\\ 1&1 \end{array}\right)$, as we always require it  to be invertible, see \eqref{wenres1}.
 }
The relative filling fraction $\nu_\alpha$ is given by 
 \be{relfill}
  \nu_\alpha = N_\alpha/N_\Phi = n_\alpha \nu = K_{\alpha\beta}^{-1}t_{\beta} \, .
  \ee

In most cases one takes $k=n$ \ie uses as many fields as there are layers, but this is not necessary. Even for a given $k$ the factorization is not unique, as shown in the example below. One can think of $\mathbf Q$ as the matrix square root of the $\mathbf K$; in the Laughlin case where $\mathbf Q$ is a scalar, we have $\mathbf Q = \sqrt{K} = \sqrt m$. The multi-particle wave functions will be built by products of factors like,
\be{cftmulti}
\av{ V _\alpha (z_i) V_\beta(z_j)  } = (z_i - z_j)^{K_{\alpha\beta}}    \nonumber
\ee
and of course also the proper gaussian factors.
As an example, we show how this works\footnote{ The CFT vertex operators for the Halperin states were first constructed by \textcite{moore1991nonabelions}.} for the Halperin $(m,m,n)$ states  \cite{halperin1983theory} with $m>n$.  The corresponding  $\kb$ can be factorized as,
\be{halperin}
\kb &=& \left(\begin{array}{cc} m & n \\ n & m\end{array}\right)   \\ 
      &=& \left(\begin{array}{cc}\sqrt m & 0                       \\ \frac n{\sqrt m} & \sqrt{\frac {m^2-n^2} m}\end{array}\right) 
              \left(\begin{array}{cc}\sqrt m & \frac n {\sqrt m} \\  0              & \sqrt{\frac {m^2-n^2} m}\end{array}\right) \nonumber \\
      &=& \left(\begin{array}{cc} \sqrt{ \frac {m+n} 2} &   \sqrt{ \frac {m-n} 2} \\  \sqrt{ \frac {m+n} 2} &  - \sqrt{ \frac {m-n} 2} \end{array}\right)
       \left(\begin{array}{cc} \sqrt{ \frac {m+n} 2} &   \sqrt{ \frac {m+n} 2} \\  \sqrt{ \frac {m-n} 2} &   -\sqrt{ \frac {m-n} 2} \end{array}\right), \nonumber
\ee
where we showed two different ways to factorize. In both cases the ground state wave function is obtained by taking an equal number of the operators $V_1$ and $V_2$ in the correlator, and it is easy to show that using the second, symmetric, factorization, only the field $\phi_1$ is related to the electric charge and requires a background. As a concrete example, the vertex operators corresponding to the first factorization are 
\be{}
V_1 &=& \, :e^{i \sqrt m \phi_1}: \nonumber \\
V_2 &=& \, :e^{i(n/\sqrt m)\phi_1 + i \sqrt{(m^2 - n^2)/m}\, \phi_2}:  \, ,\nonumber
\ee
giving the Halperin wave function
\begin{align}\label{mmn}
\psi_{mmn} = \prod_{i<j} (z_i^{\uparrow} - z_j^{\uparrow})^m  \prod_{k<l} (z_k^{\downarrow} - z_l^{\downarrow})^m  \prod_{r,s} (z_r^{\uparrow} - z_s^{\downarrow})^n
\end{align}
with ${\uparrow}$ and ${\downarrow}$ denoting the (pseudo)spin up and down species, respectively.

It is also possible to generalize Laughlin's plasma analogy to the multicomponent case, and thus to extract the topological properties of the states, as long as the equivalent multicomponent  Coulomb plasma is in a screening phase \cite{qiu1998phases}. 


\subsection{Nonabelian quantum Hall states and the Moore-Read conjecture  } \label{ssec:MRconjecture}

That the Laughlin wave functions can be represented as correlators in a CFT suggests that other CFTs might be used to generate valid quantum Hall wave functions.  Indeed, in their groundbreaking paper, \textcite{moore1991nonabelions}  suggested that quantum Hall states might generally be constructed in this way.    
The very close relationship between CFT and TQFT both allows an immediate identification of the topological properties associated with that quantum Hall state, and provides a natural way to describe the edge of both abelian and nonabelian states.  Not every CFT will produce acceptable quantum Hall states. 
One important condition is that the CFT is rational, which in this context means that there is a finite number of inequivalent particle types, and that the ground state on a torus has a finite degeneracy in the thermodynamic limit.   Further the CFT should be unitary \cite{read2009nonabelian}, as a nonunitary theory will inevitably result in a gapless state.   
However, it seems unlikely that these two conditions alone are sufficient to guarantee a well behaved quantum Hall state in many cases \cite{Jackson}.  

The physics of the quantum Hall/CFT connection, and the resulting study of  a number of nonabelian quantum Hall states, was reviewed in some depth by \textcite{nayakreview}. 
In this review we shall mainly restrict ourselves to the simplest of  these states, namely the Moore-Read, or Pfaffian state. 


\subsubsection{Electronic wave functions as conformal blocks} \label{wfasconfblock}
In order to implement the Moore-Read scheme, one  starts by choosing a CFT to work with.  There will always be a $U(1)$ charge sector given by a chiral boson,  as we explored above for the case of the Laughlin wave function.  
Multiple $U(1)$ sectors can be present if there are  multiple species of electrons, as in the multicomponent liquids  in section \ref{multicomponent}, or the hierarchies to be discussed in \ref{sect:expwfs} below.  For now, we assume only one $U(1)$ sector. The electrical charge of a particle is determined entirely by its $U(1)$ charge.  
In addition there may be a neutral sector of the CFT which might be a more complicated CFT, such as the Ising CFT in the case of the Moore-Read state. 
 Those who find the following general discussion too abstract, are referred to Sec. \ref{sub:MRstate} where the example of the Moore-Read state is treated explicitly. 

Within the CFT, we must choose some simple current field, which we will call $\psi_e$ to represent the electron (see Appendix \ref{app:CFT} for a definition of simple current).  The electron must also be a fermion, meaning that it has half-odd integer conformal weight (or scaling dimension).  Or, if one is considering the quantum Hall effect of bosons, the ``electron'' must be a boson, meaning that it has integer conformal weight.   The ground state wave function is given as the holomorphic chiral correlator or {\it conformal block}, 
$$
  \Psi(z_1,\ldots, z_N) = \langle \psi_e(z_1) \psi_e(z_2) \ldots \psi_e(z_N) {\cal O}_{bg} \rangle 
$$
where ${\cal O}_{bg}$ is the appropriate background field operator such that the correlator does not vanish.  Except for the gaussian factors, this wave function is fully holomorphic and is appropriate as a LLL wave function.   For fermionic fields $\psi_e$ the wave function is totally antisymmetric and for bosonic fields it is symmetric. 

We can further consider fields $\psi_{qh}$, yielding wave functions such as 
\be{confblock1}
  \Psi(\grupp[\eta]1 M;\sgrupp [z] ) &=& \langle \psi_{qh}(\eta_1) \psi_{qh}(\eta_2)\ldots \psi_{qh}(\eta_M)   \nonumber\\
  &\times&  \psi_e(z_1) \psi_e(z_2) \ldots \psi_e(z_N) {\cal O}_{bg} \rangle\nonumber
\ee
which represent quasiholes at positions $\eta_1$ \ldots $\eta_M$ inserted into the prior ground state wave function. In general there may be several different species of quasiholes, represented by distinct operators.    

In order to be a valid electron wave function, these correlators must be single valued, \ie  the electron operators must braid trivially with the quasihole operators and no singularities in the wave function must result.    
This condition corresponds to the following.   Consider taking a quasihole to the same position as an electron.  The combination will form some other type of quasiparticle excitation $\psi_f$, which we call the {\it fusion} of $\psi_e$ and $\psi_{qh}$.  (In  CFT language, we say $\psi_e \times \psi_{qh} = \psi_f$).    If the scaling dimensions of these operators are $h_e, h_{qh}$, and $h_f$ respectively, then examining the operator product expansion in Eq.~\eqref{eq:OPE} tells us that we must have
\begin{equation}
\label{eq:hinteger}
 h_f - h_e - h_{qh} = {\rm integer} \ge 0 
\end{equation}
for the wave functions to remain single valued (no branch cuts) and non-singular (no poles) as an electron approaches the position of the quasihole.  This provides a very strong constraint on the type of quasihole operator that can exist in our theory. 
The simplest example is that of a hole in a $\nu = 1/m$ Laughlin state. The operators \eqref{vo} are  fused using the operator product expansion $ \lim_{\eta\rightarrow z}V(z)\times H(\eta) \sim (z-\eta) :e^{i (m+1) \phi (z) /\sqrt{m} }: $, so, recalling that $V_\alpha = \, :e^{i\alpha\phi}:$ has the conformal dimension $h_\alpha = \alpha^2/2$ (\cf Appendix \ref{app:CFT}), we get $h_f - h_e - h_{qh} = \half (m+1)^2/m - \half m - \half 1/m = 1$ which fullfills the  relation \eqref{eq:hinteger}.


\subsubsection{Bulk-edge correspondence} \label{bulkedgecorr}
\label{subsub:bulkedge}

The massless boson CFT that describes the bulk of the Laughlin state is in fact the same massless bose theory that describes the Laughlin quantum Hall edge that we considered earlier in section \ref{sub:edge}.   Even the operators that create charge are, in this case, identical (compare the quasiparticle creation operators \eqref{hole} to \eqref{eq:vertexonedge}).    This similarity, which is a manifestation of what is known as bulk-edge  correspondence, is not a coincidence, but follows naturally from deep mathematical results on the quantization of Chern-Simons theories \cite{witten,cappelli1993conformal}. The general idea is that the Chern-Simons description of the quantum Hall state is essentially an isotropic $3=2+1$ dimensional theory.  If we ``cut" the 3-manifold   in a spacelike slice (at fixed time) we end up with the $2+0$ dimensional CFT that describes the ground state wave function\footnote{ In the CS description, particles are represented by Wilson lines and the statistical phase factors, abelian or nonabelian, can be extracted from the correlation functions of such lines. At a fixed time surface, the Wilson lines correspond to a number of points or ``punctures''. In his fundamental paper on the subject, Witten showed how to canonically quantize the CS theory in the presence of such punctures \cite{witten}. The outcome is that the corresponding \emph{ finite dimensiononal} Hilbert space is precisely the space spanned by the  blocks of a conformal field theory. Thus, identifying the coordinates in the conformal block as the (holomorphic) coordinates of the particles, it is quite natural to interpret the blocks themselves as \emph{bona fide} representative wave functions for the state in question, since the monodromies of these functions will properly represent the statistics of the particles, be they electrons or abelian or nonabelian anyonic quasiparticles. }.  On the other hand, if we cut the 3-manifold in a time-like slice (a fixed 1-dimensional curve crossed with time) we obtain a 1+1 dimensional CFT describing an edge.  

Also in the general case of hierarchy  and/or nonabelian states, there is a correspondence between the bulk CFT, whose correlators describe the stationary 2+0 dimensional wave function of the state, and the edge CFT that describes the dynamics of the 1+1 dimensional quantum Hall edge.     As discussed in section \ref{sub:edge}, there is always at least one chiral boson mode describing the  dynamics of the charged edge modes.   However, in addition there can also be   dynamical neutral modes corresponding to the non-boson part of the CFT. While there is often no simple representation of the  Lagrangian, the  mathematical power of CFT can be brought to bear on  this part of the problem \cite{di1997conformal}. 
For example, using CFT methods one can determine fine details of the dynamical neutral mode spectrum without ever having to write a Lagrangian \cite{roni2009experimental,Cappelli2010chiral,stern2010interference,bonderson2010coulomb}.


\subsubsection{Vector space of conformal blocks, braiding, and monodromy}
\label{subsub:vecspace}

As in the case of Eq.~\eqref{cftconn}, the wave functions need not be single valued in the coordinates $\eta$ of the quasiholes, since these are simply parameters of the wave function, rather than physical electron coordinates. We already discussed the Laughlin case where fixing the hole position $\eta_2$, the  single component wave function $\Psi_{1/m}(\eta_1,\eta_2)$ defines a function of $\eta_1$ on a Riemann surface with $m$ sheets. In the case of a more complicated CFT, and multiple quasiholes, the wave function has many components, each with a complicated analytic structure with multiple branch cuts. During a general braiding operation (that also can amount to exchange of identical quasiparticles) the components transform into each other by a unitary matrix. To make this more concrete,  let us consider the Moore-Read state at filling fraction $\nu=1/2$, that we discuss at length in section \ref{sub:MRstate}. The  correlator of four quasiholes only (no electrons) can be calculated by standard methods \cite{Zuber1977,belavin1984infinite} with the result, 
\be{blocks}
&\lim_{\omega \rightarrow \infty} \langle \psi_{qh}(0) \psi_{qh}(\eta) \psi_{qh}(1) \psi_{qh}(\omega)  {\cal O}_{bg} \rangle \nonumber  \\
& \sim  a_+ \Psi_+(\eta,\omega) + a_- \Psi_-(\eta,\omega),\nonumber
\ee
which holds for $|\omega | \gg |\eta |, 1 $ and where $a_\pm$ are arbitrary coefficients, and 
$$
\Psi_{\pm} \sim \omega^{1/4} \sqrt{1 \pm \sqrt{1 - \eta}} \, .
$$
 Put differently, the wave function, at fixed positions of the quasiholes, is given by the vector with coordinates $(a_+,a_-)$ in the basis of the two  conformal blocks $\Psi_\pm$.  
Letting the quasiparticle at $\omega$ encircle the three others clockwise, while  keeping $\eta$ fixed,  the wave function picks up an abelian phase $e^{i\pi/2}$ in close analogy to the Laughlin case. The interesting effect occurs when $\omega$ is kept fixed, and the quasiparticle at $\eta$ encircles the one at 1. This amounts to $\Psi_\pm(\eta ,\omega) \rightarrow \Psi_\mp (\eta,\omega)$, or equivalently,
$a_\pm \rightarrow a_\mp$, \ie  to a unitary rotation $\sigma_x$ of the  two dimensional state vector when thought of as a spinor.

Including the electrons to get the full quantum Hall wave function is more complicated, and was originally done for four quasiholes by \textcite{nayakwilczek} and later for the general case of  
$M$ holes and $N$ electrons by \textcite{ardonne2010chiral}. The explicit expressions are complicated, but schematically the result is
$$
\Psi(\eta_1, \eta_2 \ldots \eta_M;  \sgrupp [z])  = \sum_\alpha c_\alpha \Psi_\alpha(\eta_1, \eta_2 \ldots \eta_M; \sgrupp [z]).
$$
where the conformal blocks, labeled by $\alpha$, form a degenerate vector space. Moving the quasiparticles around each other and then returning to the same positions leaves the $\Psi_\alpha$ unchanged, but transforms the coefficients $c_\alpha$ by a unitary matrix,  
$$
   c_\alpha  \rightarrow  U^{({\rm braid})}_{\alpha,\beta} \, c_\beta,
$$
where the {\it monodromy matrix}, $U^{({\rm braid})}$, depends on the particular ${\rm braid}$ performed on the coordinates.   This is the essence of nonabelian statistics.  By braiding quasiparticles around each other, the wave function is transformed unitarily within the degenerate vector space spanned by the $\Psi_\alpha$.   The term ``nonabelian'' stems from the fact that the unitary matrices, corresponding to different braid operations, will generically not commute with each other \cite{review,moore1991nonabelions,rowell2015degeneracy}.


\subsubsection{Holonomy=monodromy and the Moore-Read conjecture} \label{subsub:monhol}

In the Abelian case, as discussed in the above Sections \ref{sub:plasma1} and \ref{subsub:mon}, the result of adiabatically moving one quasiparticle around another is given by the sum of the  phase due to the monodromy and the Berry phase ~\eqref{eq:berry1}.
Furthermore, for wave functions extracted from conformal blocks, the Berry phase solely comes from the gaussian factor which gives the Ahararonov-Bohm phase due to the magnetic field. 
 In the present case where we have a degenerate vector space, the Berry phase becomes matrix valued. Following the notation in \textcite{read2009nonabelian} we write for the holonomy
\begin{align}\label{eq:BerryMatrix}
  B  = M {\cal P} \exp\left(\oint d\tau \left\langle \Psi_\alpha \left|\frac{d}{d\tau} \right| \Psi_\beta \right\rangle \right),
\end{align}
where we denoted the monodromy matrix with $M$, and where ${\cal P}$ is a path ordering operator which is necessary since the matrices in the exponent do not commute at different times.    

Assuming that the charge sector can be ignored,  \textcite{read2009nonabelian} explicitly calculated the nonabelian  monodromy matrices for two and four quasiparticles in the Moore-Read state.
He furthermore argued that  the Berry phase in \eqref{eq:BerryMatrix}  just  gives the Aharonov-Bohm phase times a trivial identity matrix. 

In general, the  path ordered exponential in \eqref{eq:BerryMatrix} can be calculated by the methods given in section \ref{subsub:mon}, again assuming the conformal blocks to be holomorphic. In this case the Berry matrix will again just  give the Aharonov-Bohm phase times the identity,  {\it so long as we have} orthonormality of the conformal blocks
\be{onblocks}
  \langle \Psi_\alpha | \Psi_\beta \rangle  = C\, \delta_{\alpha,\beta} \, .
    \label{eq:orthonormality}
\ee
 Thus, assuming this,  the holonomy, $B$ (the physical result of moving quasiparticles around each other) is given by the monodromy, $M$ (how the wave function explicitly transforms due to its branch cuts, as discussed in section \ref{subsub:vecspace} above) multiplied by the abelian Aharonov-Bohm phase \cite{gurarienayak,read2009nonabelian}.

While for well behaved quantum Hall states, it appears that Eq.~\eqref{eq:orthonormality} is true (and this is always assumed), proving this fact turns out to be quite difficult.  Equivalently we might ask whether there are ways to prove that the holonomy is the monodromy.     
The latter has been shown to be true numerically for the Moore-Read state several times \cite{tserkovnyak,baraban}, and then more recently for the $Z_3$ Read-Rezayi state as well \cite{wuestienne}.  
 Direct numerical tests of Eq.~\eqref{eq:orthonormality} have been done by  \textcite{barabanthesis} for the Moore-Read state.  
On the analytical side, we already mentioned the calculation by \textcite{read2009nonabelian} for two and four quasiholes in the Moore-Read state. 
An alternative derivation based on a mapping to a more complicated plasma was given by \textcite{bondersongurarie}.

\textcite{read2009nonabelian} also gave some very general arguments for the assertion holonomy = monodromy, or equivalently the orthogonality relation \eqref{onblocks}, for quasiparticle wave functions constructed from holomorphic conformal blocks. The main idea is to consider expressions like   
\be{gce}
\av{ e^{i\lambda \int d^2z\, \bar\psi_e (\bar z) \psi_e (z)} H(\eta_1) \dots H(\eta_M) {\mathcal O}_{bg} } \, , \nonumber
\ee
where the electron operator has been factorized in a charge and a topological part {\it viz.} 
$\psi_e = e^{( i / \sqrt{\nu}) \phi} \, \psi$. For $\lambda = 0$ the above formula is just a CFT correlator of some primary fields, and will have power-law dependence on the quasiparticle coordinates $\eta_i$. For nonzero $\lambda$, expanding the exponential, we see that the expression resembles the grand canonical sum over different numbers of electrons, in addition to the quasiparticles. With a suitably chosen background, this will describe an $N$ electron state with negligibly  small charge fluctuations. Alternatively, we
can consider the operator $\bar\psi_e\psi_e$ as a  perturbation on the original CFT. For the charged sector, one can use the Laughlin plasma analogy to infer screening of the electric charge. 

To demonstrate  ``generalized screening'', \ie that {\it all} correlations fall exponentially, which implies  holonomy = monodromy, one must show that also the topological sector is in a massive phase. \textcite{read2009nonabelian} gives a series of arguments for this assertion, but also points out that it is hard to draw firm conclusions, since scaling arguments based on the conformal dimensions of the perturbing operators are not to be trusted since it can be argued that the renormalization group flow occurs for large ($\sim 1$) values of $\lambda$.

We now summarize the results of the preceding sections in what we refer to as the Moore-Read conjecture:

{\it Representative wave functions for quantum Hall ground states, as well as  their quasihole excitations, can be expressed in terms of conformal 
blocks of primary fields in (rational,  unitary\footnote{ The importance of unitarity was pointed out in a later paper by  \textcite{read2009nonabelian}}) CFTs. 
The holonomies, or statistical braiding matrices, equal the monodromies of these blocks. 
Furthermore, the same CFTs yield a minimal\footnote{ Many edge CFTs can correspond to a single 2d topological bulk.   This nonuniqueness is discussed in detail in \textcite{cano2014bulk}. } dynamical theory for the edge of the quantum Hall liquid. }

This conjecture, which has been a crucial  guiding principle for later work,  applies directly to all one-component states, as well as to multi-component states of the Halperin type. 
 As formulated, it however does not apply to hierarchy states where some of the electrons are described by descendant fields. This will be a main topic in section \ref{sect:expwfs}. 


\subsubsection{Example: the Moore-Read state}
\label{sub:MRstate}

As a well known example, we consider in detail the extensively studied Moore-Read state \cite{moore1991nonabelions}.  The corresponding CFT is based on the Ising CFT\footnote{More correctly, it is based on the \emph{chiral} part of the Ising CFT. For a pedagocial discussion of the full Ising theory, we refer the reader to \cite{belavin1984infinite}, as well as to \cite{di1997conformal}. }, which has three primary fields: ${\bf 1}, \psi, \sigma$ with  conformal weights $0, \frac{1}{2}$, and $\frac{1}{16}$ respectively.  The fusion rules (see appendix~\ref{app:CFT} for a crash course in CFT) are given by the table
\begin{align}\label{eq:fusiontable}
{\bf 1} \times {\bf 1} &= {\bf 1} & {\bf 1} \times \psi &= \psi & {\bf 1} \times \sigma &= \sigma \nonumber \\
& &\psi \times \psi &= {\bf 1}  &  \psi \times \sigma &= \sigma \\
&&&&\sigma \times \sigma &= {\bf 1} + \psi\,.\nonumber
\end{align}
That there is a fusion rule with two fields on the right hand side (the last  rule) indicates that this is a nonabelian theory.   This Ising  CFT is particularly simple in that $\psi$ is just a free Fermi field, so Wick's theorem applies, and we can evaluate  correlators such as
\be{pfaff}
 \langle \psi(z_1) \ldots \psi(z_N) \rangle &=& {\cal A}\left(\frac{1}{z_1 - z_2} \, \frac{1}{z_3 - z_4} \, \ldots \frac{1}{z_{N-1} - z_N}\right)  \nonumber\\
 & \equiv& {\rm Pf}\left( \frac{1}{z_i - z_j}\right), \nonumber
\ee
where $N$ is even and  $\cal A$ denotes an antisymmetric sum over all pairings of coordinates.  This expression defines a Pfaffian and we denote it as ${\rm Pf}$.   Correlators including $\sigma$ fields (which are not free so Wick's theorem does not apply) can be calculated analytically as well, although the results are far more complicated \cite{belavin1984infinite,nayakwilczek,ardonne2010chiral,bondersongurarie}.
This Ising CFT must be supplemented with a $U(1)$ CFT to make a proper quantum Hall state.   We consider the $\nu=1/2$ state of fermions, and correspondingly define
\be{eq:MRe}
\psi_e (z)= \psi(z) V(z) = \psi(z) :e^{i \sqrt{2}  \phi(z)}:
\ee
where $V(z)$ is a bosonic vertex operator in the $U(1)$ sector.  Recalling that the scaling dimension of $:e^{i \alpha \phi}:$ is $\alpha^2/2$ (see appendix \ref{app:CFT}), we note that  the scaling dimension of the electron field is $h_e = h_\psi + h_{V} = 1/2 + 1 = 3/2$ which  is a half integer, and therefore appropriate for a fermionic wave function.\footnote{Indeed, the scaling dimension of 3/2 indicates that this is actually a superconformal field theory, but we will not need this fact \cite{di1997conformal}.}   The ground state wave function is then obtained by  multiplying together the results of the Ising and $U(1)$ sectors
\begin{align}
 \Psi_{MR} &= \langle \psi_e(z_1) \ldots \psi_e(z_N)\mathcal{O}_{bg} \rangle= {\rm{Pf}}\left( \frac{1}{z_i - z_j} \right) \prod_{i< j} (z_i - z_j)^2.\nonumber
\end{align}
Note that this wave function is nowhere singular. However, the Pfaffian allows particles to `pair', \ie the wave function vanishes only as a single power, when two particle  positions coincide.

We can pause for a moment to consider the rather surprising fact that we have constructed a very interesting wavefunction based on noninteracting fermion operators (obeying Wick's theorem) and noninteracting free boson operators, and  yet this wavefunction has built into it some extremely complicated correlations.

We can also construct quasihole operators by
\begin{align}\label{eq:qh}
 \psi_{qh} = \sigma (\eta) H(\eta) = \sigma (\eta):e^{i  \phi(\eta)/(2 \sqrt{2}) }:.
\end{align}
The coefficient in the exponent of the vertex operator is  chosen to give  the lowest possible charge while still satisfying Eq.~\eqref{eq:hinteger}.
 To see this, note that $\psi_{qh} \times \psi_e = \sigma :e^{i 5 \phi/(2 \sqrt{2}) }:\, = \psi_f$.     The scaling dimensions are $h_{qh} = h_\sigma + h_{H}  = 1/16 + 1/16 = 1/8$  and similarly $h_f = 1/16 + 25/16 = 13/8$. Thus with $h_e = 3/2$ as above, we obtain $h_f - h_e - h_{qh} = 0$ which assures no singularity in the electron wave function.   
There are also other operators that are nonsingular but they can all be obtained by fusing a number of  elementary holes \eqref{eq:qh}. 

We get for the  Moore-Read wave function
with $N$ electrons and $M$ quasiholes,  
\be{MRhole1}
 \Psi_{\alpha} &=& \langle \sigma(\eta_1) \ldots \sigma(\eta_M) \psi(z_1) \ldots\psi (z_N) \rangle_{\alpha}  \nonumber\\
  &\times& \langle H(\eta_1) \ldots H(\eta_M) V(z_1) \ldots V(z_N) \mathcal{O}_{bg} \rangle \nonumber \\
 &=& \langle \sigma(\eta_1) \ldots \sigma(\eta_M) \psi(z_1) \ldots\psi (z_N) \rangle_{\alpha} \nonumber\\
   &\times&\prod_{\gamma< \mu}^M (\eta_\gamma - \eta_\mu)^{1/8} \prod_{i=1}^N  
   \prod_{\gamma=1}^M (\eta_\gamma - z_i)^{1/2} \prod_{i<j} \jasf i j 2 \, . \nonumber
\ee
For the correlator not to vanish, all the Ising fields have to fuse to the identity, and from Eq.~\eqref{eq:fusiontable} it follows that $M$ must be even. 
As described in section \ref{subsub:vecspace}, correlators involving multiple quasiholes (and thus multiple $\sigma$ fields) will generally be multivalued and  span a vector space rather than giving a single wave function.  For $2n$ quasiholes on a spherical or planar geometry, the corresponding vector space is $2^{n-1}$ dimensional. 
The most suitable basis for this vector space is given by the conformal blocks \cite{di1997conformal}, labeled by the index $\alpha$ in the above equation.
The charge of the quasihole is set by the repulsion of the electrons from the quasihole position, {\it i.e.}, the $(\eta - z)^{1/2}$ factor.   Note that although this has a fractional power, the product of all terms on the right is guaranteed, by Eq.~\eqref{eq:hinteger}, to be nonsingular and without branch cuts for electron coordinates.    The half power can be viewed as 1/2 of a Laughlin quasihole, which is obtained by inserting a flux quantum and has electrical charge $+e \nu$.  Thus, the elementary quasihole \eqref{eq:qh} has charge $\nu/2 = +e/4$. For a discussion of the corresponding braiding properties the reader is referred to  \textcite{nayakreview}.


\subsection{Nonabelian effective Chern-Simons theory}
Just as for the abelian states, there are effective low energy gauge theories also for the nonabelian states, and not surprisingly, they contain nonabelian Chern-Simons terms. These theories can be derived in  different ways, and there can be different candidate theories for the same state. 


\subsubsection{Construction based on the anomaly}
Starting from the  bosonic version of the Moore-Read state at $\nu =1$, the edge theory consists of a Majorana fermion $\psi$ and  a chiral boson $ :e^{i   \phi(z)}:$. This boson can be fermionized to a Dirac fermion, or equivalently two Majorana fermions. Consequently the edge theory has  three Majorana fermions. From these one can form three current operators that transform as a spin one representation of an
$SU(2)_2 $ Kac-Moody algebra, where the subscript denotes the level.  This algebra  is also generated by a chiral $SU(2)_2$ Wess-Zumino-Witten model, which is thus the purely bosonic incarnation of the edge theory.  To proceed from here, we recall the anomaly approach to the bulk-edge correspondence described in section \ref{sub:csedge}. Starting from the bulk theory, we derived an anomalous edge theory where the non-conservation of the edge charge was due to inflow from the bulk.
Here we  turn the argument around. Knowing the edge theory we can find the anomaly, and then ask what bulk theory will have a non-gauge invariant boundary term that exactly cancels this anomaly. The answer is 
a nonabelian $SU(2)_2$ CS theory.   We refer to the original article by \textcite{fradkin1998chern} for details of the derivation, and also for  a similar, albeit more complicated, construction for the $\nu =1/2$ fermionic Moore-Read state. (In a later paper, a Ginzburg-Landau approach, similar to the one described in section \ref{sec:GLCS},  is constructed for the bosonic case \cite{fradkin1999landau}.) 


\subsubsection {Parton construction}\label{ssec:parton}
An alternative way to derive effective  Chern-Simons theories for the nonabelian states is the {\it parton construction} due to 
\textcite{wen1999projective}. The main idea is to introduce a redundant set of free fermions, or {\it partons}, that transform under an $SU(N)$ ``color" group. The electron operator is a singlet under this group, and the projection on the singlet sector is achieved by gauging the symmetry. By integrating out the redundant parton degrees of freedom one is left with an effective nonabelian theory for the gauge field. We illustrate this with an  example given by \textcite{wen1999projective}. We start from a set of fermionic, charge  $e/2$ parton fields, $\psi_{a\alpha} $, with the Lagrangian, 
\be{part}
{\cal L}_{part}  = i \psi^{\dag}_{a\alpha} \partial_t  \psi_{a\alpha} + \frac 1 {2m}
\psi^{\dag}_{a\alpha} \left(\partial_i -i \frac{e}{2} A_i \right)^2 \psi_{a\alpha}\nonumber
\ee
 where $\alpha$ is a spin half index and $a=1,2$ an $SU(2)$ ``color" index, also in the spin half representation. By combining two partons we 
 can form a charge $e$ spin one {\it bosonic} electron,
 \be{elaspart}
\Psi_{m}(z)= \psi_{a\alpha}(z)\psi_{b\beta}(z) \epsilon_{ab}C^m_{\alpha\beta} \, ,\nonumber
\label{Psipsi}
\ee
where $m$ is a spin one index and  $C^m_{\alpha\beta}$ the pertinent Clebsch-Gordan recoupling coefficients. 
 Next we assume that the only physical excitations are bosons with spin one and unit electric charge. This can be achieved by coupling the color current to a dynamical  $SU(2)$ gauge field, $a_\mu^k$,
 \be{projpart}
    {\cal L}^{(Proj)}_{part} &=& i \psi^{\dag}_{a\alpha} (\delta_{ab} \partial_t -i a_0^k t^k_{ab} ) \psi_{b\alpha}   \\ 
    &+&  \frac 1 {2m}
\psi^{\dag}_{a\alpha} \left((\partial_i-i \frac{e}{2} A_i )\mathbf{1}  -i \vec{a}_i \cdot \vec{\mathbf t} \,  \right)^2_{ab} \psi_{b\alpha}\, , \nonumber
\ee
where $\vec{\mathbf t}= \half  \vec \sigma$ with $\sigma^k$ being the Pauli matrices.
 In this theory only the color singlet operators are physical. This can be seen in different ways. First, variation with respect to the gauge fields shows that the color currents vanish identically. Alternatively one can  think of adding a Maxwell term $\frac 1 {4g^2} Tr F^k_{\mu\nu}F^{k, \mu\nu}$ to the action and note that Eq.~\eqref{projpart} corresponds to the strong coupling (confining) limit $g \rightarrow \infty$. Since the color singlet is antisymmetric the spin wave function must be symmetric by the Pauli principle, so the theory indeed describes spin one bosons. 
To establish that the state is nonabelian, one integrates the parton field to get a low energy effective action. Assuming that the  partons are at a density corresponding to two filled  Landau levels, the resulting low energy action is 
 again a nonabelian $SU(2)_2$  CS theory, which is the spin one counterpart of the spin zero theory derived using the anomaly approach.
 
In a later work \textcite{barkeshli2010effective} carried out a parton construction for the $Z_k$ Read-Rezayi  parafermion states at filling factor $\nu = k/(kM+2)$, 
for which they find  the nonabelian \emph{fermionic} bulk theory  to be $[U(M)\times Sp(2k)]_1$.


\subsection{CFT quasiparticles and hierarchies --- a summary}  \label{shortsum}

The task of generalizing the CFT construction to even the abelian hierarchies has turned out to be quite difficult, and we shall pursue this in more depth in the coming sections. Since these by necessity will be somewhat more technical, we here provide a short summary of the main results with references to the proper sections. 

Before doing so, however, we note that as long as  we are concerned only with universal topological properties (and do not aim at  obtaining accurate wave functions) there is a relatively straightforward path to be followed \cite{moore1991nonabelions}.   Following the idea of a pseudo wave function (Eq.~\eqref{hiwf}), we can for example, construct a $\nu=2/7$ wave function as follows (compare Eq.~\eqref{eq:hier2/7}).
\be{readform1}
   \Psi &=& \int \prod_{\alpha=1}^M  d^2 \eta_\alpha  \left\langle \hat V(\bar \eta_1)\ldots \hat V(\bar \eta_M) {\cal O}_{bg} \right\rangle   \nonumber\\
   &\times&  
   	\left\langle  H(\eta_1)\ldots H(\eta_M)   V_1(z_1) \ldots V_1(z_N) {\cal O}_{bg}  \right\rangle \nonumber
\ee
with $M=N/2$. Here $V(z)$ and $H(\eta)$ were defined in Eq.~\eqref{vo} (with $m=3$ in this case),   and
\be{vo2}
\hat V(\bar \eta) &=& : e^{ i \sqrt{2+ 1/3} \,\, \phi (\bar \eta ) } : \, .\nonumber
\ee
  Note that the correlator of the $\hat V$ is actually {\it anti-}holomorphic, and corresponds to fields with opposite chirality in order to properly cancel the phases in the  holomorphic factors $(\eta_\alpha-\eta_\beta)^{1/3}$ that arise from the second  correlator.   Ignoring, for simplicity, the background charges and the resulting gaussian exponential factors,  we obtain almost the same  wave function as in Eq.~\eqref{eq:hier2/7} -- the only difference is that the fractional power of the factors $|\eta_\alpha - \eta_\beta|$ changes from 1/3 to 2/3.  In section \ref{quasilocal} we will argue that this is a short-distance effect that does not influence the topological properties.  

 Topologically the  above construction is valid, but, as stressed in section \ref{sub:difficulties}, it is  very difficult to evaluate the integrals for a reasonable number of particles.  Also, if one tries to describe condensates of quasielectrons by  naively replacing the hole operators $H(\eta)$ with their inverse $H^\star (\bar\eta) = e^{-i \phi(\bar\eta)}$, the correlators will have singularities $\sim (z_i - \bar\eta)^{-1}$. Again, this should not change the long distance topological properties, but to get  acceptable wave functions one would need not only to handle these singularities  in the integrals over the positions $\bar\eta_i$, but also to project the result on the LLL for the electrons. 

 Physically the singularities come about since introducing a vortex of strength $m$   changes the angular momentum of all the electrons by $m$ units with respect to the position of the vortex. The problem occurs when one tries to make an anti-vortex, \ie a vortex with negative $m$, at a position where there is already an electron. This electron cannot lower its angular momentum while still staying in the LLL, and this is what is manifested in the singularities $\sim (z_i - \bar\eta)^{-1}$ encountered when naively trying to form a quasielectron using the operator $H^\star$.  Leaving the LLL, the angular momentum can be lowered by including a factor $(\bar z_i - \bar\eta)$ which after projection onto the LLL  provides the rationale for Laughlin's proposal that going from a quasihole to a quasielectron amounts to 
 $\prod_i (z_i - \eta) \rightarrow \prod_i (2\partial_i  - \bar\eta)$. The resulting multi quasielectron wave functions are however equally difficult to handle in the hierarchy integrals as their quasihole counterparts. 
 
 Fortunately, there is a way out of these problems, and the key insight is that calculating correlators of $H$, or $H^\star$, is not the only way to represent quasiparticles using the CFT formalism.  To explain the alternative method, we first notice that in the Laughlin ground state, the $i^{th}$ electron is surrounded by a ``correlation hole'', $\prod_j (z_i - z_j)^m$, that we can think of as $m$ elementary holes on top of each other. The charge counting, with respect to the ground state, now works as follows. Introducing $m$ elementary holes amounts to electric charge +$e$, and the actual electron at position $z_i$ adds charge $-e$, meaning that the electron together with its correlation hole is neutral. This is as it should be, since adding an electron to the incompressible quantum Hall liquid does not change the local charge density, but just expands the droplet a tiny bit at the edge. This  provides the crucial clue for how to introduce quasiholes or quasielectrons --- {\it we just modify a number of electron 
operators by contracting or expanding their correlation holes by one unit.} 
These words can be translated into CFT equations; contracting amounts to fusing an electron operator $V(z)$ with an inverse hole, i.e. the operator $H^\star$. Just as in section \ref{wfasconfblock} fusion  means making an operator product expansion, but in this case the first term turns out to vanish, so the leading term in the expansion will be a {\it descendant} which means that it includes a derivative. In the simplest case of the $\nu = 1/(2p+1) $ Laughlin state, fusing the electron operator \eqref{vo} with the conjugate of the hole \eqref{hole}
 we get  $\tilde V(z) = (H^\star \otimes V(z))  \sim\,  :\partial_z e^{i2p/\sqrt{2p+1}\phi (z)}:$.
  We can now form states with an arbitrary number of quasielectrons simply by replacing a number of the original electron operators, $V$, with the new operators $\tilde V$.\footnote{
 As discussed in detail in \ref{quasilocal} this is not entirely true; as written the operator $\tilde V$ is anyonic, and has to be augmented by another vertex operator to give fermionic commutation relations.}
  This will give rise to a state with increased density at the second level of the hierarchy. In particular, if half of the original electrons are replaced by $\tilde V$, one obtains the $\nu = 2/(4p+1)$ Jain state. 
 By constructing the relevant hole operators in this new state the procedure can be iterated.
 
 In complete analogy one can introduce quasiholes by expanding the correlation holes around the electrons; this is done by fusing $V$ with an appropriate hole operator. \footnote{Perhaps surprisingly, the pertinent hole operator is {\it not} the naive Laughlin hole $H$; we get back to this technical detail in section \ref{ssec:qe}.} 
In this case, modifying a fraction of the original electrons results in a second level hierarchy state with charge density lower than the parent state. In particular, this procedure will give the states of the negative Jain sequences. That some of the Jain states were obtained in these simple cases is not a coincidence;  we shall show that  all the Jain composite fermion states are exactly reproduced by this procedure. 
 
 Note that in the above there was no reference to the hierarchy integrals, or even states of several {\it localized} quasiparticles. Such states can however be constructed by inserting a number of {\it quasi-local} operators ${\mathcal H}(\eta)$ and  ${\mathcal P}(\bar\eta)$, representing quasiholes and quasielectrons respectively, into the correlators. The exact definition and  properties of these operators are discussed in section \ref{quasilocal}. A most remarkable property of the resulting multi-quasiparticle wave functions is that when inserted in the integral expressions for  the hierarchy wave functions,  {\it the integrals over the quasiparticle coordinates can be carried out exactly, thus giving closed form expressions for the wave function of the daughter state.} These are just the wave functions discussed in the previous paragraphs. The details of this are explained in section \ref{CFvshier}.  An upshot of this is also that {\it all composite fermion wave functions can be exactly written in a 
manifestly hierarchical form. } 

In section \ref{subsub:monhol} we emphasized the importance of the plasma analogy for proving the ``monodromy = holonomy" conjecture that the topological properties of the quantum Hall states could be obtained  directly from the monodromies of the conformal blocks. This turns out to be more difficult in the case of the hierarchy states, which is discussed in section \ref{hiplasma}. 


\section {Explicit hierarchy wave functions for abelian quantum Hall states } \label{sect:expwfs}  
In section \ref{ssec:hierarchy} we gave the general form of a hierarchy wave function expressed as integrals over quasiparticle coordinates, and in \ref{sect:CF} the explicit composite fermion wave functions in the Jain series. As already mentioned, the latter have successfully been compared with experiments, while the former are usually too hard to evaluate. In this section we shall tie the two pictures together using the CFT approach. Following \textcite{bergholtz2008quantum}, and \textcite{suorsa2011quasihole,suorsa2011general}, we first present explicit vertex operators whose correlators produce trial wave functions for any abelian hierarchy state. These are consistent with  Wen's topological classification, and reduce to Jain's wave functions whenever these exist. We then proceed to explain how to construct excited states with quasielectrons and quasiholes. Finally we show that  these wave functions do form a hierarchy in that they can all be written in the Halperin form \eqref{hiwf}, or in other words, they can all be obtained by successive condensations of quasiparticles.  For sake of readability, we shall often omit the background charge operator \eqref{multibg} from the correlators in this section.


\subsection {The chiral hierarchy}  {\label{ss:chiral}}

For simplicity, we first consider states with a positive semi-definite  $K$-matrix which describe fully chiral states, where all the modes of the minimal, \ie unreconstructed, edge move in the same direction. These states form a subhierarchy within the full hierarchy that will be considered in  section \ref{ss:full}. 


\subsubsection {The need for many electron operators}
In order to describe a state at level $n$ in the hierarchy, we shall use $n$ different electron
 operators \cite{wen1995topological,read1990excitation}, see also Section \ref{ssec:effectiveCS}.
In the edge theory, these are CFT vertex operators.
 A natural assumption, consistent with the Moore-Read conjecture, is that the bulk should be described
 by the same CFT (\cf Section IIA in  \cite{suorsa2011quasihole}). 
Given this, it is natural that the description of the bulk states involves one electron (vertex) operator for each level of the hierarchy.

The $n$ electron operators will differ by their conformal spin, which has a natural interpretation  as orbital spin (\cf sections \ref{wenzee}, \ref{sc:effresp} and \ref{othergeom}). This however does not mean that the electrons come in different types.  They are indistinguishable, and the electronic wave functions are fully antisymmetric as required by the Pauli principle. 
For heuristic purposes, it is however sometimes useful to think in terms of distinct types of electrons. 
For instance, we will show below  that in all cases where composite fermion wave functions exist, the CFT approach leads to identically the same wave functions. In these cases, the existence of $n$ different ``kinds" of electrons finds its natural counterpart in the CF language, where the different electron operators correspond to composite fermions in different effective ``$\Lambda$ levels".


\subsubsection {Chiral ground states: wave  functions from topological data} \label{sec:chiralgs}
 We now present the general form of ground state wave functions for chiral states. These are hierarchy states resulting from quasielectron condensation only. They can be expressed as holomorphic correlators of local chiral operators, and their edge states all have the same chirality. These wave functions can be derived via an explicit hierarchical condensation procedure, which will be discussed in Section \ref{sec:cfthier}. Here we directly present the upshot -- a set of vertex operators whose correlators give wave functions that are consistent with Wen's topological classification  \cite{bergholtz2008quantum,suorsa2011general}. 

The vertex operators at level $n$ of the hierarchy closely resemble those of the multicomponent states, 
\be{ver1}
V_{\alpha}(z) =\,  : \partial_z^{\alpha-1} e^{i\sum_{\beta} Q_{\alpha \beta}\phi^\beta(z)} : \ \ \ \  \alpha = 1, 2 \dots n \, ,
\ee 
where we recall from Eq.~\pref{kfact} that ${\mathbf K} = {\mathbf Q} {\mathbf Q}^T$. 
$V_\alpha$ has  conformal spin $s_\alpha = \half K_{\alpha\alpha} + \alpha - 1$, where the two last terms come from the derivatives, and the first term is 
just the conformal dimension of the corresponding primary field.  
These are exactly the entries in the hierarchy spin vector \eqref{chirs}, which should be interpreted as the  {\it orbital spins} of the 
electrons. 
The  derivatives appear naturally from the condensation construction presented below, and are necessary  for  the resulting wave functions not to vanish under antisymmetrization.    In our scheme we divide the electrons into $n$ groups $I_1 \ldots I_n$, assigning a different electron operator $V_\alpha$ to each of the groups, and then sum over all ways of  choosing the groups. The ground state wave functions become
\be{genstate}
\Psi  &=& {\cal  A}  \langle\prod _{\alpha=1} ^{n}\prod_{i_\alpha\in I_\alpha} V_\alpha (z_{i_\alpha})  {\cal O}_{bg} \rangle    \nonumber \\
 &=& {\cal A} \{ (1-1)^{ K_{11} } \partial_2 (2-2)^{K_{22}}\dots  \partial_n^{n-1}(n-n)^{K_{nn}}  \nonumber \\
&\times&  (1-2)^{K_{12}} (1-3)^{K_{13}} \dots ((n-1)-n)^{K_{n-1,n}} \}   \nonumber  \\
&\times& e^{-\sum_i |z_i|^2/(4\ell^2)},
\ee
where  $\mathcal A$ is the antisymmetrization operator. The derivatives act on the Jastrow factors, but not on the gaussian, and  we use the short-hand notation, 
\be{shorthand}
\partial_\alpha^k&\equiv&  \prod_{i_\alpha \in I_\alpha}  \partial_{z_{i_\alpha}}^k  \nonumber\\
 (\alpha-\alpha)^{K_{\alpha\alpha}} &\equiv& \prod_{i<j\in I_\alpha}(z_i-z_j)^{K_{\alpha\alpha}}  \nonumber
\\
(\alpha-\beta)^{K_{\alpha \beta}} &\equiv&  \prod_{i_\alpha\in I_\alpha} \prod_{i_\beta\in I_\beta}(z_{i_\alpha}-z_{i_\beta})^{K_{\alpha \beta}} \, .
\ee
The groups $I_\alpha$ have the same partial filling fractions $\nu_\alpha = K_{\alpha\beta}^{-1}t_{\beta}$, as the multicomponent states in section \ref{multicomponent}, and the neutralizing background \eqref{multibg} is also the same. The size of each of the groups is determined by requiring homogeneity. For details about how to impose the homogeneity condition, we refer to the original paper by \textcite{bergholtz2008quantum}. 
The vertex operators in Eq.~\eqref{ver1} are by no means unique.
In principle, one needs to consider more general expressions, where the derivatives act only on parts of the vertex operator, such as
\begin{align}\label{spelop}
:e^{ia\phi^\alpha}\partial^{n}e^{ib\phi^\beta}:\, .
\end{align}
The freedom in placing the derivatives reflects the various possibilities to regularize  operator products between electron and quasiparticle operators, see \eg \cite{hansson2009quantum, suorsa2011quasihole,suorsa2011general}. 
Numerical studies indicate that on the plane the precise placement of the derivates does not matter, \ie it does not change any of the topological features of the state. 
This is widely used in numerical computations in order to find alternative descriptions of the state that can be implemented more efficiently on a computer, see \eg  \cite{JainKamillaLLL,JainKamillaLLL1,MollerSimon,rodriguez2012quasiparticles,bonderson2008fractional}, even though there is no rigorous proof of the equivalence of these various alternatives as of yet. 
However, not all choices  are allowed in Eq.~\eqref{spelop}, as some  would give singular wave functions by having too many derivatives acting on a Jastrow factor to a broken power like $(z_i - z_j)^{p/q}$. 
On the plane, a simple prescription that avoids this is to always move all derivatives to the left so that they act on fully holomorphic functions. 
For instance, the operators \eqref{ver1}, as well as $\mathcal P$ and $\mathcal H$ in section \ref{quasilocal} are defined such as to give this prescription when used in a hierarchy construction.  
However, it was noted by \textcite{kvorning2013quantum} that  it does not give correct vertex operators for the spherical geometry, and one is forced to use the more general expression \eqref{spelop} in order to obtain wave functions that are non-vanishing, see section \ref{sect:sphere}.

By evaluating the wave functions \eqref{genstate} on a thin cylinder,  \textcite{bergholtz2008quantum} reproduced exactly the same charge density wave patterns as those found by Bergholtz and Karlhede in the TT-limit (\cf  section \ref{ssec:otherhierarchies}). 
It is both reassuring and encouraging  that the CFT wave functions precisely reproduce the exact ground states found in a non-trivial model of interacting electrons, and constitutes an important consistency check of the former.

To make this discussion less abstract, let us consider the example of the quantum Hall state at filling $\nu=2/5$, and factorize  $\kb$ as in the first alternative in the expression \eqref{halperin},
\be{2/5K} 
\kb=\left(\begin{array}{cc} 3&2\\2&3\end{array}\right) = \left(\begin{array}{cc} \sqrt 3&0\\ \sqrt{\frac 4 3}&\sqrt{\frac 5 3}\end{array}\right) 
\left(\begin{array}{cc} \sqrt 3 & \sqrt{\frac 4 3} \\ 0 &\sqrt{\frac 5 3}\end{array}\right)  .\nonumber
\ee
Eq.\eqref{ver1} then gives the two vertex operators, 
\begin{align}
V_1 &=\, :e^{i\sqrt 3\phi_1}:   \, ,&V_2&=\, :\partial_z e^{i \left( \sqrt{4 /3} \phi_1+ \sqrt{ 5/ 3}\phi_2\right)} :\,  . \nonumber
\end{align}
Homogeneity requires the two sets, $I_1$ and $I_2$, to be of equal size, \ie $N/2$, where $N$ is the number of electrons, and the wave function becomes
\begin{align}\label{2/5wf}
\Psi_{2/5}=\mathcal{A}\left\{ (1-1)^3\partial_2(2-2)^3(1-2)^2  \right\}  e^{-\sum_i |z_i|^2/(4\ell^2) }\, . 
\end{align} 
This wave function is identical to the one obtained below in Eq.~\eqref{eq:hierarchylevel2}  by condensing quasielectrons on top of the Laughlin $1/3$ state, and in fact also identical to the $\nu=2/5$ Jain wave function. 

 At level $n$ of the hierarchy there are $n$ elementary quasiholes,
 \be{}
H_{\alpha}(\eta) =\,  :e^{i \sum_{\beta}Q_{\beta\alpha}^{-1}\phi^{\beta}(\eta)} :\,  ,\nonumber
\ee
where ${\bf Q}^{-1}$ is the inverse of the matrix $\bf Q$ in the electron operators (\ref{ver1}).\footnote{If $\bf Q$ is not a square matrix, one instead chooses $\bf Q^{-1}$ as the Moore-Penrose pseudo-inverse, which (as $n$ is the rank of $\mathbf K$) fulfills $\sum_{\beta}Q_{\alpha \beta}Q^{-1}_{\beta\gamma}=\delta_{\alpha\gamma}$. }
In Wen's language, a generic quasihole is some composite of these, as encoded in the $\bf l$-vector defined in section \ref{ssec:effectiveCS}. For example, a fundamental hole created by  the above operator $H_{\alpha}$ corresponds  to $l^q_{\alpha} = \delta_{q\alpha}$.  
A Laughlin hole, on the other hand, amounts to a unit vortex in all condensates, and is created by the operator $H_L = \prod_{\alpha} H_{\alpha}$ and represented by the $\bf l$-vector $(1,1,1...,1)$.
The simple connection between the $K$-matrix and the above hierarchy wave functions is manifest only in  the ``symmetric basis"   where the $\bf t$-vector equals $(1,1,...,1)$ (see section \ref{ssec:effectiveCS}).    The statistics of any  quasihole follows from Eq.~\eqref{wenres3}. 

The conformal spin of the operator $H_\alpha$ is calculated as $2s_\alpha = \sum_\beta Q_{\beta\alpha}^{-1} Q_{\beta\alpha}^{-1}  
= \sum_\beta (Q^{-1})^T_{\alpha\beta} Q_{\beta\alpha}^{-1} = K^{-1}_{\alpha\alpha}$ in accordance with \eqref{chirK}, if we again 
identify the conformal spin with the orbital spin in the Wen-Zee formalism. The holes obtained using  $H_\alpha$ do not in general have
the same charge, but there is always one with the minimal charge $e/q$ for a state with $\nu = p/q$. We believe that at level $n$ in the 
hierarchy one can always find $n$ distinct hole operators with the same charge and that only differ by integer units of 
conformal spin.\footnote{We have not tried to prove this in general, but we have tested several nontrivial cases of $K$-matrices of 
rank two and three.} Since the $n$ holes are topologically the same, it is fair to ask if all are needed. The tentative answer is yes, but
is based on arguments that go beyond topology. Just as the $n$ electron operators are needed in order to construct a ground state
wave function at level $n$ in the hierarchy, we presumably need all $n$ quasihole operators to  describe a general quasihole
wave function. This belief is based on calculations of the particle entanglement spectrum, and is briefly outlined in 
section \ref{sect:edgesandentanglement}.

While the above description of  quasi{\it holes} was rather straightforward, the situation is more complicated for quasi{\it electrons}. A naive approach leads to singular wave functions, and there is no local operator that creates acceptable quasielectron wave functions.
A heuristic way to understand what goes wrong was given in section \ref{shortsum}.
 In section \ref{ssec:qe} we  explain how this problem can be circumvented  \cite{hansson2009conformal,hansson2009quantum}, and discuss in some detail how to construct operators for quasielectrons.

\subsection {Ground states of the full abelian hierarchy }
\label{ss:full}
Having, as above, positive definite $K$-matrices is a special case. 
For general hierarchy states,  containing quasihole condensates and, correspondingly,  antichiral edge modes, a description in terms of purely holomorphic conformal blocks is not possible. \textcite{suorsa2011quasihole,suorsa2011general} generalized the construction in the previous section to  general hierarchy states. The idea is to split the $K$-matrix as 
\be{ksplit}
\bf K = \bkappa -\bar \bkappa , 
\ee   
where both $\bkappa$ and $\bar \bkappa$ are positive semi-definite and are associated  with the chiral and antichiral sector, respectively. This decomposition is in general not unique, but all such choices are believed to represent the same topological phase. In particular, filling fraction, quasiparticle charges and statistics only depend on the full $K$-matrix, \cf section \ref{ssec:effectiveCS}. There might be ways to restrict the freedom in splitting the $K$-matrix, for instance by demanding that the resulting wave function should have  the correct thin torus (TT) limit just as the chiral states discussed earlier. 
The freedom of decomposing the $K$-matrix will be illustrated in examples below.

 Given a splitting of the $K$-matrix, we parallel the decomposition \eqref{kfact} in section \ref{multicomponent}, and write $ \bkappa = q q^T$ and $  \bar\bkappa = \bar q\bar q^T$.\footnote{Recall that $q$ and $\bar q$ are matrices.} With this notation, the background operator \eqref{multibg} generalizes to
\be{fullmultibg}
{\mathcal O}_{bg} = \exp \left( -i \int d^2 r\, \rho(\vec r)  n_\alpha ( q_{\alpha\beta} \phi^\beta +  \bar q_{\alpha\beta} \bar \phi^\beta)  \right)      ,
\ee
and the  generalized version of the vertex operators \eqref{ver1}  is
\be{ver2}
V_{\alpha} = \, :\partial_z^{\sigma_{\alpha}}\partial_{\bar z}^{\bar \sigma_{\alpha}} e^{i\sum_{\beta} q_{\alpha \beta}\phi^\beta}e^{i\sum_{\beta} \bar q_{\alpha \beta}\bar\phi^\beta}:,
\ee
where the powers of derivatives are related to the spin vector, $s_\alpha = \half K_{\alpha\alpha} + \sigma_\alpha - \bar \sigma_\alpha$.  In a minimal hierarchy construction we get a single derivative at each level, so $\sigma + \bar\sigma = \alpha - 1$. 
By adding extra derivatives, it can be generalized to states with a shift \eqref{eq:shift} that differs from the minimal one obtained by the condensation procedure to be described in section \ref{sec:cfthier}, but this has so far no application to experimentally or numerically observed quantum Hall states. 

The properly antisymmetrized correlators of products of these operators give the functions,
\begin{align}\label{eq:antichiral}
  &\Psi (\{\xi_i\},\{\bar\xi_i\}) = {\cal A} \Big\{ \prod_\alpha \partial_{\alpha}^{\sigma_{\alpha}} \partial_{\alpha}^{\bar \sigma_{\alpha}}  
  (\alpha-\alpha)^{\kappa_{\alpha\alpha}   }
  (\bar\alpha - \bar\alpha)^{\bar\kappa_{\alpha\alpha} } \nonumber \\
  &\times    \prod_{\alpha<\beta}
  (\alpha-\beta)^{\kappa_{\alpha\beta}} 
  (\bar\alpha-\bar\beta)^{\bar\kappa_{\alpha\beta}}  \Big\}  e^{-\sum_i |\xi_i|^2/(4\ell^2)} \nonumber \\
 & = \Psi_{\bkappa}(\{\xi_i\}) \times \Psi_{\bar \bkappa}(\{\bar\xi_i\})  \, , 
\end{align}
where  we have used the same  short-hand notation as in Eq.~\eqref{shorthand}, except that the $z$'s are replaced by $\xi$'s and the bar denotes  complex conjugation. 
They nicely factorize into a chiral and antichiral part (except the gaussian, which is distributed equally between these two parts).
Clearly the expressions in \eqref{eq:antichiral} are not  valid (holomorphic) LLL wave functions, but they can be interpreted as wave functions in a coherent state representation, \ie the $\xi$'s are considered as coordinates for the guiding centers of the electrons. The 
coherent state wave functions are related to LLL wave functions by the transformation,
\be{LLLprojection}
\Psi(\sgrupp [z]) &=& \int [d^2\xi_i] \langle \sgrupp [z]  | \sgrupp [\xi]\rangle \Psi_{\bkappa}(\{\xi_i\})   \nonumber \\
&\times& \Psi_{\bar \bkappa}(\{\bar\xi_i\}).
\ee
Here $\langle \sgrupp[z] | \sgrupp[\xi] \rangle = \prod_{i=1}^N \bracket {z_i} {\xi_i}$, with $ \bracket z \xi = \exp [-(|\xi|^2 -2 \bar\xi z + |z|^2)/(4\ell^2)]$ is the coherent state kernel, describing $N$ particles maximally localized to the points $(\xi_1, ..., \xi_N)$; up to a phase, $ \bracket z \xi $ equals $\exp[-|z-\xi|^2/(4\ell^2)]$. 
Technically, this transformation can also be interpreted as a LLL projection, as the coherent state kernel and the gaussian from $\Psi$ combine to a LLL delta function,  
\begin{align}\label{eq:LLLdelta}
\delta_{LLL}&= e^{-(|\xi|^2 - \bar\xi z)/(2\ell^2)}.
\end{align}
 A third interpretation is that the wave function $\Psi(z_1, z_2, ..., z_N) $ can be viewed as a CFT correlator of {\it quasi-local} electron operators 
\be{eq:curlyV}
{\cal V}(z) = \int\frac{d^2 \xi}{2\pi} e^{-(|\xi|^2 - 2\bar\xi z + |z|^2)/(4\ell^2)} \, V(\xi, \bar\xi).
\ee
This latter view is appealing in that it displays the inherent ``fuzziness"  of particles in the LLL, \ie that  they cannot be localized more precisely than the magnetic length. The concept of quasi-locality will be central below when we discuss how to explicitly build a hierarchy of quasielectron and quasihole condensates.

The ``sewing together" of the chiral and antichiral sectors amounts to a two-fluid picture in which the chiral component consists of the bare electrons and all quasielectron condensates, and the antichiral sector consists of the hole condensates. The actual particles are built as composites of the positively and negatively charged parts, with their coordinates  identified \cite{suorsa2011quasihole}.

 With this background, we can now propose a generalized  version of the Moore-Read conjecture: 

{\it
Representative wave functions for ground states of hierarchical quantum Hall phases, as well as their quasiparticle excitations, can be constructed from antisymmetrized products of chiral and antichiral conformal blocks of primary and descendant fields in a CFT, in the basis of coherent states $| \xi_1, ... , \xi_N\rangle $. These wave functions obey generalized screening, and have  minimal edge theories  given by the same CFTs that describe the bulk wave functions. 
}

The assumption about generalized screening, which is equivalent to postulating holonomy = monodromy, is discussed further in section \ref{hiplasma}.
So far we only considered abelian phases, and presented a concrete construction that yields explicit wave functions for the full hierarchy.  In section \ref{sect:nahier} we generalize to non-abelian states, some of which again have explicit wave functions, while others only can be expressed in terms of integrals over quasiparticle coordinates. 

\bigskip
\noindent
{\bf Examples:} \\
The coherent state formulation and the freedom in decomposing the $K$-matrix, can be illustrated already at the level of the Laughlin wave function. The $K$-matrix for the standard $\nu = 1/m$ state is simply $\kb=m$. However one can imagine a more general form
\be{}
\kb =  \bkappa - \bar\bkappa = (m+k) - k.\nonumber
\ee
This corresponds to the modified Laughlin wave functions introduced by  \textcite{girvin1984formalism},
\be{eq:Laughlinvariational}
\Psi(\sgrupp[z]) &=& \int [d^2\xi_i] \langle \sgrupp[z] | \sgrupp[\xi] \rangle 
 \prod_{\alpha<\beta} (\bar\xi_\alpha-\bar\xi_\beta)^{k}   \nonumber \\
 &\times& \prod_{\alpha<\beta}(\xi_\alpha-\xi_\beta)^{k+m} e^{-\sum_i |\xi_i|^2/(4\ell^2)}  \\
 &=& \int \prod d^2\xi_i \, \delta_{LLL}(z_i, \xi_i)
 \prod_{\alpha<\beta} |\bar\xi_\alpha-\bar\xi_\beta|^{2k} \nonumber \\ 
 &\times& \prod_{\alpha<\beta}(\xi_\alpha-\xi_\beta)^{m} e^{-\sum_i |\xi_i|^2/(4\ell^2)}\, . \nonumber
\ee
The second line emphasizes that this can be considered as the LLL projection of the Laughlin wave function with additional correlation factors $ \prod_{\alpha<\beta} |\bar\xi_\alpha-\bar\xi_\beta|^{2k} $. 
For integer $k$, the explicit form of the LLL projected wave function is
\be{}
\Psi(\sgrupp [z]) = \prod_{i<j}  (\partial_i - \partial_j)^k  \prod_{i<j} (z_i - z_j)^{k+m} e^{-\sum_i |z_i|^2/(4\ell^2)} . \nonumber
\ee
The pertinent vertex operator for obtaining this generalized Laughlin wave function as a CFT correlator is
\be{}
V_{(m,k)}(\xi, \bar\xi) =\,: e^{i\sqrt{m+k}\phi(\xi)} \, e^{i \sqrt{k}\bar\phi(\bar\xi)}:.\nonumber
\ee
Girvin and Jach proposed Eq.~\eqref{eq:Laughlinvariational} as a particular scheme to make Laughlin's wave function variational, with $k$ as the variational parameter. The additional correlation factor in Eq.~\eqref{eq:Laughlinvariational} or, more generically a function  $ f\left(\prod_{\alpha<\beta} |\bar\xi_\alpha-\bar\xi_\beta|^{2k} \right) $, can be used to change the short-distance behaviour of the wave function, and thus lower the ground state energy,  without effecting the topological properties \cite{FremlingShortDistance}. 

One might wonder whether such a non-minimal  description of the Laughlin states would have implications for the edge physics. To our knowledge this question has not been investigated numerically, but the simplest guess is that the edge still just supports a single, charged, mode. To see this, we use a non-minimal vertex operator,
\be{}
\tilde V_{(m,k)}(\xi, \bar\xi) =\, :e^{i\sqrt{m}\phi(\xi)} \, e^{i \sqrt{k}\phi_1(\xi)}  \, e^{i \sqrt{k}\bar\phi_1(\bar\xi)} :      \, ,\nonumber
\ee
where $\phi$ is the same field as in the original description of the Laughlin state --- \cf Eq.~\eqref{vo}. 
The corresponding edge Lagrangian contains kinetic terms for the left moving charged field, $\phi$,  as well as for the neutral  right- and left-moving  fields  $\phi_R$ and $\phi_L$. 
However, it generically also contains a term $\sim \cos(\phi_R - \phi_L)$ which opens a gap in the neutral sector, while leaving the edge unchanged, up to microscopic details. 

As a second example, consider $\nu = 2/3$. It can be viewed as the particle-hole conjugate of the $1/3$ Laughlin state or alternatively as a hole condensate in the $\nu = 1$ quantum Hall state. In the language of composite fermions it is described as a ``reverse flux attachment" CF state. Its $K$-matrix is 
\be{}
{\bf K}  = \left(
\begin{array} {cc}
1 & 2 \\
2 & 1 
\end{array} \right).\nonumber
\ee 
There is no consistent purely chiral description of this state, since $\bf K$ has a negative eigenvalue, corresponding to a counter-propagating edge mode. This can also be seen at the level of wave functions, in that the naive (112)-state has a very poor overlap with the exact Coulomb state, and it does not have the correct thin torus limit  \cite{suorsa2011quasihole,suorsa2011general,bergholtz2008quantum}. So one has to introduce an antichiral component. Again there are many possible ways of splitting $\bf K$. In particular, all integer-valued splittings 
\be{}
{\bf K} = \bkappa - \bar\bkappa 
= \left(\begin{array} {cc}
1+k & 2+l \\
2+l & 1+k 
\end{array} \right) - 
 \left(\begin{array} {cc}
k & l \\
l & k 
\end{array} \right) \nonumber
\ee
are `allowed', in the sense that $\bkappa$ and $\bar\bkappa$ are both positive (semi-)definite, provided $k \geq l+1$ ($k,l\geq 0$). The special case ($l=0, k=1$) reproduces Jain's $\nu = 2/3$ ``reverse flux attachment" wave function  \cite{wu1993mixed},
\begin{multline}
\psi_{2/3}^{Jain} = (\bar 1 - \bar 1) \bar\partial_2 (\bar 2 - \bar 2) \, (1-1)^2(2-2)^2(1-2)^2 \\ \times e^{-\sum_{j} |z_j|^2/(4\ell^2)}.\nonumber
\end{multline}
This wave function has been tested numerically (in spherical geometry) and is very close to the exact Coulomb ground state \cite{wu1993mixed}. It would be interesting to perform a more systematic numerical comparison of states with different splittings of the $K$-matrix. 
Again, note that the derivatives do not follow from the form of the $K$-matrix itself, but rather from the explicit condensation procedure explained below, and they are needed in order to get the correct shift. 


\subsection{Local and quasi-local quasiparticle operators} \label{quasilocal}
\label{ssec:qe}
Our `recipes' for writing down representative ground state wave functions for general abelian quantum Hall ground states were stated in sections \ref{ss:chiral} and \ref{ss:full}, without further explanation.  In this  section we discuss the CFT description of fractionally charged excitations, in particular quasielectrons. 

Recall Laughlin's  wave function for a quasihole in a $\nu=1/m$ quantum Hall state, localized at $\eta$, 
\be{}
\psi_{Lqh}(\eta,z_1,...z_N) = \prod_{i=1}^N (z_i - \eta) \psi_L(z_1, ..., z_N)\nonumber
\ee
and correspondingly for quasielectrons, 
\be{lqe}
\psi_{Lqe}(\bar \eta, z_1, ..., z_N) = \prod_{i=1}^N (2\partial_i - \bar\eta) \psi_L(z_1, ..., z_N)\,.
\ee
While the former agrees with the composite fermion quasihole, the CF quasielectron differs quite substantially from Laughlin's  \cite{jain2007composite}. The CF expression for a quasielectron localized at $\eta$  is
\begin{multline}\label{cfqe}
\psi_{CFqe}(\bar \eta ; \sgrupp[z]) = \sum_i (-1)^i e^{- q_h(|\bar\eta|^2-2\bar\eta z_i)/(4\ell^2)}   \\
 \times \prod_{j<k; j,k\neq i}\,\,(z_j - z_k)^m \partial_i \prod_{l \neq i} (z_l - z_i)^{m-1} e^{-\sum_{j} |z_j|^2/(4\ell^2)}\,.
\end{multline}
Numerical studies  \cite{kjoensberg1999charge,jeon2003fractional} have shown that the CF quasielectrons do better than Laughlin's  in that the former exhibits well-defined fractional statistics for braiding of two well-separated quasielectrons, while this is not the case for the latter. 
 This  issue was revisited by   \textcite{jeon2010thermodynamic}, who found that also Laughlin's quasielectron has the correct braiding statistics as long as the ratio $A_{\mbox{\tiny path}}/A\rightarrow 0$, where  $A_{\mbox{\tiny path}}$ is the area  enclosed by the braiding path and $A$ is the total area of the quantum Hall droplet. 
However, if $A_{\mbox{\tiny path}}$ encloses a finite fraction of the droplet, then the braiding phase of Laughlin's quasielectron becomes ill-defined,  due to the $\mathcal{O}(1/N)$ corrections to the Aharonov-Bohm phase found by  \textcite{kjoensberg1999numerical}.

There is an inherent asymmetry between the traditional descriptions of quasiholes and quasielectrons that is reflected also in the CFT formulation. When writing wave functions as CFT correlators, the expression for a Laughlin quasihole at $\eta$ is reproduced by inserting a local hole operator, $H(\eta)=\exp{[i\phi(\eta)/\sqrt m]}$. 
However, the naive guess for a quasielectron, using the inverse hole operator $H^{-1}(\eta)$ does not work, as this leads to unwanted singularities of the form $\prod_j (z_j-\eta)^{-1}$ in the wave function. The physical reason behind this was discussed in section \ref{shortsum}, where it was also pointed out that the problem can be solved by noting that since every electron in the liquid is surrounded by a correlation hole ($m$-fold in the $\nu=1/m$ state), it is `safe' to contract the liquid by contracting the correlation hole around any electron. Formally, this is obtained by modifying one of the electron operators; in the simplest case of a $1/m$ Laughlin state one introduces
\be{Pz}
P(z) = \partial_z :e^{i(\sqrt m - \frac{1}{\sqrt m})\phi(z)}:.
\ee 
This is essentially an inverse quasihole  fused with an electron, using  a particular regularization scheme to avoid the singularities mentioned above while retaining the correct fractional charge and anyonic statistics of the inverse hole. Replacing one of the electron operators $V(z)$ in the ground state correlator by $P(z)$ and antisymmetrizing over particle coordinates exactly reproduces the expression for a CF quasielectron at the center. It is then possible to, by hand, write an expression for a {\it localized} quasielectron as a coherent superposition of such a quasielectron in different angular momentum states. This is however not fully satisfactory, as one would wish for one single operator ${\cal P}(\bar\eta)$ which, when inserted into the ground state correlator, directly produces such a localized quasielectron state -- in analogy to the quasihole operator $H(\eta)$. 
Ideally, such an operator should make the anyonic properties --  fractional charge and anyonic braiding statistics -- manifest. It turns out that requiring manifest braiding statistics, \ie an anyonic monodromy factor and zero Berry's phase as discussed in Section \ref{sub:MRstate},  is quite non-trivial, and has so far not been achieved. It is, however, possible to instead use bosonic or fermionic representations of the quasiparticles. The monodromies are then bosonic or fermionic, and the statistics is hidden in the Berry phase. 
Before presenting this operator, let us shortly explain why obtaining explicit anyonic monodromies is problematic for quasielectrons.

The root of the problem can already be seen from the expression \eqref{Pz}; trying to use it to create multi-quasielectron states results in factors $\sim (z_i - z_j)^{4/3}$ for each pair of $P$'s. This is unacceptable since the wave function, although being a many-quasielectron state, describes a fermionic many-body state in the LLL and thus must be analytic and antisymmetric. 
To remedy this problem, one introduces (chargeless) auxiliary fields that make the braiding of the quasielectrons trivial (either bosonic or fermionic), while leaving their braiding properties vis-a-vis the electrons and quasiholes unchanged.  For instance, for a Laughlin state at filling $1/m$ one can use 
\be{Pz2}
P(z) = \partial_z :e^{i \frac{m-1}{\sqrt m}\phi(z)+i\sqrt{\frac{2m-1}{m}}\chi(z)}:
\ee
instead of the operator \eqref{Pz}, which amounts to fusing a bosonized inverse quasihole, $H_b^{-1}=\, :e^{-i\sqrt{\frac{1}{m}}\phi+i\sqrt{\frac{2m-1}{m}}\chi}:$, with an electron operator. Even though this  operator has fermionic monodromy, it can be used to create proper many-quasielectron states that exhibit the expected anyonic braiding phase.  
 
To get a better feel for this, let us briefly recall the case of quasiholes. In their original form, $:e^{i\phi/\sqrt{m}}:$, the quasihole operator produces factors $\sim (\eta_\alpha-\eta_\beta)^{1/m}$ for any pair of quasiholes. In this representation, the holonomy equals the monodromy, \ie the fractional braiding phase is fully displayed in the many-hole wave function, as explained in section \ref{ssec:MRconjecture}.
Such a representation -- commonly believed to be connected to the `minimal' CFT description --  is not the only viable choice. 
Alternatively, we can change the monodromy of $H$, \eg to bosonic ones, by introducing auxiliary bosonic fields as shown above. 
This does not, of course,  change the physics, but amounts to `shuffling'  the fractional statistics from the monodromy into the Berry phase. 
For quasiholes, we can therefore freely change the monodromy factor,  while for quasielectrons we are limited to either bosonic or fermionic monodromy, in order to avoid branch cuts in the electron coordinates. 
In this latter case the different choices will not just be a matter of normalization, but in fact correspond to wave functions that differ by factors $(z_i - z_j)^n$ where the coordinates are those of the quasielectrons. Although such factors clearly are important when the quasielectrons come close to each other, they do not influence the rest of the electrons. They are also negligible when the quasiparticles are far apart, and thus do not change the topological properties. 

Here we should also recall two other sources of short-distance ambiguities related to the CFT wave functions that were mentioned earlier. The  first occurred in the first example in section  \ref{ss:full}, and   amounts to introducing Jastrow factors like $\prod_{i,j}|\xi_i - \xi_j|^k$, or even more general functions of the distance between the particles. Since such factors do not introduce any angular momentum, they will not influence the density of the liquid and most likely\footnote{  
In spite of this reservation, we strongly believe in this statement.}  
neither will they influence  any other topological property. The second source of short-distance ambiguities was discussed in section \ref{ss:full}, and relates to  the particular ordering of the derivatives that are present in the LLL wave functions.   We should stress that this, rather large, freedom in changing the short-distance details of the wave functions without altering the topological properties, is in fact a virtue.  It might be used in a variational sense to find good wave functions with specified topological properties for realistic Hamiltonians. 
In this context we should also recall that within the composite fermion picture, similar short-distance ambiguities can occur. 
A prominent example is the approximation to the  LLL projection that was devised by \textcite{JainKamillaLLL1} and shows very good agreement with the exact projection.


\subsubsection{Quasilocal quasiparticle operators} \label{ss:qlqp}
Let us now return to the question of how to actually write down an operator ${\cal P}(\bar \eta)$ that directly creates a quasielectron at an arbitrary position $\vec \eta$. The idea is that the surplus of charge is generated by contracting the correlation holes of electrons in the vicinity of $\vec\eta$.  This localization is accomplished by means of a gaussian weight which localizes the quasielectrons on the scale of the magnetic length; thus the term ``quasi-local". The quasielectron operator has the same charge and conformal dimensions as $H^{-1}$. As indicated earlier, it involves ``fusing" an inverse quasihole with an electron in an appropriate way. In order to avoid the branch cuts that rendered Eq.~\eqref{Pz} unacceptable, we need to use a quasihole operator with trivial -- either fermionic or bosonic -- monodromy factor. The two choices differ in their short-distance properties, \ie how exactly the charge is localized. 
Depending on the task at hand, one or the other might be more convenient and we will use both possibilities in the following discussion. 
In analogy with the quasihole case, a correlator of the type $\langle {\cal P}(\bar\eta) \prod_i V(z_i)\rangle$ gives a wave function for a quasielectron localized around $ \vec\eta$. Moreover, multi-quasielectron states can be generated by multiple insertions of $\mathcal{P}(\bar\eta)$.

To be more concrete, the operator for a quasielectron with charge $-q_he$ is given by\footnote{For simplicity we are here somewhat cavalier about the the subtle issue of the background charge. See  \cite{suorsa2011general} for a precise discussion of this point.}
\begin{align}\label{qe}
{\cal P}(\bar \eta) = \int d^2w \, e^{- q_h/(4\ell^2)(|w|^2 + |\eta|^2 - 2\bar\eta w)} \, \left( H^{-1} \bar\partial J \right)_n(w) \, .
\end{align}  
Note the similarity to the quasi-local electron operators \eqref{eq:curlyV} discussed in section \ref{ss:full}. 
The exponential factor indeed localizes the quasielectron around the position $\vec\eta$, as it equals the gaussian $\exp[(-q_h/4\ell^2)|w-\eta|^2]$ up to a pure phase.  Note that this exponential factor is fully determined by this condition together with the requirement that the resulting gaussian factors for the electrons should come out correctly.  This is also equivalent with demanding that the gaussian factor in $\bar\eta$ in Eq.~\eqref{qe} is precisely the one expected for a charge $q_h$ particle. 
$J(w)= i \partial \phi(w)/\sqrt{m}$ is the holomorphic $U(1)$ current related to the electric charge. Its divergence, $\bar\partial J (w)$, vanishes except at the position of the charges. Thus, it can effectively be replaced by a sum of $\delta$-functions centered at the electron positions  \cite{hansson2009quantum}. 
$(...)_n$ denotes a generalized normal ordering, which we get back to below. Its purpose is to regularize the fusion of the inverse hole with the electron operators. 

A very appealing property of the quasielectron operator \eqref{qe} is its general applicability. It is not limited to Laughlin states, but can be used for any abelian, or even nonabelian, quantum Hall state, as long as one makes sure to use the appropriate choice for the inverse quasihole. 
For hierarchy states at level $n$, only one of the $n$ quasiholes gives a non-vanishing result when inserted into  Eq.~\eqref{qe}. Thus, for any level there is only a {\it single} quasielectron, in accordance to findings by \textcite{jain2007composite} for the CF series.
The example of nonabelian quasielectrons in the Moore-Read state is discussed in more detail in section \ref{sect:nahier}.
Finally, we stress that in spite of the quasi-local nature of the operator $\mathcal{P}$, the resulting density profile of the corresponding quasielectron is only marginally larger than that of the quasiholes in the same state (see for instance chapter 8.5 in \cite{jain2007composite}).


\subsubsection{A case study --- quasielectrons in the $\nu = 1/3$ Laughlin state} 
To illustrate what happens when computing correlators of the type $\langle {\cal P}(\bar\eta) \prod_i V(z_i)\rangle$, we consider the simplest case, quasielectrons in the $\nu = 1/3$ Laughlin state, so that $V=\, :\exp(i\sqrt 3 \phi_1):$ . Without  changing the topological properties, we can use  a bosonized inverse quasihole,  
$$H_b^{-1} = \, : e^{-i\frac{1}{\sqrt{3}}\phi_1+i\sqrt{\frac{5}{ 3}}\phi_2 }: \, .$$
As  mentioned above, current conservation implies that    $\bar\partial J(w)$ has support only at positions of charges. This  leads to delta functions, $\sum_i \delta^2(w-z_i)$ where $z_i$ are the electron positions, and thus, 
\be{}
\langle {\cal P}(\bar\eta) \prod_i V(z_i)\rangle &=& 
\langle \int d^2w \, e^{-(q_h/4\ell^2)(|w|^2 + |\eta|^2 - 2\bar\eta w)} \nonumber  \\  
&\times& \left( H_b^{-1} \bar\partial J \right)_n(w) \prod_i V(z_i)\rangle  \nonumber\\
&=& 
\sum_k (-1)^k  \, e^{-(q_h/4\ell^2)(|z_k|^2 + |\eta|^2 - 2\bar\eta z_k)} \nonumber \\
&\times& \langle \left( H_b^{-1}  V \right)_n(z_k) \prod_{i \neq k} V(z_i)\rangle\, ,\nonumber
\ee
where $q_h=1/3$ denotes the quasielectron charge. 
In the simplest case of Laughlin states, the normal ordering is the conventional one,
\be{no}
\left( H_b^{-1}  V \right)_n(z_k) &=\,  & \nonumber\frac{1}{2\pi i}\oint \frac{dz}{z-z_k}H_b^{-1}(z) V(z_k)  \\
&=& \partial :e^{\left(i\frac{2}{ \sqrt 3} \phi_1(z_k) +i\sqrt{\frac{5}{3}} \phi_2(z_k)\right)}:,
\ee
where the operator product expansion of $H_b^{-1}V$ was carried out  to next to leading order before integrating, thus keeping the first descendant (here just meaning first derivative) of the vertex operator.  
The result is nothing but the operator $P(z_k)$ in Eq.~\eqref{Pz2}. Note that the derivative is a direct consequence of the normal ordering.\footnote{
For more general hierarchy states, involving several bosonic fields, one has to apply a generalized normal ordering (which coincides with the expression \eqref{no} in the Laughlin case),
$$
(AB)_n(w) \equiv \oint_w dy \,T(y) \oint_w dz \,A(z) B(w)
$$ 
where $T$ is the energy-momentum tensor. This normal ordering, including the definition of the integration contours, is explained in detail in  \cite{hansson2009quantum}. Here we just mention that the $y$-integration amounts to taking the first descendant of the operators to its right, which again produces the pertinent derivatives.    }
So, to summarize, one gets
\be{1qesum}
\langle {\cal P}(\bar\eta) \prod_i V(z_i)\rangle &=& \sum_k  (-1)^k e^{-(q_h/4\ell^2)(|z_k|^2 + |\eta|^2 - 2\bar\eta z_k)}   \nonumber \\
&\times&  \langle P(z_k) \prod_{i \neq k} V(z_i)\rangle \, , 
\ee
with the announced gaussian-weighted sum over ``shrinking the correlation hole" (\ie turning $V$'s into $P$'s) at the various electron positions. 
This expression is identical to the one for a CF quasielectron localized at $\eta$ in Eq.~\eqref{cfqe}. However, the expressions for several quasielectrons do differ slightly between the CF and CFT construction  \cite{hansson2009quantum}.   


\subsubsection{Quasilocal quasiholes and local quasielectrons}  \label{nonlocalop}
 In fact, the description  of quasielectrons and quasiholes is not as asymmetric as it seems at first sight. It is possible  to reverse the above logic and construct a {\it quasilocal quasihole} operator ${\cal H}(\eta)$ as a gaussian-weighted sum of {\it expanding} the correlation holes around the electrons in the vicinity of some position $\eta$  \cite{suorsa2011general}. Technically, this involves fusing the inverse of a {\it local} quasielectron operator (which in fact corresponds to Laughlin's original quasielectron \eqref{lqe}) with a suitable representation of the electron operator. This is well-defined within the coherent state formalism discussed in section \ref{ss:full}.
A single insertion of a hole operator $\langle {\cal H}(\eta) \prod_i {\cal V}(z_i) \rangle$ produces an alternative to Laughlin's quasihole wave function which actually is slightly better numerically than the latter  \cite{jeon2005composite}. It is appealing that quasielectrons and quasiholes can be put on the same footing in this way. Moreover, it is this description of quasiholes, rather than local Laughlin holes, one has to resort to when carrying out an explicit hierarchy construction for condensates of quasiholes.
Also note the similarity with the quasi-local electron operator in \eqref{eq:curlyV}. This  emphasizes the point made in passing at the end of section \ref{sec:genform}:  Because of the cyclotron motion, all charged particles acquire a size in the LLL, so  condensing quasiparticles into a hierarchy state might not be conceptually very different from condensing  electrons into a Laughlin state.


\subsection {CFT wave functions as hierarchy states } \label{sec:cfthier} 
 \textcite{hansson2009quantum}  proved that the chiral states of section \ref{ss:chiral} can be written in the Halperin form, and thus form a hierarchy. The extension to the full hierarchy, including hole condensates, was given by \textcite{suorsa2011general}. We shall here give the argument for the simplest case of the $\nu = 2/5$ state in the positive Jain series that is obtained by condensing quasielectrons as densely as possible in the $\nu=1/3$ state. The starting point is Eq.~\eqref{hiwf}
\be{}
\Psi_{n+1} ( \vec r_1\dots \vec r_{N}) &=&   \int  d^2\vec \eta_1 \dots   \int  d^2\vec \eta_M    \, \Phi_n^\star(\vec \eta_1 \dots \vec \eta_M )  \nonumber\\
&\times&  \Psi_n (\vec \eta_1 \dots \vec \eta_M ; \vec r_1 \dots \vec r_{N})  \nonumber \, ,\nonumber
\ee
describing the $\nu=2/5$ wave function as a coherent superposition of quasielectrons in the parent Laughlin state $\Psi_1$ at filling $\nu=1/3$. For this densest possible condensate, the number of quasielectrons, $M$, equals half the number of electrons in the parent state. In general, the relation between the number of quasielectrons and electrons is determined from demanding homogeneity of the final state  \cite{hansson2009quantum}. One constructs multi-quasielectron states by multiple insertions of the quasielectron operator,
\be{}
\Psi_{1}(\grupp[\eta] 1 M ; \sgrupp[\vec r])=   \langle {\cal P}(\bar\eta_1) {\cal P}(\bar\eta_2) \ldots {\cal P}(\bar\eta_M) \prod_{i=1}^N V_1(z_i)  \rangle \nonumber
\ee
where $V_1= \, :e^{i\sqrt 3 \phi_1}:$. The evaluation of this correlator is a direct generalization of the derivation of Eq.~\eqref{1qesum}, with one set of gaussian factors for each quasielectron coordinate.
Since the final polynomial has to be antisymmetric, one has to choose a symmetric pseudo wave function for the Laughlin-like state of the quasielectrons. Its generic form is
\be{}
\Phi^*(\vec\eta_1, ... \vec\eta_M) = \prod_{i<j} (\eta_i - \eta_j)^{2k} e^{-(q_h/4\ell^2)\sum_i|\eta_i|^2} \, , \nonumber
\ee
where in this case $q_h=1/3$, and $k=0$ for the densest possible condensate. The gaussian factors in the pseudo wave function combine with those from the many-quasielectron correlator into holomorphic delta functions \eqref{eq:LLLdelta},
so that the integration over the $\eta$'s becomes trivial. 

The result is, for general $k$,
\begin{multline}\label{eq:hierarchylevel2}
\Psi_2(\sgrupp[\vec r])={\cal A} \left[  \langle \prod_{k=1}^{M} V_2(z_k) \prod_{l=M+1}^{N} V_1(z_l) \rangle \prod_{i<j=1}^{M} (z_i - z_j)^{2k} \right] \\
	=(1-1)^3 (2-2)^{2k}\partial_2 (2-2)^{3} (1-2)^2 e^{-\sum_{j} |z_j|^2/(4\ell^2)} \,, 
\end{multline}
with $M=N/(2+2k)$   and 
$$V_2= \, : e^{i\sqrt{2k}\phi_3}\partial e^{i2/\sqrt{3}\phi_1+i\sqrt{5/3}\phi_2}:\, .$$ 
Note that the precise form of the vertex operator depends both on the details of the normal ordering \eqref{no}, as well as on the choice of the quasihole operator. For instance, using the simplest {\it fermionic} quasihole (combined with a fermionic pseudo wave function), yields a vertex operator
$$V_2=\, : e^{i\sqrt{2k+1}\phi_3}\partial e^{i\sqrt{4/3}\phi_1+i\sqrt{2/3}\phi_2}: \, .$$
For $k=0$, \pref{eq:hierarchylevel2} is simply the Jain 2/5 wave function, while the less dense condensate with $k=1$ gives a trial wave function for filling fraction $\nu=4/11$  \cite{bergholtz2007microscopic}. This procedure naturally generalizes to the entire chiral hierarchy, and gives the wave functions shown in Eq.~\eqref{genstate}.  Similarly, quasihole condensation can be carried out using multiple insertions of the quasilocal quasihole operator ${\cal H}(\eta)$ (discussed in section \ref{nonlocalop}) to construct the required many-quasihole states  \cite{suorsa2011quasihole}. 


\subsection{Topological properties and plasma analogy}  \label{hiplasma}

We already mentioned that it, perhaps  surprisingly,  has turned out to be quite difficult to extract the topological content from even explicitly given abelian hierarchy wave functions. Superficially, the problem resembles the one we touched  upon in section \ref{subsub:monhol} in the context of nonabelian states. There we saw that correlators describing states with many quasiparticles, are not uniquely defined, but depend on the fusion channels of the quasiparticles, and are generally to be thought of as a vector in the space of conformal blocks. For the abelian hierarchy wave functions, there is no such degeneracy, but instead the wave functions are given by sums of conformal blocks; these sums are needed in order to satisfy the Pauli principle.  
In neither of the cases, there is a simple plasma analogy as in the case of the Laughlin states, or  the multi-component  states discussed in section \ref{multicomponent}. An additional complication in the hierarchy case is that the operators are not all primary 
fields, but descendants involving derivatives. At this point it is important to stress that  although a hierarchy state  is an antisymmetrized (in the $z_i$'s) sum of correlators   $\langle \prod_k{\cal H}(\eta_k)  \prod_l{\cal P}(\eta_l)  \prod_i {\cal V}(z_i) \rangle$, each term has the same monodromy. 
Thus it is meaningful to talk about the monodromy of a hierarchy  state, and assuming generalized screening as discussed in section~\ref{subsub:monhol}  we posited that  holonomy = monodromy also for these states. We now discuss the status of this conjecture.

Most attempts to understand the topology of the abelian hierarchy states boil down to the claim that they essentially can be described by a multi-component Coulomb plasma. In section \ref{ssec:hierarchy} we mentioned  the arguments for this given by \textcite{read1990excitation}. These are based on the assumption that quasiparticle  wave functions  are orthogonal if the particle positions differ much more than a magnetic length, and while quite reasonable, at least for quasiholes,  it is not so easy to prove. In the case of quasielectrons, there are extra complications due to the need to regularize singularities like $(z_i - \bar\eta)^{-1}$.

In section \ref{subsub:monhol} we discussed various attempts to find plasma mappings also for nonabelian states. 
Here we should mention that  
although the main focus in  \cite{read2009nonabelian} is on the nonabelian states, the analysis is  also applicable to abelian multicomponent states, and the Halperin $(m,m,n)$ state discussed in section \ref{multicomponent} is in fact taken as an example. Because of the need of anti-symmetrization between inequivalent groups of electrons, it is far from clear that this kind of analysis would apply to the hierarchy states, but it is certainly worth further considerations. \textcite{bondersongurarie} 
also consider hierarchy states, but  again it is not clear that their analysis properly takes into account the need for anti-symmetrization. }

To  appreciate the problem, consider the simplest possible case --- the normalization integral for the $\nu = 2/5$ ground state wave function \eqref{2/5wf}
$$
\Psi =\mathcal{A}\left\{ (1-1)^3\partial_2(2-2)^3(1-2)^2  \right\}  e^{-\sum_i |z_i|^2/(4\ell^2) }\, . \nonumber
$$
For a $2N$-particle state, the two groups each contain $N$ particles, and recalling that the function is fully antisymmetric within the groups,  the sum due to the antisymmetrizer $\mathcal A$ contains $(2N)!/(N!)^2$ distinct terms. Up to an overall constant, the normalization integral is, 
\be{normasz}
\sum_P \int \prod d^2 z_i \,  \Psi^\star ( z_1 \dots z_N) \Psi (z_{P(1)} \dots z_{P(N)} ) 
\ee
where $P$ denotes a permutation of the coordinates. Let us first consider those permutations that only reshuffle the coordinates {\it within } the groups. Disregarding, for the time being, the derivatives, the integrand will  be 
$$ \prod_{i <j = 1}^N  |z_i - z_j|^6 \prod_{i <j = N+1}^{2N}  |z_i - z_j|^6 \prod_{i =1}^N \prod_{j = N+1}^{2N} |z_i - z_j|^4 $$
and can directly be written as the partition function for a two-component Coulomb plasma, just as in the multicomponent case discussed earlier. 
 
However, for general terms in the sum \eqref{normasz} the coordinates will not match, since a particle at $z_i$ belonging to group 1 together with another  at $z_j$ belonging to group 2 will give the complex factor  $(\bar z_i - \bar z_j)^3 (z_i - z_j)^2$.
Thus, the Hamiltonian for the equivalent plasma will, in addition to the usual ``electric" interaction $\sim \ln|z_i - z_j|$ contain a term $ \sim i \mathrm{ Arg} (z_i - z_j)$   which is interpreted as the interaction between a charge and a vortex  \cite{nienhuis1984critical}.   Thus, still disregarding the derivatives, the terms in the sum in the expression \eqref{normasz} have the interpretation of a Coulomb plasma with electric charges, and a number of vortices. It is rather straightforward to include quasiparticles, so the question is whether all of the terms (or at least the dominant ones) correspond to a screening plasma. Also, 
including the derivatives is far from trivial, but the hope is that  they can be mapped onto impurities in the plasma. It remains an open question whether or not these expressions can be  understood well enough to be useful in the quantum Hall context.


\subsection{Composite fermions and the hierarchy  } \label{CFvshier}

Recall that a strength of the hierarchy construction is that it encodes the topological properties of a hierarchy state, while these are less obvious in the CF description. On the other hand, the CF formulation gives a very appealing, and phenomenologically successful, picture of the prominent states in the Jain sequences as systems of essentially free fermions, while the hierarchy picture of  these states, is rather complicated ($\nu = 4/9$, for example,  amounts to  three successive quasielectron condensates on top of a Lauglin $\nu=1/3 $ state). It is also far from clear why any of the schemes should lead to good candidates for ground states.  The argument in the CF case is that the introduction of a ``flux-attachment" factor that keeps the electrons apart, will vouch for a low interaction energy; but this is true only provided that the projection to the LLL is not too important, which is by no means intuitively obvious. In the hierarchy picture one expects low energy states since the daughter states 
are built by correlating  the low energy excitations of the parent state, {\it viz.} the quasiparticles, in an energetically favorable Laughlin-like configuration. It is however  not clear why the good correlations in the original parent liquid are not totally destroyed by the sequence of condensations. 

Thus, at the level of heuristics, both pictures have their pros and cons, and the precise relation between them has been a topic of debate. 
\textcite{read1990excitation} has argued that both schemes have the same topological content, and thus essentially describe the same physics,\footnote{
The same conclusion was reached by \textcite{bergholtz2008quantum} by studying the TT limit of the hierarchy, and also by    \textcite{bonderson2012hierarchical}. }
 while \textcite{jain1990theory,jain2007composite} has argued  that the two schemes are distinct. This issue has been difficult  to settle; while the CF approach has provided explicit wave functions that can be tested numerically to a great precision, there has been no viable numerical scheme to evaluate the hierarchy wave functions. 

The results of the previous sections shed considerable light on the relation between the hierarchy and the composite fermion states. 
First, it shows that Jain's very successful wave functions fit beautifully into the hierarchy scheme, which strongly suggests that they have the expected topological properties. Secondly, it shows that  the hierarchy construction of ``condensates within condensates" in the end produces wave functions which intuitively have very low interaction energy since they build in a large number of Jastrow factors in the electron coordinates, in a way that  is compatible with the filling fraction and the shift.  Thirdly, and for the same reason, it provides an intuitive understanding of why the Jain states are so good in spite of the LLL projection. 


\section{CFT description of nonabelian hierarchies}  \label{sect:nahier}
 While the  physics pertinent to the LLL is very well understood in terms of the Haldane-Halperin hierarchy described in the previous sections, this is not true for the second  LL. The latter is potentially very interesting, as numerical studies have shown that  certain nonabelian quantum Hall states could be stabilized. 
One of the underlying reasons is the nodal structure of single-particle orbitals within the second LL. 
This causes the matrix elements,  even for purely repulsive interactions, in the second Landau level to have a `hollow-core' ---  meaning that the effective interaction between two electrons can be attractive at certain length scales. 
This favors pairing or clustering of electrons, which is a common feature of nonabelian candidate states. 
 From numerical work,  \eg by \textcite{wojs2001electron}, one expects clustering to be important in the second LL, which in turn may stabilize nonabelian quantum Hall liquids \cite{greiterwenwilczek, moore1991nonabelions,read1999beyond}. One naturally may wonder  if there is an underlying, governing principle for the fractions in the second LL, similar to the Haldane-Halperin hierarchy in the LLL. 
There have been several attempts to generalize the Haldane-Halperin hierarchy to nonabelian states. 
The main difficulty in such an endeavor is the nonabelian statistics of the quasiparticles. 
It is far from clear how to make sense of the integration over quasiparticle positions in Eq.~\eqref{hiwf},  as the many-quasiparticle states form a vector space and moving quasiparticles around changes not just an overall phase, but the actual state. 

Below, we review three different possible solutions to this conundrum that lead to quite different predictions as to what filling fractions can be obtained, and what properties the resulting quantum Hall liquids  have. 
For the sake of simplicity, we restrict the  discussion to using the Moore-Read state as parent state. 
But all the hierarchies can be applied to general nonabelian states, in particular the Read-Rezayi series  \cite{read1999beyond}. 
Note that hierarchy states on top of the anti-Pfaffian \cite{AntiPfaffian1,AntiPfaffian2} -- the particle-hole conjugate of the Moore-Read Pfaffian state --  can  be obtained by simply taking the particle-hole conjugate of hierarchy states on top of the Moore-Read state. 
While we are of course interested in these hierarchies because of their relevance for the second LL, we will phrase the wave functions (and filling fractions) within the LLL. Note that you can always map between different LLs as long as the interaction Hamiltonian is properly adjusted (see Appendix \ref{app:QHE}).


\subsection{The Bonderson-Slingerland hierarchies} \label{sec:bs}
\label{ssec:BS}
The hierarchy  by \textcite{bonderson2008fractional} (BS) is essentially an abelian hierarchy on top of a nonabelian state. 
The basic idea behind this hierarchy is that for a sufficiently large density of the nonabelian quasiparticles, the interactions among them lift the degeneracy between the different fusion channels. 
Thus, it may be energetically favorable that pairs of quasiparticles form essentially abelian excitations, which may then condense in the same fashion as for abelian quantum Hall states \eqref{hiwf}.
Using the CFT description of the Moore-Read state  in Section \ref{sub:MRstate}, these  abelian quasihole operators can be written either as 
\begin{align}\label{eq:BSqh1}
\psi_{qh}&=\mathbf 1 \, :e^{i\sqrt{1/2}\phi_1(\eta)}:  
\end{align}
or as 
\begin{align}\label{eq:BSqh2}
\psi_{qh}&=\psi(\eta) \, :e^{i\sqrt{1/2}\phi_1(\eta)}:\,,
\end{align}
\ie they combine the usual Laughlin quasihole of an abelian, bosonic $\nu=1/2$ state with the fields $\mathbf 1$ or $\psi$ depending on which of the fusion channels is energetically preferred. 
For quasielectrons, one  instead combines a Laughlin or CF quasielectron with $\mathbf 1$ or $\psi$. 
The two possibilities for the preferred fusion channel give distinct hierarchies. In particular, the series of filling fractions, as well as the shift at a given filling fraction differ for the two cases.  
Note that condensing abelian quasiparticles cannot change the nonabelian part of the CFT, but only the charge part. 
In particular, daughter states in the BS hierarchy on top of the Moore-Read state  always have quasiparticles of Ising type. 
Consequently, the torus ground state degeneracy for such hierarchy states is always given by $3q$ for even number of electrons and $q$ for odd number of electrons, where $q$ denotes the denominator of the filling fraction. 
The first factor $3$ is due to the Pfaffian, while the second factor $q$ encodes the (trivial) center-of-mass degeneracy of fractional quantum Hall states on the torus, see Eq.~\eqref{eq:gsdegeneracy} or alternatively Section~\ref{sect:torus}.

Let us first focus on the condensation of pairs in fusion channel $\mathbf 1$, where the discussion is slightly simpler. 
As we are essentially condensing Laughlin quasiholes or quasielectrons, we can use the same pseudo wave function as  for the abelian hierarchy. 
The resulting daughter wave functions have a particularly simple  form,
\begin{align}
\label{eq:BSwf}
\Psi_{BS}&=\mbox{Pf}\left[\frac 1 {z_i-z_j}\right]\times \Psi_{\nu}^{(b)}(z_1,\ldots,z_N) , 
\end{align}
where $\Psi_{\nu}^{(b)}$ is a bosonic state in the abelian hierarchy over the  Laughlin $1/2$ wave function,  and the possible filling fractions for $\Psi_{BS}$ are given by Eq.~\eqref{eq:fillingfraction} with $m=2$. Its shift $S$ differs from the one for $\Psi_{\nu}^{(b)}$ due to the Pfaffian, which reduces the highest possible angular momentum by one. Thus,  $S=S_{\nu}+1$, where $S_{\nu}$ denotes the shift of  $\Psi_{\nu}^{(b)}$ as given in Eq.~\eqref{eq:shiftK}.
The expression \eqref{eq:BSwf} is an example of building a hierarchy by multiplying two known wave functions, as discussed in Section \ref{ssec:otherhierarchies}.  

To give a concrete example we consider the first hierarchy level on top of the Moore-Read state. 
In this case, we can  use the techniques of the abelian hierarchy in section \ref{sect:expwfs} to obtain explicit wave functions. 
For instance, condensing quasielectrons with  maximal  possible density gives a model wave function for filling fraction $\nu=2/3$,
\begin{align}
\label{eq:BSqe1}
\Psi_{BS}&=\mbox{Pf}\left[\frac 1 {z_i-z_j}\right]\times \Psi_{2/3}^{(b)}(z_1,\ldots,z_N)\nonumber\\
&=\mbox{Pf}\left[\frac 1 {z_i-z_j}\right]\times\mathcal{S}\left\{ (1-1)^2\partial_2(2-2)^2(1-2)  \right\}  \nonumber  \\
&\times e^{-\sum_i |z_i|^2/(4\ell^2)}, 
\end{align}
where $\mathcal{S}$ denotes an overall symmetrization.
Note that  $\Psi_{2/3}^{(b)}$ is simply the bosonic version of the wave function~\eqref{2/5wf}. 
This state has two electron operators, both of which are combinations of $\psi$ with the vertex operators of the $\nu=2/3$ bosonic CF state,
\begin{align}
\psi_{e1} (z)&=\psi(z):e^{i\sqrt{2}\phi_1(z)}:\nonumber\\
\psi_{e2}(z)&=\psi(z)\partial_z :e^{i\sqrt{1/2}\phi_1(z)+ i\sqrt{3/2}\phi_2(z)}:\,. \nonumber
\end{align}
The quasihole spectrum is generated by two  operators, both with charge $e/3$,
\begin{align}
\psi_{qh1} (\eta)&=\sigma(z)\,:e^{i\sqrt{1/8}\phi_1(\eta)+i/(2\sqrt{6})\phi_2(\eta)}: \nonumber\\
\psi_{qh2} (\eta)&=\, :e^{i\sqrt{1/2}\phi_1(\eta)+i \sqrt{1/6}\phi_2(\eta)}:\, .\nonumber
\end{align}
$\psi_{qh1}$ is the straightforward analog to the Moore-Read quasihole. 
The  vertex operator is determined by requiring it to reproduce $(z-\eta)^{1/2}$  in correlation functions with the vertex operators of both $\psi_{e1} $ and $\psi_{e2} $, \ie the quasihole-electron correlations are identical to the ones in the Moore-Read state.
$\psi_{qh2}$, on the other hand, is the usual abelian quasihole that lives entirely in the charge sector. 
The state \eqref{eq:BSqe1} has the same $ K$-matrix as the abelian bosonic $\nu=2/3$ state, \ie 
\be{eq:bsK1}
\mathbf K&=\left (\begin{array}{cc} 2&1\\ 1&2 \end{array}\right),\nonumber
\ee
 shift $S=4$, and a ground state degeneracy of $9$ on the torus for even number of electrons. 

Condensing abelian quasiparticles of type $\psi(\eta): e^{i\sqrt{1/2}\phi_1(\eta)}:$ is not conceptually more difficult than the case above, but the expressions become much more complicated, as the quasiparticle coordinates enter in the Pfaffian. 
Also, note that the extra fermion field $\psi$ in Eq.~\eqref{eq:BSqh2} changes the statistics of the quasiholes compared to that  in Eq.~\eqref{eq:BSqh1}.
To compensate for this, the pseudo wave function acquires an extra Jastrow factor in the quasihole positions, which in turn alters the $ K$-matrix of the daughter states. 
As the  filling fractions in the BS-hierarchies are solely determined by the $K$-matrix, one finds a different series of filling fractions compared to the case where the condensed quasiparticle pairs are in the trivial fusion channel. 
The shifts of the daughter states also differ, and are given by  
$ S=S_\nu+2-\sum_{j} (K^{-1})_{1j}$, where $S_\nu$ is determined by the $K$-matrix via Eq.~\eqref{eq:shiftK}. 
To give a concrete example, let us again consider the maximal density quasielectron condensate at filling $\nu=3/5$, which is described by the $K$-matrix
\be{eq:bsK2}
\mathbf K&=\left (\begin{array}{cc} 2&1\\ 1&3 \end{array}\right). \nonumber
\ee
This state has shift $S=13/3$  and a torus ground state degeneracy of  $15$ for even number of electrons. Note that even though the shift is fractional, $\nu  (N_\Phi+S)$ in Eq.~\eqref{eq:shift} is an integer.


\subsection{The Levin-Halperin hierarchies} \label{levinhalperin}

The  hierarchy by \textcite{levin2009collective} is simplest to describe for condensing nonabelian quasiholes. 
As already mentioned, this is in general not possible because of the branch cuts at the quasiparticle positions that make the integrals \eqref{hiwf} ill-defined. The way out devised by Levin and Halperin is to choose a pseudo wave function which combines with the original multi quasiparticle wave function to form a correlator of full CFT fields, \ie fields containing both holomorphic and antiholomorphic parts. Such a correlator is  guaranteed not to contain any branch cuts and can thus be integrated over with impunity. The pseudo wave function can, in fact, be interpreted as the proper many-body wave function for the nonabelian anyons and is given by 
\begin{align}
\Phi^\star_\alpha(\vec \eta_1,\ldots, \vec \eta_M)&=\langle \prod_{j=1}^M \bar\sigma(\bar \eta_j)e^{i\sqrt{2m+1/8}\,\bar\phi(\bar\eta_j)}\mathcal{O}_{bg}\rangle_\alpha \nonumber
\end{align}
for the Moore-Read state. 
Here, $M$ is the number of anyons, $\alpha$ labels the conformal block of the $\sigma$ correlator, and $m$ is a positive integer that sets the density of quasiholes in Eq.~\eqref{hiwf}. 
Note that the bosonic field $\bar\phi$ in the pseudo wave function is independent of the original bosonic field in Eq.~\eqref{eq:MRe}. 
The product of the pseudo wave function and the many-quasihole wave function in  Eq.~\eqref{hiwf} now becomes an inner product, as both wave functions are vectors in the $2^{M-1}$ dimensional vector space,
\begin{multline}
\label{eq:LHwf}
\Psi_{1} ( \vec r_1\dots \vec r_{N})  =   \int  d^2\vec \eta_1 \dots   \int  d^2\vec \eta_M    \\
 \,\sum_{\alpha} \Phi^\star_\alpha (\vec \eta_1 \dots \vec \eta_M )  \Psi_{MR}^{\alpha} (\vec \eta_1 \dots \vec \eta_M ; \vec r_1 \dots \vec r_{N})\, .
\end{multline}
As promised, the integrand is  well-defined in the $\bar\eta$ coordinates, \ie  it  no longer has branch-cuts. 
This can readily be seen from the decomposition of the full $\sigma$-correlator into holomorphic and anti-holomorphic blocks,
\begin{align}
\langle \prod_{j=1}^M \sigma (\eta_j,\bar \eta_j)\ldots \rangle&=\sum_{\alpha}\langle \prod_{j=1}^M \bar\sigma (\bar \eta_j)\ldots \rangle_\alpha \times\langle \prod_{j=1}^M \sigma (\eta_j)\ldots \rangle_\alpha,
\end{align}
and using that the full correlator has no branch cuts. 
This hierarchical construction yields candidate wave functions for filling fractions $\nu={8m}/({16m+1})$. It turns out that these states are abelian, despite being daughters of a nonabelian state. 

Evaluating Eq.~\eqref{eq:LHwf}  is in principle possible for small numbers of electrons, but there is no closed expression for an arbitrary number of electrons. 
We can, however, still deduce the quasihole properties by requiring that the corresponding operators are {\it local} with respect to the electron operators and the full quasihole operator, \ie  the combination of both the holomorphic and anti-holomorphic part given by
\be{}
\sigma(\eta,\bar\eta):e^{i\phi(\eta)/(2\sqrt 2) +i\sqrt{2m+1/8}\bar\phi(\bar\eta)}:\,.\nonumber
\ee
The above locality conditions ensure that the full expression in Eq.~\eqref{eq:LHwf} has no branch cuts or poles, and, thus, gives a valid electronic wave function. 
 From this constraint one can deduce that the quasihole spectrum is generated by a single  quasihole operator with electric charge $ e/( {16m+1} )$ and abelian exchange phase $\theta=\pi {(16m-1)}/{(16m+1)}$. 
 As the first level  daughter states are abelian, further condensation of quasiholes will, of course, only yield abelian candidate states. 
 Interestingly, it was pointed out by \textcite{levin2009collective} that the series of states \eqref{eq:LHwf} can alternatively be obtained by an abelian hierarchy starting from the strong-pairing $\nu=1/2$ state, which is a $\nu=1/8$ Laughlin state of strongly bound electron pairs that was originally discussed by \textcite{halperin1983theory}. The two approaches yield wave functions with the same topological data, and should therefore be in the same universality class.

The simplest example for the Levin-Halperin hierarchy is the state with the highest density of quasiholes, \ie  $m=1$, which yields a candidate state at filling $\nu=8/17$. Even though this state is abelian, it is distinct from the abelian Haldane-Halperin hierarchy state at this filling fraction; the latter is obtained with $q=8$ and $p=1$ in the Jain series in Eq.~\eqref{jainwf}.
This can most easily be seen from their different  shifts; the Levin-Halperin hierarchy state has shift $S=5/2$, while the abelian Haldane-Halperin hierarchy state has shift $8$. 
Candidate states for this filling can also be obtained by the other hierarchies, but only at  higher levels. 

Levin and Halperin also suggested model states for condensing nonabelian quasielectrons, which in the simplest case yields a candidate state at $\nu=7/13$. 
This seems not as natural as the quasihole condensates. In particular, the quasielectron wave functions contain unphysical poles, and it is  not clear if these can be regularized while keeping the nonabelian braiding statistics explicit. 
The proposal for a state at filling fraction  $7/13$ is still very interesting because its particle-hole conjugate state at $\nu=2+6/13$ has been seen in experiments by \textcite{kumar2010nonconventional}, suggesting that this quantum Hall state may be a daughter state of the anti-Pfaffian (see also the discussion in \ref{sss:fractions}). 
Note that this filling fraction also occurs in the BS and Hermanns hierarchies.
However, neither of these provides a natural explanation for its occurrence.
The BS hierarchy contains this state at the second level, but it also predicts a state at filling $4/9$ as more prominent.  The latter was, however, not observed in \textcite{kumar2010nonconventional}. 
The Hermanns hierarchy discussed below contains this filling in the third level of hierarchy as a daughter state of $4/7$. 
The latter was again not observed. 
 Thus, so far only the Levin-Halperin hierarchy provides a natural explanation for $\nu=2+6/13$. 


\subsection{The Hermanns hierarchies}
\label{ssec:NAC}

In contrast to the other nonabelian hierarchies discussed in the previous sections, the daughter states in this hierarchy have {\it nonabelian statistics different from their parents.  }
Before getting to the actual construction, we first devote some time to explain how nonabelian quasielectrons can be represented in terms of CFT operators without unphysical singularities. 
We focus on the condensation of quasielectrons. 
One can also study quasihole condensation, but the wave functions will no longer have such a simple form and interpretation.

\subsubsection{Quasielectrons in the Moore-Read state}

Before discussing the nonabelian quasielectrons in the Moore-Read state, let us first recall again the essential steps in the definition of the abelian quasielectron \eqref{qe}.
The quasielectron operator is obtained by attaching an inverse quasihole to the electrons in such a way that the resulting correlator has no branch cuts and no poles in the electron coordinates. 
The latter is ensured by regularizing the resulting operators using a generalized normal ordering. 
The former is achieved by using a quasihole operator which has trivial  monodromies.
In the abelian case, the monodromies of the quasihole can be made trivial simply by adding auxiliary, chargeless, boson fields. 
This is not possible for the nonabelian quantum Hall states, within the usual description in terms of the Ising CFT, as the quasiholes are manifestly nonabelian. 
Braiding quasiholes around each other does not solely give an additional phase factor, but in general also changes the state. 
As a result, the quasihole of the Ising CFT description cannot be used in the definition of the quasielectron operator, as it necessarily leads to branch cuts in the electron coordinates. 

A way around this problem is to use an equivalent description of the Moore-Read state that does allow for trivial monodromies of the quasiholes. In the following, we use the multi-layer description of   \textcite{cappelli2001parafermion}. The bosonic Read-Rezayi  state at level $k$ is described by $k$ layers  of bosonic Laughlin $\nu=1/2$ states with a subsequent symmetrization over the layer index.
For example, the bosonic Moore-Read wave function can alternatively be written --- using the shorthand notation \eqref{shorthand} --- as
\be{Capelli}
\Psi_{MR}&=&\mathcal{S}\left\{ (1-1)^2 (2-2)^2\right\}  e^{-\sum_i |z_i|^2/(4\ell^2)} \,    ,\nonumber
\ee
where we divided the particle coordinates into two equal-sized groups, $I_1$ and $I_2$, and $\mathcal{S}$ denotes the overall symmetrization over all such partitions. 
This faithfully reproduces the Moore-Read ground state. Quasihole excitations are obtained by inserting Laughlin quasiholes in each of the layers. At level $k$,   one thus artificially introduces $k$ quasihole operators, which are then identified by the symmetrization. The vector space of $n$ quasihole states is spanned by the various orderings of the $k$ types of operators at positons $\eta_1,\ldots,\eta_n$, and is identical to the space obtained using the Ising  CFT description  \cite{cappelli2001parafermion}. 
However, in this description {\it the nonabelian part of the braiding statistics resides entirely in the Berry's phase}. 
The monodromies can then be made trivial by introducing $k$ additional bosonic fields in the same way as in the abelian case. 

As an example, the fermionic Moore-Read state --- phrased in the  two-layer description --- has vertex operators 
\begin{align}\label{CapelliVertex}
V(z)&=V_+(z)+V_-(z)\nonumber\\
V_\pm(z)&=\, :e^{\pm i\phi(z)}e^{i\sqrt{2} \chi_1(z)}:\nonumber\\
H_\pm (\eta)&=\, :e^{\pm \frac{i}{2}\phi(\eta)+i\sqrt{1/8}\chi_1(\eta)-i\sqrt{ 3/8}\chi_2(\eta) \mp (i/2) \chi_3(\eta)}:
\end{align}
where $\pm$ denotes the two layers  \cite{cappelli2001parafermion} and the manifest monodromies of the quasiholes were chosen to be fermionic. 
Using the quasiholes of Eq.~\eqref{CapelliVertex}  in Eq.~\eqref{qe} one can construct the two corresponding quasielectron operators, $\mathcal{P}_+(\bar\eta)$ and $\mathcal{P_-(\bar\eta)}$, with the correct quasi-local excess charge  $q_e=-q_h$. 
Just like the quasiholes, the quasielectrons must be inserted in pairs, otherwise the correlator vanishes because of charge neutrality in the $\phi$ sector. 
Inserting $2n$ quasielectrons, one finds $2^{n-1}$ linearly independent orderings of $\mathcal{P}_+(\bar\eta)$ and $\mathcal{P_-}(\bar\eta)$ in complete analogy to the quasiholes  \cite{hansson2009conformal,hansson2009quantum}.
An appealing, alternative interpretation of these operators is possible in the bosonic Moore-Read state. 
There, they can  be seen as the usual abelian quasielectrons in each of the layers, and the nonabelian properties arise due to the symmetrization over the layers. 
As shown by \textcite{rodriguez2012quasiparticles}, this allows for a numerically very efficient way to study   the properties of quasieletrons as well as  of the neutral excitations. 
\subsubsection{Condensing quasielectrons}
We now proceed to generalize the Haldane-Halperin hierarchy using the nonabelian quasielectrons introduced above. 
We restrict to the Moore-Read state as parent state, but mention briefly, at the end of this section, how the construction can be adapted to  other nonabelian states, such as the Read-Rezayi series. 

Consider a quasielectron wave function for  an insertion of $2M$ quasielectrons using  an equal number of $\cal{P}_+$ and $\cal{P}_-$ at positions $\eta_1,\ldots,\eta_{2M}$. 
Such a multi-quasielectron state is for instance given by 
\begin{multline}
\Psi_1(\vec{\eta}_1,\ldots,\vec{\eta}_M;\vec r_1,\ldots,\vec r_N)=\\ \langle\prod_{j=1}^M \mathcal P_+(\eta_j) \prod_{j=M+1}^{2M} \mathcal P_-(\eta_j)   
  \prod_{j=1}^N V(z_j)\mathcal{O}_{bg}\rangle, \nonumber
\end{multline}
but any other ordering of $\cal P_+$ and $\cal P_-$ gives an equally valid wave function.
 Note that the order of the quasielectron operators is, in general, important. In particular, 
\begin{multline}
\langle \mathcal P_+(\eta_1)\mathcal P_+(\eta_2)\mathcal P_-(\eta_3)\mathcal P_-(\eta_4 )\ldots \rangle \\
 \neq \langle \mathcal P_+(\eta_1)\mathcal P_-(\eta_2)\mathcal P_+(\eta_3)\mathcal P_-(\eta_4 )\ldots \rangle,  \nonumber
\end{multline}
even though $ \mathcal P_+$ and $ \mathcal P_-$ are indistinguishable by any local measurement.

Only $2^{M-1}$ of these orderings are linearly independent, but that still leaves a macroscopic number of parent states to choose from. 
It turns out that it is unimportant which of these states is chosen in Eq.~\eqref{hiwf}, as the integration over the quasielectron coordinates can be interpreted as an average over all positions, and the final result is independent of which of the multi-quasielectron states was used as a starting configuration. 
As the monodromies of the quasielectrons are fermionic, the simplest appropriate pseudo wave function is a fermionic Laughlin state for each of the layers, 
\be{pseudoNA}
\Phi^\star(\vec \eta_1,\ldots,\vec \eta_{2M})=\prod_{i<j\in I_1}(\eta_i-\eta_j)^{2p-1} \nonumber \\
\times \prod_{a<b\in I_2}(\eta_a-\eta_b)^{2p-1}e^{-q_h\sum_{j=1}^{2M}|\eta_j|^2/4\ell^2},\nonumber
\ee
where $I_1$ contains $\eta_1,\ldots,\eta_{M}$, \ie all quasihole positions appearing in $\cal P_+$, and $I_2$ contains the rest. 
The integer $p$ is fixed by requiring homogeneity of the daughter state, $M/N=1/(4p)$. 
The integral in Eq.~\eqref{hiwf} is straightforward to evaluate, and the daughter state at filling fraction $\nu=4p/(8p-1)$ can be compactly written as,
\begin{multline}
\label{daughterNAcomp}
\Psi_n(\vec r_1,\ldots,\vec r_N)= \mathcal{S}\Big\{ \Psi_{\tilde \nu}(\vec r_1,\ldots,\vec r_{N/2})  \\ 
\times \Psi_{\tilde\nu}(\vec r_{N/2+1},\ldots,\vec r_{N})\Big\}\prod_{i<j} (z_i-z_j), 
\end{multline}
where $\Psi_{\tilde \nu}$ is a {\it bosonic} hierarchy state at filling fraction $\tilde\nu=2p/(4p-1)$ and $\prod_{i<j} (z_i-z_j)$ denotes a {\it full} Jastrow factor containing all the electrons. 
Note that in writing the wave function in this form, we are cavalier about the exact placement of the derivatives, which now act solely within  $ \Psi_{\tilde \nu}$. 
While this changes the wave function slightly, it is widely believed  not to have an impact on its nonabelian properties (see also the discussion  below Eq.~\eqref{spelop}).
The benefit of writing the expression \eqref{daughterNAcomp} in this form lies in its natural interpretation  as symmetrized products of hierarchy (or CF) wave functions (times a Jastrow factor)-- very similar to the above interpretation of the Moore-Read state as a symmetrized product of  two Laughlin $1/2$ states (again times a Jastrow factor). 
Successive condensation of quasielectrons yields a hierarchy of model states with filling fractions $\nu=2\tilde\nu/(1+2\tilde \nu)$, where $\tilde \nu$ denote the filling fractions of the chiral (bosonic) hierarchy over the Laughlin 1/2 state.

 An exciting feature of this hierarchy is that the nonabelian statistics changes from parent to daughter state, in contrast to the hierarchy described in Section~\ref{ssec:BS} and without making the nonabelian statistics trivial as in Section~\ref{levinhalperin}. 
From Eq.~\eqref{daughterNAcomp}, however,  it is not apparent what the topological properties of the states are, as all the nonabelian properties are hidden in the Berry phase.
The easiest way to proceed is to consider the bosonic version of Eq.~\eqref{daughterNAcomp} and use the techniques introduced by \textcite{ardonne2008degeneracy} and  \textcite{ardonne2009domain} for the thin torus limit  \cite{bergholtz2008quantum}. 
This yields the ground state degeneracy and the quasihole fusion rules, but not the braiding properties of the state. 
For instance, for the subset of wave functions obtained by successively condensing quasielectrons with maximal density, 
\begin{multline}
\Psi_n(\vec r_1,\ldots,\vec r_N)=\mathcal{S}\left\{ \Psi_{n/(n+1)}(\vec r_1,\ldots,\vec r_{N/2}) \right. \nonumber\\ 
\times \left.\Psi_{n/(n+1)}(\vec r_{N/2+1},\ldots,\vec r_{N})\right\},\nonumber
\end{multline}
one finds a ground state degeneracy on the torus of $(n+1)(n+2)/2$ (for even number of particles) and $su(n+1)_2$ type fusion. 
The properties of the corresponding fermionic filling fractions can then be deduced by incorporating the Jastrow factor again. 
In fact, for the highest density condensate ($p=1$) in Eq.~\eqref{daughterNAcomp} at filling fraction $\nu=4/7$, one can even identify the CFT that makes braiding and fusion properties explicit in the monodromy factor by noting the close resemblance to the nonabelian spin-singlet state proposed  by \textcite{ardonne1999new} and \textcite{ardonne2001nonabelian}. 
This analysis determines the pertinent CFT to be $su(3)_2/u(1)^2$  \cite{gepner1987new}, as shown by \textcite{hermanns2010condensing}.
The generalization of this hierarchy to the Read-Rezayi states by taking $k$ layers instead of two is straightforward.  
Numerical tests on these type of wave functions have been performed by \textcite{sreejith2011bipartite} and \textcite{sreejith2013tripartite}, showing their relevance as possible candidate states in the second LL.


\section{CFT hierarchy wave functions in other geometries } \label{othergeom}
\newcommand {\s}{ {S}}
There are several reasons for studying quantum Hall liquids on geometries other than the plane, and in particular on closed geometries. Although these are, for obvious reasons, not experimentally realizable  (we do not have magnetic monopoles at our disposal in the lab), they are very interesting both from  theoretical and numerical point of view.  
In particular, closed geometries do not suffer from edge/boundary effects. On the disk geometry, these can be substantial   for the small system sizes possible to probe with numerical techniques.

The simplest example of a closed manifold is the sphere, which after the pioneering work by \textcite{haldane1983fractional}  has not only become the favorite geometry for numerical simulations, but also has been instrumental for understanding how the quantum Hall liquids respond to spatial curvature. 
The other example that we shall treat is the torus. Here the geometry is locally flat, but the manifold is topologically nontrivial which is manifested in a characteristic ground state degeneracy. 
In both these cases, a key point is to understand how  the magnetic translations, which are the proper symmetries of an infinite plane penetrated by a constant magnetic field, are manifested in other geometries. In the following we will outline how this is done and how hierarchy wave functions on the sphere and the torus can be obtained using CFT techniques. 


\subsection{The sphere} \label{sect:sphere}

In a suitably chosen gauge (Dirac gauge with the string entering at the north pole) solutions of the Landau problem on the sphere are monopole harmonics \cite{jain2007composite,stonebook}, and the multi particle solutions can be organized in angular momentum multiplets, with homogeneous states, such as  ground states, having $L=0$. Stated differently, the state should be invariant under {\it magnetic rotations}. We already saw that in the CFT formulation the physical condition of homogeneity of the quantum Hall liquid translated into charge neutrality enforced by coupling to a  background charge density. On a curved surface this is not sufficient for having a consistent description, since the electrons and the quasiparticles carry (orbital) spin and thus respond to curvature. As we shall see, what is needed is both an extra piece in the action and a modification of the electron operators. 

Before turning to the general hierarchy wave functions, it behooves  us to stress that in many cases one does not have to worry about these complications. For the ``simple" wave functions, \ie those that can be written as   correlators of primary fields only, it is straightforward  to move from the expressions on the plane to those on the sphere. In essence, this recipe amounts to expressing the $z$ coordinates in the polynomial part of the wave functions in terms of the spherical coordinates $(\theta, \phi)$, by the stereographic mapping  \cite{ReadRezayi1996}, 
\be{ster}
z \rightarrow 2 R \tan(\theta/2)e^{i \phi}  \, ,
\ee
from the infinite plane to  a sphere with radius $R$. At the same time, the gaussian factors have to be changed as $\exp(-|z|^2/4\ell^2) \rightarrow 1/(1 + |z|^2/4R^2)^{1+N_\Phi /2} $. 
Choosing the number of electrons $N_e$ and the number of flux quanta $N_\Phi$ as to implement the correct shift \eqref{eq:shift}, one obtains a valid wave function \cite{haldane1983fractional}.  
A similar, although somewhat more complicated scheme for moving wave functions from the plane to the sphere, has also been devised for the composite fermion states in the Jain series \cite{jain2007composite}.  
For the general hierarchy states, which cannot be written as a projected version of a number of filled Landau levels, these simple methods do not work.  

We now turn to the general case, and  review the main results from the original paper by \textcite{kvorning2013quantum}.
Naively, one would expect that expressing hierarchical wave functions on the sphere is straightforward: one simply has to evaluate the corresponding CFT correlators (e.g. Eq.~\eqref{genstate} for the fully chiral states) on the sphere. 
It would indeed be that simple, if the correlators contained only proper CFT operators (primaries or descendants), but the background charge makes it more complicated. 
So far we referred to the insertion ${\mathcal O}_{bg}$ in Eq.~\eqref{fullmultibg} as a background operator. An equivalent way, which is more appropriate for the present purpose, is to view it as a part of the action for the fields $\phi^\beta$. 
\textcite{kvorning2013quantum} showed that using
\begin{align}
\label{sback} 
S_{bg} =  -i    \frac{\left(K^{-1}\right)^{\alpha\beta}}{2\pi}\int dS\, \left( e\,t_{\alpha}B + s_\alpha {\mathcal R } \right) \,\left( q_{\beta\gamma}\phi^\gamma +  
\bar q_{\beta\gamma}\bar\phi^\gamma \right)  
\end{align}
 in the action when evaluating the correlators, indeed gives wave functions that are invariant under magnetic rotations. Note that adding the term $\sim \mathcal R$ is the standard way to include curvature, while the term $\sim B$ describes the neutralizing background. 
 Eq.~\eqref{sback} is quite natural in light of the combination 
 $ t_\alpha  eA_\mu \epsilon^{\mu\nu\sigma} \partial_\nu a_\sigma^\alpha + { s_\alpha}  \omega_i \epsilon^{i \nu\sigma} \partial_\nu a_\sigma^\alpha$
 appearing in the Wen-Zee Lagrangian \eqref{wenlag}. Making a partial integration to rewrite it in terms of the dual electromagnetic field strength $^\star F^\mu = \epsilon^{\mu\nu\sigma} \partial_\mu A_\nu$
and the corresponding $^\star R^\mu = \epsilon^{\mu\nu\sigma} \partial_\mu \omega_\nu$ where $\omega_i$ is  the spin connection,
the $a_0^\alpha$ coefficient is precisely the  combination $\left( e\,t_{\alpha}B + s_\alpha {\mathcal R } \right)$ in  Eq.~\eqref{sback}. 

The remaining difficulty for obtaining the hierarchy wave functions on the sphere is related to the precise form of the vertex operators, when they contain derivatives. 
In section \ref{sect:expwfs}, we already pointed out that there is a freedom in distributing the derivatives acting in the wave functions.
On the disk, a valid (in fact the simplest!) prescription is to move all derivatives to the left, which ensures that the derivatives always act on fully holomorphic functions (or antiholomorphic in the case of $\bar\partial$). 
 However, this prescription does not work on the sphere (and likely not on any other closed surface).

The reason is that convoluting the resulting coherent state wave functions with the spherical version of the coherent state kernel,  as in Eq.~\eqref{LLLprojection},    gives zero, since it basically amounts to integrating a total derivative over a closed manifold. We must thus resort to  consider more general  expressions for the vertex operators of the type \eqref{spelop} 
$$
:e^{ia\phi^\alpha}\partial^{n}e^{ib\phi^\beta}:\, =\, :D^{n}[b\phi^\beta] e^{ia\phi^\alpha} e^{ib\phi^\beta} : \ ,
$$
 where $D^{n}[f]=e^{-if}\partial^{n}e^{if}$, so $D^{n}[b\phi^\beta] $ is nothing but a polynomial in $\phi^\beta$.  
For  details on how to adapt the operator $D$ to the sphere, and how to use it to  construct a set of electron operators that give well behaved representative wave functions for all states in the hierarchy, we refer the reader to the original article by \textcite{kvorning2013quantum}.

\subsection{The torus}  \label{sect:torus}
The torus is interesting for several reasons. 
First, the ground state wave function is degenerate, where the degeneracy is identical to the number of particle typesin the CFT.  
This ground state degeneracy is in fact a smoking gun for a topologically ordered phase.
From a computational point of view, the torus is interesting since candidate wave functions with different shifts can be directly compared with each other (on the sphere,  wave functions with different shifts require different number of flux quanta and, therefore, cannot be directly compared). 
In fact, it was at first not clear in what way states with different shift  would differ on the torus.
Through work by \textcite{avron1995viscosity} and  \textcite{read2009nonabelian} it became clear that the shift is related to the Hall viscosity, which is a transport coefficient characterizing a non-dissipative flow possible only in 2d systems violating time reversal symmetry.    
As explained in section \ref{ssec:modular} below, the Hall viscosity can be probed by deforming the geometry of the torus, and can  be  extracted from  
the CFT wave functions, again evoking assumptions about screening similar to those described in section \ref{subsub:monhol}.
Last, but not least,  studying quantum Hall wave functions on the torus provides an independent test of the holonomy = monodromy conjecture, as will be discussed in \ref{sec:hallvisc} below. 
Note that in this section we shall only consider chiral abelian states.

\subsubsection {QH wave function on the torus } \label{torusQHwf}
There are various ways to  obtain the proper torus wave functions for the ``simple''  quantum Hall states, \ie those built from   CFT correlators involving only primary fields. 
The torus version of the Laughlin wave function was first derived by \textcite{Haldane1985periodic}, using the vanishing properties of the state as well the constraints imposed by the magnetic translations.
The latter implies that proper torus wave functions need to fulfill certain quasiperiodic boundary conditions.  
These two properties are sufficient to uniquely specify the set of $m$ degenerate ground state wave functions for $\nu=1/m$ Laughlin states. 
In fact, not only is the ground state $m$-fold degenerate, but so is the entire spectrum \cite{haldane1985many}. 
This center-of-mass degeneracy occurs, because the magnetic translations along the two cycles of the torus do not commute.

A similar approach as described above for the Langhling states can be used for the Halperin $mmn$ states \eqref{mmn} \cite{wen2012modular} and even the Read-Rezayi series,  see \eg \textcite{greiterwenwilczek} and  \textcite{chung2007explicit} for the Moore-Read ground state and quasihole states, respectively. Note that for nonabelian states, there is an extra ground state degeneracy in addition to the kinematic center-of-mass degeneracy \cite{ReadRezayi1996}.

Another way to derive  the pertinent wave functions is to  compute the relevant CFT correlators on the torus.
Then, just as on the sphere, one can extract  wave functions  by calculating correlators of vertex operators to get the conformal blocks, and then impose the boundary conditions, see  \cite{hermanns2008quantum}. 
For technical reasons, one needs to consider the full fields (holomorphic and antiholomorphic components) and separate the correlator into a sum over conformal blocks  \cite{di1997conformal}. When doing this, it is also important to correctly handle the Gaussian factor, which amounts to distributing it symmetrically. 
A detailed derivation of these wave functions, keeping the full $\tau$-dependence, was given for the Laughlin case by \textcite{read2009nonabelian} and for  the multilayer states   by \textcite {fremling2014hall}. 
In both cases, the ground state wave functions are fully determined by the topological data; in the Laughlin case  this is only the filling fraction, $\nu$, while for the multilayer states it is both the $K$-matrix,  and the spin vector.

One would naively assume that this approach has a straightforward generalization to the hierachical states. 
However, the presence of derivatives in the wave functions makes the problem much harder, as they destroy the quasiperiodic boundary conditions  of the holomorphic blocks. Thus, one needs to identify the proper torus analog of derivatives. To do this, we must understand the modular properties of the torus wave functions, which is the topic of the next subsection.


\subsubsection{  Modular properties of CFT wave functions } \label{ssec:modular}

The geometry of a (flat) torus is determined by a complex number,  the {\it modular parameter}, $\tau$, which defines the periods under which the coordinates are to be  identified.  We imagine starting with a complex plane and drawing a segment from 0 to 1 on the real axis, and another segment from 0 to $\tau$.  Forming a parallelogram from these two segments and identifying opposite edges, we obtain a torus.   The boundary conditions on the wave functions are  defined by specifying how they transform under  $z \rightarrow z+1$ and $z \rightarrow  z+ \tau$. The relation between the geometry of the torus and the modular parameter is  not one-to-one; different $\tau$ can describe the same torus  ({\it i.e.}, they enforce the same periodicity on the plane). The different values are related  via {\it modular transformations}, which form a group that is generated by the two transformations,
\be{defmodtrans}
{\mathcal T} : \ \ \ \tau \rightarrow \tau + 1  \nonumber \\
{\mathcal S} : \ \ \ \tau \rightarrow  -\frac 1\tau \, .\nonumber
\ee
It will be important in the following that the $\mathcal S$ transformation involves a rotation of the torus. Since values of $\tau$ that differ only by a modular transformation describe the same physics, the space spanned by the ground states  must be invariant, meaning that the states must transform into each other by a unitary transformation.

For the CFT hierarchy construction, we shall need the  modular transformations of the multicomponent wave functions, which can be derived using the techniques employed by \textcite{read2009nonabelian} for the Laughlin case.  Schematically, they can be written as\footnote{
The complete formulae, which include gauge transformations, $\tau$-independent phase factors, and offsets in the arguments of the modular $\mathcal S$ and $\mathcal T$ matrices, also involve changes in the boundary conditions. 
These can be periodic or anti-periodic in the cycles of torus. 
The $\mathcal S$ transformation effectively swaps $x \leftrightarrow y$  and consequently also their boundary conditions, for details see  \cite{fremling2014hall}.
}
\begin{eqnarray}\label{modtrans}
  \psi_s &\stackrel{\mathcal{S}}{\to}&  \left(\frac{\tau}{|\tau|}\right)^{ h_{tot}}
      \sum_{s^\prime=1}^m \mathcal S_{s,s^\prime} 
  \psi_{s^\prime}  \nonumber\\
  \psi_s &\stackrel{\mathcal{T}}{\to}&
   \sum_{s^\prime=1}^m\mathcal T_{s+s^\prime}\psi_{s^\prime}   \, ,
\end{eqnarray}
where $\psi_{s} $ are the (degenerate) ground state wave functions, and $h_{tot}$ is the {\it total conformal weight} of all the electron operators. 
In the $\mathcal S$ transformation we also let $z \rightarrow z^\prime = |\tau|z/\tau$, while in the $\mathcal T$ transformation, $z$ is unchanged. 
Note that the only $\tau$-dependence in Eq.~\eqref{modtrans} is in the  $\mathcal S$ transformation. This is easy to understand: If we write $\tau = |\tau |e^{i\alpha}$ the $\mathcal S$ transformation is associated to a rotation of the coordinate $z$  by an angle $\alpha$ in the negative direction. Thus, an operator with conformal dimension $h_\psi$, which here equals the  conformal spin, acquires a phase factor $e^{i \alpha h_\psi}$. The factor $(\tau/|\tau|)^{h_{tot}}$ in Eq.~\eqref{modtrans} is precisely the product of these phase factors for all of the electrons. 

We are now ready to treat the  hierarchy states.  But before showing how the general CFT formalism in section \ref{ss:chiral}  can be implemented on the torus,  we describe some  work specially aimed at the Jain states.

\subsubsection {Composite fermion wave functions on the torus}
For obtaining the Jain series on the torus, one can take a different route, by using the fact  that the product of two single-particle states at fluxes $N_{\Phi_1}$ and $N_{\Phi_2}$ is a valid single particle state at flux $N_{\Phi_1}+N_{\Phi_2}$ --- independent of the underlying geometry. 
Thus, the Jain wave functions (including the Laughlin states) on the torus can be computed by using the torus version of Eq.~\eqref{jainwf}. 
Note that this expression only contains integer quantum Hall wave functions, which are known on the torus.  
The only remaining difficulty lies in the projection to the LLL,  for which explicit expressions were derived by \textcite{hermanns2013composite}. 
This approach is numerically inefficient and one can only handle small system sizes for two reasons. 
First, there is no analog of the fairly simple projection scheme of \textcite{girvin1984formalism} on the torus. 
Second, due to the periodic boundary conditions, momentum is not conserved when multiplying two single particle wave functions (it is only conserved modulo $N_\Phi$), which makes the  evaluation of the projection much more numerically costly than \eg on the sphere. \\

Note that evaluating Eq.~\eqref{jainwf} on the torus always gives a unique ground state wave function, in contrast to the expected ground state degeneracy on the torus. As momenta are only defined modulo $N_\Phi$, the flux attachment of Eq.~\eqref{jainwf} does not conserve the total momentum of the state  and the resulting wave function is, in general, a linear combination of the different degenerate ground states. The set of degenerate ground states can be recovered by acting with the center-of-mass translation operators or by allowing non-trivial boundary conditions around the two handles of the torus for the individual IQH wave functions. This construction even works  well  for the MR state, when using the ``layer description'' by \textcite{cappelli1999unified,cappelli2001parafermion}. For the general RR $Z_k$ series, however,  the full set of degenerate ground (and excited) states can only be generated by amending this construction as discussed by \textcite{repellin2015projective}. 


\subsubsection{Chiral hierarchical  wave functions on the torus }\label{ssec:modular}
The first attempt to use CFT methods for obtaining hierarchy states was made by \textcite{hermanns2008quantum}, where the derivatives were replaced by finite translations. While these were chosen to locally have the same effect as derivatives, the resulting wave functions did not transform properly under  modular transformations, and worked well only for tori of certain shapes. 
These difficulties were resolved by  \textcite{fremling2014hall} who conjectured
that the transformations \eqref{modtrans}, which  were derived for the multicomponent case, should hold true also for the hierarchy states. 

The main achievement of Fremling {\it et al.} was to replace the derivatives with combinations of finite differences in a way such that Eq.~\eqref{modtrans} still holds, but with the total conformal dimension $h_{tot}$ adjusted  to include the contributions from the derivatives. The construction is rather technical, but the upshot is that the torus generalization of the chiral hierarchy wave functions \eqref{genstate} is given by
\begin{equation}\label{finwf}
  \tilde\psi_s = \mathcal{A}\prod_{\alpha=1}^n \mathbb D_{(\alpha)}^{\alpha - 1}\psi_s \,,
\end{equation}
where $\psi_s$ is the toroidal wave function of the corresponding multicomponent state with the same $K$-matrix. The product is over the $n$ distinct electron operators, and the generalized shift operator $\mathbb D_{(\alpha)}$ is a sum over finite magnetic translations on a lattice with $N_\Phi^2$ points.\footnote{This lattice provides a natural way to define coherent states on the torus \cite{haldane1985many,fremling2013coherent}. } The actual expression for  $\mathbb D_{(\alpha)}$ is quite complicated but,  up to a single choice of a sign, it is fully determined by requirement that the wave function \eqref{finwf} transforms as \eqref{modtrans} with the appropriate conformal spins. The simplest example of a wave function \eqref{finwf}, namely the  second level Jain state at $\nu = 2/5$, shows an excellent agreement over the whole $\tau$-plane with the numerically obtained Coulomb wave function for eight electrons. It should be stressed that the overlap, {\it as a function of} $\tau$, is obtained without  
fitting any parameter.


\subsubsection{The Hall viscosity  } \label{sec:hallvisc}
\newcommand{\etah}{\eta^H}

\textcite{avron1995viscosity} showed that the Hall viscosity, which is a transport coefficient that determines the non-dissipative 
 response to a {\it strain rate}, can be obtained 
by adiabatically changing the modular parameter $\tau$, much like the Hall conductance $\sigma_H$ can be obtained from an adiabatic change of the flux though the holes in the torus \cite{niu1985quantized}. 
In analogy with this case, we define the Berry potentials related to the modular parameter $\tau = \tau_1 + i\tau_2$ by ${\mathcal A}_\tau = i \bracket{\Psi(\tau)}{\partial_\tau \Psi(\tau)}$, and ${\mathcal A}_{\bar\tau} = i \bracket{\Psi(\tau)}{\partial_{\bar\tau} \Psi(\tau)}$,
and the corresponding field strength 
${\mathcal F}_{\tau \bar\tau} = i \partial_{\bar\tau} {\mathcal A}_{\tau}- i \partial_\tau {\mathcal A}_{\bar\tau}$.
In terms of this quantity, the result of \textcite{avron1995viscosity} reads,
\begin{equation}\label{hvisk}
  \etah = -\frac {2\tau_2^2} A {\mathcal F}_{\tau\bar\tau}\, ,
\end{equation}
where  $A$ is the area of the system. 

 \textcite{avron1995viscosity} used \eqref{hvisk} to calculate $\etah$ for the filled LLL level where the wave function is known. 
Later  \textcite{read2009nonabelian} applied the same methods to  the Laughlin and Moore-Read states, using  the representative CFT wave functions.\footnote{
This paper also discusses  earlier work on the Hall viscosity, and gives relevant references.}
The result of these calculations could be summarized in the formula,
\be{readform}
\etah = \frac \hbar 2 \rho \bar s  = \frac \hbar 4 \rho S  \, ,
\ee
where $\rho$ is the electron density, and $\bar s$ the average conformal spin of the electrons, $S$ is the shift on the sphere (see Eq. \ref{eq:shiftK}), and where we reinserted $\hbar$. 
Read  has argued that Eq.~\eqref{readform} should hold also for general quantum Hall states.

\vskip 2mm \noindent
i. \emph{Adiabatic calculation  for the Laughlin states} \\
Here we outline the main ingredients used by \textcite{read2009nonabelian} to calculate $\etah$ for the Laughlin states. 
To determine the Berry potentials ${\mathcal A}_\tau$ and ${\mathcal A}_{\bar\tau}$,  and thus the field strength, one must determine the full $\tau$-dependence of the normalized
wave functions. In this case this can be done by generalizing  the plasma analogy described in section \ref{sub:plasma1} to the case of a torus with arbitrary $\tau$. Technically it amounts to keeping the full $\tau$-dependence in the calculation of the conformal blocks used to construct the wave function as already discussed. 
The  calculation of ${\mathcal A}_\tau$ then proceeds similarly to the derivation of \eqref{berryresult} --- using that the normalization constant is $\tau$-independent, the only contribution comes from the non-holomophic (in $\tau$) piece in the wave function which turns out to be determined by the orbital spin. In fact this non-holomorphic piece is directly related to the conformal spin of the electron operators. 
So, as long as holonomy = monodromy, the relation \eqref{readform} will hold for any wave function which is just a product of conformal blocks up to Gaussian factors.

\vskip 2mm \noindent
ii. \emph{Other approaches to the Hall viscosity} \\
The most direct  way to use \eqref{hvisk} is to simply compute  ${\mathcal F}_{\tau\bar\tau}$ numerically from a normalized wave function, that can be either a model wave function, or one obtained numerically from a model Hamiltonian.  Such calculations for the $\nu=1/2$ bosonic and $\nu=1/3$ fermionic Laughlin, and the numerically found Moore-Read state,   were done by \textcite{read2011hall}, who found good agreement with \eqref{readform}. They also checked the stability of the result for the $\nu=1/2$ state by adding a perturbation to the exact model Hamiltonian.

 It was pointed out by \textcite{hoyos2012hall} and \textcite{bradlyn2012kubo} that -- asuming Galilean invariance -- the Hall viscosity can be extracted from the $q^2$ term in the momentum dependent 
 Hall conductance $\sigma_H (q)$, which also can be read of from the term $\sim \bar s Ad\omega$ in the  effective response action \eqref{resplag} in section \ref{sc:effresp}. This is perhaps not very surprising given that, according to \eqref{readform}, the Hall viscosity is proportional to the average orbital spin $\bar s$.

\vskip 2mm \noindent
iii. \emph{Hall viscosity for abelian hierarchy states} \\
Using the same techniques as in \cite{read2011hall}  \textcite{fremling2014hall} calculated $\etah$ for the $\nu=2/5$ Jain state, and found good agreement, both with the formula \eqref{readform} and with  the corresponding results from the numerically obtain Coulomb wave functions. To appreciate the significance of this, let us again  emphasize  that the strong arguments for the  formula \eqref{readform} 
were given under the assumption that our wave functions are normalized such that no extra Berry phases are incurred  during the 
adiabatic change of $\tau$, \ie that all relevant $\tau$-dependence is explicit in the wave functions. For some of the states that can be written as correlators of primary fields, this can be convincingly argued using the exact mapping onto  a classical plasma. 
The formula \eqref{readform} will, however,  apply to hierarchy states only if the reasoning based on the plasma analogy holds. Thus, the mere fact that the numerical results for $\etah$ are in accordance with it, gives strong  support for the assertion that $holonomy =  monodromy$ in the case of adiabatic changes of the $\tau$ parameters. Although this does not {\it per se} lend support to the corresponding claim for adiabatically moving quasiparticles, it certainly lends indirect support. 

\vskip 2mm \noindent
iv. \emph{ Hall viscosity in realistic systems} \\
Here we comment on the significance of the Hall viscosity when we move away from a isotropic and  clean system. 

Concerning anisotropies, \textcite{haldane2011geometrical,park2014guiding} 
have argued that when rotational invariance is broken, \eg by a tilted magnetic field, the relation \eqref{readform} between Hall viscosity and orbital spin has to be modified. Also on a sufficiently deformed torus, \ie coming close to the TT limit,  Eq.~\eqref{readform} no longer holds \cite{fremling2015analytical}, in much the same way as  braiding phases of quasiparticles are well defined only when they are sufficiently separated. Stated differently, we can only expect the topological properties to emerge in a suitable scaling limit \cite{frohlich1991universality}.

As far as we know, there is no conclusive answer to whether or not the Hall viscosity has any meaning for weakly disordered systems, and the same holds for the shift on the sphere. 
In Ref. \cite{read2011hall} it is claimed  that in the presence of disorder ``the shift ceases to have significance, due to the loss of rotational invariance on the sphere. The same will be true for the Hall viscosity''. In a strict sense, this is presumably true, but we believe that for sufficiently weak disorder, the shift ought to be robust, since changing the shift amounts to changing the number of particles in the ground state at given flux. Seen from the undisturbed system, this amounts to exciting quasiparticles that will be pinned to the impurity centers, but this will be energetically advantageous only if the strength of the impurity is comparable to the gap. We know of no similar argument for the Hall viscosity, and find it quite likely that the quantization of this quantity is lost even for weak disorder. However, even in this case the average value of the orbital spin of the constituent electrons might have a meaning, at least for weak disorder. It is however completely unclear if this is an observable that could even in principle be measured. 

Having said this, we want to stress that both from a conceptual, and a computational point of view the Hall viscosity is quite interesting quantity, as should have been clear from the above discussion.

\section{Summary and future directions}
The main message of this review is that to address the plethora of observed  states, and the multitude of theories, models and schemes, there are powerful theoretical tools, and deep physical principles, that  help us  organize our knowledge of the vast and fascinating field of quantum Hall physics. 

Hierarchy is such a principle. It comes in many varieties, but the basic idea of starting from a parent state and generating a sequence of topological liquids, by successive condensation of quasiparticles, is common to all of them. If the parent state is abelian, so is the hierarchy, while a nonabelian parent state can give rise to qualitatively different hierarchies depending on which quasiparticles condense, and the details of how it is done. In particular certain types of condensations give rise to abelian offspring, while others can result in nonabelian states either similar to or different from that of the parent. 

Conformal field theory has emerged as a unifying theoretical tool in quantum Hall physics. Originally its relation to quantum Hall physics was via Chern-Simons theory. Expectation values of the Wilson loops in these gauge theories, which encode the topological properties of quasiparticles, are closely related to the monodromies of the conformal blocks of the corresponding CFT. In the quantum Hall context, the CFT plays the dual role of providing representative bulk wave functions, and being the dynamical theory of the edge modes. In this review we also showed that the CFT techniques are not restricted to states that can be expressed as correlators of primary fields, such as the Laughlin, Moore-Read and the Read-Rezayi states, but can also successfully be applied to hierarchy states. Although we did not mention it prominently, there is a very general, and in a sense minimal, mathematical description of topological phases,  phrased in terms of tensor categories \cite{wangbook,bondersonthesis,kitaev2003}.

There are many intriguing questions concerning the relationships between different hierarchies, and between hierarchy constructions and other organizing principles. Approaches based on effective field theories for composite fermions and bosons, naturally lead to hierarchy schemes, and effective field theories based on Chern-Simons gauge fields are general enough to encompass all the hierarchies discussed in this review. The composite fermion wave functions also fit nicely in a hierarchy scheme, although they were originally organized using quite a  different logic.

In spite of the successes of the various hierarchy schemes, there are many unanswered questions and several promising lines of research. Clearly there is still no comprehensive theoretical understanding of the nonabelian hierarchies, and we are still waiting for a universally accepted experimental signature for any nonabelian state. At the more detailed level, there are many remaining theoretical problems. We already mentioned the need for proofs of the expected topological properties of quasiparticles in the hierarchy states. At the theoretical level this would require new tools, such as much more refined plasma analogies, while convincing numerical work would require developing novel methods. The extension of the matrix product state methods \cite{zaletel2012exact,EstienneMPS} to hierarchy states might be the way to go forward. Another challenging problem is to find  a direct and precise link between the microscopic physics of the quantum Hall effect, and the description in terms of CFT correlators,  which 
is not based on 
case-by-case numerical verifications of the Moore-Read conjecture. 

The biggest question is, perhaps, to what extent the methods discussed in this work will pertain to other non-quantum Hall physical systems. 
The quantum Hall  effect is clearly quite special as the only (for dimensions $d>1$) experimental system that has convincingly demonstrated topological properties, such as charge fractionalization.
Nonetheless, there is an enormous amount of research activity exploring other systems that have nontrivial topological properties \cite{TopInsulators,Kallin16chiral,QSLreview}  and many such materials have recently been found.   
While no experiment has observed any hierarchy physics in non-quantum Hall systems, there is nonetheless substantial influence of theoretical ideas from the quantum Hall world on these research topics \cite{Ashvin1,Ashvin2}.   
It is very likely that our deep understanding of quantum Hall hierarchies will continue to enlighten a broad range of future research directions.

 {\bf ACKNOWELDGMENTS:}   THH, MH and SV thank Emil Bergholtz, Chia-Chen Chang, Mikael Fremling,  Jainendra Jain, Anders Karlhede, Nicolas Regnault and Juha Soursa for collaboration on various studies of the CFT wave functions.  
In addition, we all thank Eddy Ardonne, B. Andrei Bernevig, Parsa Bonderson, and Joost Slingerland for many enlightening discussions. Special thanks to Thomas Kvorning for helpful comments on section VII.A. and to Steve Kivelson and Nick Read for many discussions and very constructive comments on the original manuscript. 
 THH is supported by the Swedish Research Council, MH by the Emmy-Noether program of the DFG, SV by the  Research Council of Norway and  SHS is funded by EPSRC.   
 Statement of compliance with EPSRC policy framework on research data: This publication reviews theoretical work that does not require supporting research data.

\appendix

{\begin{center} \bf APPENDICES \end{center}}


\section{Super-abbreviated quantum Hall basics} \label{app:QHE}

The bare basics of quantum Hall physics will be given here for reference. For further details we refer the reader to various books on the subject \cite{ezawa2008quantum,stonebook,prangebook,chakrabortybook}.

We consider a particle in two dimensions, of mass $\mu$ and charge $q$ in a magnetic field $B$ perpendicular to the plane.  The single particle energy spectrum breaks up into degenerate ``Landau levels" with energies $E=\hbar \omega (n+1/2)$ where $\omega=q B/\mu$ is the cyclotron frequency.   Working in symmetric gauge so that the vector potential is $\vec A = (B/2)(y,-x,0)$, we can construct a set of ladder operators
\begin{eqnarray*}
	  a = \frac{1}{\sqrt{2}}( z /2  + 2 \bar \partial)  & ~~~~~~~~~~&   a^\dagger  = \frac{1}{\sqrt{2}}(\bar z/2  - 2 \partial) \\
	  	  b = \frac{1}{\sqrt{2}}(\bar z/2  + 2 \partial)  & ~~~~~~~~~~&   b^\dagger  = \frac{1}{\sqrt{2}}(z/2  - 2 \bar \partial) 
\end{eqnarray*}
with $z=x + i y$ and where $\partial \equiv \partial/\partial z= \frac{1}{2}(\partial/\partial x - i \partial/\partial y)$ and $\bar \partial\equiv\partial/\partial \bar z = \frac{1}{2}(\partial/\partial x + i \partial/\partial y)$ and the magnetic length scale   $
 \ell = \sqrt{\hbar/e B} $  has been set to unity (a convention that we adopt often).   Note that $\partial^\dagger = -\bar \partial$.  These operators satisfy $[a,a^\dagger]=[b,b^\dagger]=1$ with other commutators vanishing.    The single particle Hamiltonian and the $z$ component of angular momentum take the form
\begin{equation}
\label{eq:Hnonint}
  H = \hbar \omega (a^\dagger a + 1/2)
 ~~~~~~~;~~~~~ L_z = \hbar (b^\dagger b - a^\dagger a).
\end{equation}
Thus, $a^\dagger$ is a Landau level raising operator, which changes the energy eigenvalue and $b^\dagger$  increases the angular momentum one unit within the Landau level. 

Defining a fiducial state  $|0,0\rangle$, which is annihilated by both $a$ and $b$, we  can construct a single particle eigenstate with energy eigenvalue $\hbar \omega(n +1/2)$ and $L_z$ eigenvalue $\hbar(m-n)$ 
$$
 |n,m\rangle \sim (a^\dagger)^n (b^\dagger)^m | 0,0\rangle \,.
$$
Explicitly, we have the real-space wave function form of the fiducial state
$
  \langle z | \psi_{00} \rangle  \sim e^{-|z|^2/(4\ell^2)}
$ 
so that the LLL basis states become
$$
\langle z | \psi_{0m} \rangle \sim (b^\dagger)^m e^{-|z|^2/(4\ell^2)} \sim z^m e^{-|z|^2/(4\ell^2)}.
$$
 Here the $m^{\rm th}$ state is shaped like a ring of radius $r^2 = x^2 + y^2 = 2 m\ell^2$.   The density of single particle eigenstates (and hence the density of a filled Landau level) is thus $m / (\pi \,2 m \ell^2 ) = 1/(2 \pi\ell^2)$.    The most general LLL wave function for a single particle is  $f(z) e^{-|z|^2/(4\ell^2)}$ for any analytic function $f$.

Unless otherwise stated, we will always be interested in the physics of a partially filled LLL.   We define the filling fraction $\nu=\rho \phi_0/B$ where $\rho$ is the particle density and $\phi_0 =2 \pi \hbar/e$ is the flux quantum.   For fermionic particles, $\nu$ can also be understood as the fraction of the orbitals available in the Landau level that are occupied.  When the LLL is partially filled there is an enormous degeneracy corresponding to the choice of which orbitals should be filled and which should be left empty.\footnote{For bosons, the degeneracy remains even for $\nu$ an integer.}  This degeneracy is broken only by the interactions between the particles, since within a single Landau level, the kinetic energy $H$ from Eq.~\eqref{eq:Hnonint} is simply an uninteresting constant.  

When there are $N$ identical particles in a fractionally filled LLL, a many particle wave function must be of the form
$$
  \Psi = f(z_1, \ldots, z_N) \prod_{i=1}^N e^{-|z_i|^2/(4\ell^2)}
$$
where $f$ is a symmetric (antisymmetric) analytic function when the particles are bosons (fermions).   If the Hamiltonian is rotationally symmetric and the ground state is nondegenerate, then it should be an angular momentum eigenstate, which means the function $f$ should be a homogeneous polynomial with an overall degree that is set by the total angular momentum of the system.   Given a function $f$, one can determine the filling fraction, by finding the highest power $L_{max}$ of $z_1$  (or of any given $z_i$). This determines the maximum radial extent of the droplet.  Thus, the filling fraction is given by 
$$
 \nu \rightarrow N/L_{max}
$$
as both $N$ and $L_{max}$ become large.  For finite $N$ we have a relation $L_{max} = \frac{N}{\nu} -  S$ where $S$ is a constant known as the {\it shift}, see Eq.~\eqref{eq:shift}. 

Which analytic function $f$ should be chosen is determined by the particular inter-particle interaction so as to minimize the interaction energy at the appropriate filling fraction in order to find a ground state wave function.    Laughlin's original ansatz for $\nu=1/m$ given by
$$
 f = \prod_{i < j} (z_i - z_j)^m
$$
was motivated by the  intuition that forcing the wave function to vanish rapidly as two particles approach each other, will essentially keep the particles further apart, and is therefore  prone to lower the  energy due to a  short ranged repulsive interaction.     

In fact, the Laughlin wave function can be shown to be the exact ground state of a particular ultra-short range interaction which gives positive energy when the wave function vanishes  slower than $m$ powers as two particles approach each other \cite{haldane1983fractional,trugman,Pokrovsky}.   The Laughlin  $\nu = 1/m$ ground state is the highest density zero energy state of this interaction.  The quasihole states, obtained by multiplying the wave function by  factors $\prod_i (z_i - \eta_a)$ are also zero energy states of the Hamiltonian, but are at lower electron density.   Similarly, edge excitations can be obtained by multiplying the Laughlin ground state by any symmetric polynomial. 

Let us very quickly comment on  the case of a partially filled higher Landau level.  The easiest way to handle such higher Landau level wave functions is to use the one-to-one mapping to LLL wave functions, via
$$
 |n^{\rm th} \mbox{ LL wave function}\rangle = \prod_{i} (a_i^\dagger)^n  |\mbox{LLL wave function} \rangle
 $$
where $a^\dagger_i$ is the Landau level raising operator for the $i^{\rm th}$ particle.   
 It is worth noting that the application of $a^\dagger$ operators puts $\bar z$ factors in front of the gaussian, and typically the $n^{\rm th}$ Landau level wave functions will have terms with $n$ such factors of each $\bar z$.     
A wave function can be projected \cite{girvin1984formalism} to the LLL by moving all $\bar z$ coordinates to the left and replacing them with $2 \partial_z$.

For many filling fractions, the system will have a gap in the excitation spectrum above the ground state --- this is a sign of incompressibility.  If the system is also fluid (in the sense of not having broken translational symmetry) then this is known as a quantum Hall ground state. The stability of the quantum Hall state is determined by the size of the excitation gap. There are two types of collective excitations in quantum Hall systems. The low-energy excitations are intra-Landau level quasiparticle-quasihole pairs, which are commonly called magnetorotons. Their energy gap is set by the Coulomb interaction, and it is this gap that is relevant for the stability of the quantum Hall state. The other collective mode is the cyclotron resonance mode of the center of mass, also called the `Kohn-mode'. 
At zero momentum it is just the cyclotron motion of the center of mass, so the gap is $\hbar \omega_c$ 
independent of the interactions \cite{Kohnmode1961}. 

In the absence of disorder, by Galilean invariance\footnote{The Galilean invariant result holds for strictly zero disorder, but not necessarily for the limit of disorder going to zero.}, the conductance matrix of a two dimensional system in a magnetic field is always given by $\sigma_{xx} = \rho_{xx}=0$ for the longitudinal component and $\sigma_H = \sigma_{xy} = 1/\rho_{yx} = \nu e^2/h$ for the Hall component, with $h$ being Planck's constant and $e$ the electron charge.    If we consider a fractional quantum Hall ground state with filling fraction $\nu=p/q$, when disorder is added, the Hall conductance will be fixed at  $ \sigma_{H} =  (p/q) e^2/h$ for a range of values of $\nu$ around $p/q$ and $\sigma_{xx}$ remains zero over this range as well.   This is known as a quantum Hall plateau and typically the size of the plateau, {\it i.e.} the range of $\nu$ for which the conductance is quantized, is larger for quantum Hall states with larger gaps.  If the disorder or the temperature becomes larger than roughly the gap scale, then the quantum Hall state will be destroyed: $\sigma_{xx}$ becomes nonzero, and the quantization of $\sigma_{H}$ is lost. 
\


\section{Everything you need to know about Conformal Field Theory} \label{app:CFT}

Conformal Field Theory (CFT) is a rich and beautiful subject and there are many excellent introductions to the subject itself, most notably perhaps \cite{di1997conformal}. For a pedagocical review on how it is applied to quantum Hall states, we refer the reader to \cite{nayakreview}.  Although the topic is quite complex,\footnote{Pun intended} for our purposes we will only use some of the very basic ideas. 

While CFTs exist in higher dimensions, they are particularly powerful for describing the physics of gapless 1+1 dimensional relativistic systems with spatial coordinate $x$,  time coordinate $t$ and speed of light $c$ which we usually set to unity.      
We often use the complex coordinates $ z = x + i t $  and $\bar z = x - i t$, which naturally describe  left-moving and right-moving particles.  In general, gapless 1+1 dimensional systems the two sectors decouple almost perfectly, and one can treat the holomorphic ($z$) and antiholomorphic ($\bar z$) variables almost entirely independently of each other.   In this Appendix we will  mainly focus on the holomorphic parts, so we are essentially considering a left-moving chiral theory.  The generalization to the right-moving, or anti-holomorphic sector, which is needed when considering general quantum Hall states, should be obvious. 

The use of CFT in the quantum Hall context is two-fold.  As discussed in section \ref{ssec:history} in the main text, the CFT is used to describe not only the 1+1 dimensional dynamics of the chiral quantum Hall edge, but also the bulk (2+0) dimensional wave function, which essentially is a purely holomorphic polynomial.   While CFTs in 1+1 dimension completely describe the physics of such systems, for our purposes is it easier to think of a particular CFT as being simply an operator algebra with certain nice properties, and which is used to generate holomorphic functions, and hence quantum Hall wave functions. 

While one might attempt to build a quantum Hall state, both at the edge  and in the  bulk, from any 1+1 dimensional CFT, it turns out that various subtle consistency conditions prevent most CFTs from describing sensible quantum Hall states.   Most of them turn out to be described by some version of one of the simplest CFTs imaginable -- the chiral boson -- which we will describe in some detail later.


\subsection{General CFTs}

Generally, a CFT  contains a set of  primary fields $\Phi_i(\bar z,z)$  (or sometimes just $\Phi_i$) each with scaling dimensions $\Delta_i$.
Each primary field also has a tower of corresponding descendant fields  (which are basically derivatives of the primary fields) whose scaling dimensions are higher by integer steps from the corresponding primary.   The primary fields obey a set of fusion rules\footnote{Fusion rules simply specify the possible outcomes when taking the short-distance product of two local fields. One can consider it a short-hand notation of the operator product expansion in Eq.~\eqref{eq:OPE}.    When the integer $N^k_{ij}$ is greater than one, it means that when   $\Phi_i$ fuses with  $\Phi_j$ it results in several copies of the field $\Phi_k$. This is analogous  to tensor products of group representations; \eg in SU(3), $ 8 \otimes 8 = 1 \oplus 8 \oplus  8 \oplus  10 \oplus  \bar{10} \oplus  27$, so 8 occurs twice. }
\begin{equation}
\label{eq:fusion}
 \Phi_i \times \Phi_j = \sum_k N_{ij}^k \Phi_k
\end{equation}
where the $N_{ij}^k$'s are non-negative integers. A field $\Phi_i$ is known as a {\it simple current} if fusion with this field acts as a permutation on the fields --- or in other words,
if for fixed $i$ and any $j$ we have $N_{ij}^k$ being zero for all values of $k$ except a single value of $k$ for which $N_{ij}^k=1$.

There always exists a special field known as the identity (written usually as $\bf 1$) which fuses trivially with all fields $N_{{\bf 1}j}^k = \delta_{j,k}$.  Each field $i$ has a unique  conjugate field $\bar \imath$, 
with which it can fuse to form the identity, {\it i.e.}, such that $N_{i\bar \imath}^{\bf 1} = 1$.   Note that a field's inverse may be itself.   A so-called charge conjugation matrix $C_{ij}$ is defined to be unity for two fields that are conjugates  $C_{i\bar \imath} = 1$ and zero otherwise.  The matrix $N_{ijp} = \sum_k C_{pk} N_{ij}^k$ must be fully symmetric under permutation of $i,j,p$.       

 The objects of particular interest for us are the conformal blocks, which are  correlation functions of holomorphic or
anti-holomorphic fields.  In many cases there are techniques for calculating these blocks independently, while in general they can be extracted by factorizing the correlators of the full fields, $\Phi(\bar z, z)$, which essentially are the products of the holomorphic, $\Phi(z)$, and the anti-holomorphic fields, $\bar\Phi(\bar z)$, 
\begin{multline} \label{chiralfac}\av { \Phi(z_1 \bar z_1),    \dots \Phi(z_N, \bar z_N) } \\
= \sum_\alpha \av { \bar \Phi( \bar z_1),  \dots \bar\Phi( \bar z_N)  }_\alpha \av {  \Phi(  z_1),   \dots \Phi(  z_N) }_\alpha \, ,
\end{multline}    
where $\alpha$ labels the different blocks. In simple cases the sum has only one term, but in general there are many blocks that will correpond to different quantum Hall wave functions, as discussed in sections \ref{wfasconfblock} and \ref{sect:torus}. The scaling dimensions, or ``conformal weights" of the fields $\Phi_i$ and $\bar\Phi_i$ are denoted by $h_i$ and $\bar h_i$ respectively. \footnote{Note that the conjugate of a field $i$ has the same conformal weight as the original field, i.e. $h_i$ = $h_{\bar \imath}$. } The total scaling dimension is then $\Delta_i = h_i + \bar h_i$, and the {\it conformal spin}, which in the context of quantum Hall wave functions is identified with the orbital spin, is given by $s_i = h_i - \bar h_i$.

There is a ``neutrality'' condition on any correlator $\langle \Phi_1 \Phi_2 \ldots \Phi_M \rangle$ that the  fields inside the correlator must fuse to the identity, or the entire correlator will vanish.   The correlator of two conjugate fields must take the form
\begin{equation}
\label{eq:phinorm}
 \langle \Phi_i(z) \Phi_{\bar \imath}(w) \rangle = (z-w)^{-2 h_i}
\end{equation}
by dimensional analysis.  The coefficient of unity (not written) on the right of this equation is a conventional normalization of the fields.  There is a completely algebraic approach to computing the correlators $\langle \Phi_1 \Phi_2 \ldots \Phi_M \rangle$, using fusion rules and conformal invariance, but in the cases where there is a Lagrangian description the correlators are just the usual vacuum expectation values of time (or in the Euclidian case radially) ordered products of field operators, that can be calculated using standard field theory techniques. The main examples used in this paper, namely the chiral bosons and the Ising CFT (used to construct the Moore-Read state), belong to this category.

The fusion rules and the scaling dimensions allow one to write the operator product expansion which describes what happens when coordinates of the fields approach each other
\begin{equation}
\label{eq:OPE}
 \Phi_i(z) \Phi_j(w) = \sum_k {\cal C}_{ij}^k (z - w)^{h_k - h_i - h_j} \Phi_k(w) + \ldots
\end{equation}
where the coefficients ${\cal C}_{ij}^k$ are constants that are only nonzero when $N_{ij}^k$ is.    There are also terms on the right hand side corresponding to descendant fields\footnote{Descendants are obtained by acting with derivatives on primary fields.}, but if we are considering only the leading terms in the expansion, it is sufficient just to write the primaries.    


\subsection{The chiral boson}
\label{app:chiralboson}

Perhaps the simplest example of a CFT is the chiral boson,  described by the Lagrangian \eqref{eq:edgeLagrangian1}.
We define a boson field  $\phi(z)$ as a function of the holomorphic coordinate $z$. These are free bose fields so the bosonic version of Wick's theorem applies.  The elementary two point correlator is given by
$$
  \langle \phi(z) \phi(z') \rangle = - \frac 1 k \log(z-z') \, .
$$
Note that the constant $k$ can be changed by renormalizing the fields. 
In section \ref{sub:edge} 
we have $k=1/\nu$.  In section \ref{sect:CSCFT} and onwards, we will take $k=1$.  
In the main text, chiral bosons with normalization $k=1/\nu$ are usually denoted by $\varphi$, while those with normalization $k=1$ are denoted by $\phi$.

The  {\it vertex operators} are defined as
\be{vertop}
 V_\alpha(z) = \, : e^{i \alpha \phi(z)}: \, ,
\ee
with the colons representing normal ordering.   The fusion rules are given by 
$$V_\alpha \times V_\beta = V_{\alpha+ \beta}$$
The key identity (which can be proven by use of Wick's theorem) is 
\begin{align}\label{eq:corrAPP}
 \langle V_{\alpha_1}(z_1) V_{\alpha_2}(z_2) \ldots V_{\alpha_N}(z_N) \rangle = \prod_{i < j} (z_i - z_j)^{\frac {\alpha_i \alpha_j} k}   
\end{align}
provided that $\sum_j \alpha_j = 0$, otherwise the correlator vanishes identically. 
The reason for this is that $\alpha$ is the charge of the operator $V_\alpha$ with respect to the $U(1)$ symmetry $\phi(z) \rightarrow \phi (z) + a$. 
For the correlator on the left-hand-side of Eq.~\eqref{eq:corrAPP} to be invariant under such $U(1)$ transformations, the resulting phase factor,   $\prod_{j} e^{i \alpha_j a}$, has to be $1$, thus yielding the charge neutrality conditions. 
Frequently we can be sloppy about this final constraint on the sum of $\alpha$  as we can imagine placing an appropriate neutralizing charge at infinity to force it to be satisfied if we want.  Or, as discussed in the main text (see Eq.~\eqref{eq:background}) we can use a smeared background charge operator. 

By dimension counting, the scaling dimension  of the  {\it primary field}  $V_\alpha$ is given by $\alpha^2/2k$, and combining this with 
 the operator product expansion formula \eqref{eq:OPE}, we see that, taking $k=1$, the vertex operators \eqref{vertop} are bosonic or fermionic for $\alpha$ being an even or odd integer respectively.


\subsection{The compactified boson}

In the quantum Hall context (say, for the Laughlin states) we do not want our bose CFT to contain quasiparticle operators $V_\alpha$ for all possible $\alpha$.  As discussed in connection with Eq.~\eqref{eq:hinteger}, once we define our electron operator $\psi_e  = V_{\alpha_0}$,  we must insist that all other operators in the theory are mutually local with respect to this operator --- that is, there should be no branch cuts in the electron wave function due to the insertion of quasiparticle operators.  This limits the particle content of our theory to only contain operators $\psi_n  =   V_{n/\alpha_0}$  with integer $n$. 

This constraint on the  bose vertex algebra is in fact equivalent to having what is known as a ``compactified" boson.   One begins with the same  bose Lagrangian Eq.~\eqref{eq:edgeLagrangian1}, but imposes the  periodicity relation 
$$
   \phi(z) = \phi(z) + 2 \pi R \, ,
$$ 
where $R$ is known as the compactification radius, \ie  $\phi$ is defined on a circle with  radius $R$. 
This immediately constrains the possible vertex operators since
$$
V_{\alpha}= \, : e^{i \alpha \phi}: \, = \,:e^{i \alpha (\phi + 2 \pi R)}: \,  = 
e^{2 \pi i \alpha R} V_{\alpha}
$$
which implies that $\alpha = n/R$, \ie there is a discrete set of allowed values for $\alpha$.

\section{Commutator of Edge Operators}\label{sec:commutator}
In the text we motivated the fundamental anomalous commutator, Eq.~\eqref{eq:KacMoody}, by a heuristic argument, based on the expected equation of motion for  edge excitations. Here we give two complementary field theoretic derivations. \\
{\it Approach 1: The UV perspective.} 
Here we give a derivation which is essentially the original one by \textcite{mattis1965exact} but cast in the quantum Hall context. 
Working in Landau gauge, single electron orbitals can be taken to be strips parallel to the edge, indexed by their momentum in the direction along the edge $k$, and having position perpendicular to the edge $y=k \ell^2$.  For $\nu=1$, the ground state is simply described as a Slater determinant of $k$ states filled up to a Fermi energy $k_F$ (we imagine a half-infinite space so we only consider one edge).   The density operator along the edge (with the same normalization as section \ref{ssec:edge})  is then 
\begin{equation}
\label{eq:com1app}
\rho_q = \frac{1}{\sqrt{L}} \sum_k c^\dagger_{k+q} c_{k}^{\phantom{\dagger}} 
\end{equation}
Direct calculation of the commutator yields
$$ 
[\rho_q, \rho_{q'}] =  \frac{1}{L} \sum_{k,k'} \left[  c^{\dagger}_{k'+q+q'}c_{k'}^{\phantom{\dagger}} \delta_{k,k'+q}-  c^{\dagger}_{k+q+q'}c_{k}^{\phantom{\dagger}} \delta_{k',k+q}  \right] 
$$
While it may look like the two terms on the right precisely cancel, consideration of the ultraviolet limit ($k$ far from the Fermi surface) can give something nonzero. When we are far below the Fermi surface, since all states are filled, $c^{\dagger}_{k+q+q'}c_{k}$ is zero unless $q=-q'$.   However, when $q=-q'$, then we are generally subtracting two nonzero terms from each other and we must be more careful since there are potentially an infinite number of such terms --- and depending on how they are paired up, they can leave a finite result. Thus we focus on this particular case and obtain
\begin{equation}
\label{eq:com2}
[\rho_q, \rho_{-q}] =  \frac{1}{L}\sum_{k} \left[n_{k+q} - n_k \right].
\end{equation}
where $n_k=c^\dagger_k c_k$ which is exactly unity  deep below the Fermi surface.    Now suppose we try to cut off the infinite sum in Eq.~\eqref{eq:com1app}, say by including a regularization factor $w_k$ inside the sum  which is equal to unity near the Fermi surface, but drops to zero very far away from the Fermi surface.   For any form of this cutoff function (we can consider a step function) we will find that the difference in the two terms is nonzero and gives exactly $q (L/2 \pi)$ which is just the number of $k$ states that lie between $k+q$ and $k$, thus recovering exactly Eq.~\eqref{eq:KacMoody}.  In the case of FQHE, one need only modify this derivation by assuming that far below the Fermi surface $n_k$ will be $\nu$ rather than $1$.  \\
{\it Approach 2: The IR perspective.}  

The above derivation was contingent upon subtle non-cancellations of potentially UV divergent terms, and as such appear as a UV effect. Although formally  correct, this looks strange, particularly from a condensed matter perspective where there is certainly no infinite Dirac sea. A similar conundrum is related to the axial anomaly in 1+1 dimensions.  Defining left and right moving currents to be $j_+$ and $j_-$ respectively, the vector and axial currents are  $j_V = j_+ + j_-$ and $j_A = j_+ - j_-$.    We then have $\partial_\mu j^\mu_A =  eE/\pi$, which for a chiral theory amounts to an anomaly in the fermion current, $j = ( j_A + j_V)/2$ given by $\partial_\mu j^\mu =  eE/(2\pi)$. 
This can, by the following reasoning, be understood as an IR effect:     

A constant electric field in the positive $x$-direction will give rise to a spectral flow at the Fermi level, which corresponds to creating right-moving particles and left-moving holes in such a way that the total electric charge is conserved, while the axial charge, \ie the difference between right and left movers, is not. 
For a more detailed description of this  ``infinite hotel''\footnote{
	This name seems to have been coined by H.B. Nielsen in  reference to Hilbert's infinite hotel paradox.}
derivation of the axial anomaly, see the nice presentation by  \textcite{infinitehotel1}.   

We now give a direct derivation of the anomalous commutator using only IR arguments.  
Starting with an achiral  one-dimensional system, we can construct a  chiral theory by separating  the two chiralities.  To do this we start with the usual one-dimensional (non-chiral) current and density operators in first quantization  $\rho(x) = \sum_{n=1}^N  \delta(x - x_n)$ and $j(x)=\frac{1}{2m} \sum_{n=1}^N  [p_n \delta(x - x_n) + \delta(x-x_n) p_n] $ from which we can directly derive
\be{dencurrcom}
[\rho(x),j(x')] = -i (\hbar/m) \rho(x) \partial_x \delta(x - x')
\ee
along with $[\rho(x),\rho(x')]=[j(x),j(x')]=0$.   We then assume small excitations of both the left-going and right-going Fermi surfaces. 
Then we can write $\rho=\rho_0 + \rho_+ +\rho_-$ and $j = v_F (\rho_+ - \rho_-)$ where $+$ and $-$ again indicate left and right movers, and $\rho_0$ is a constant background density which is much larger than the perturbations $\rho_+$ and $\rho_-$.   We then have
\begin{eqnarray*}
	[\rho_+(x),\rho_+(x')] &=& \frac{1}{4} [\rho(x)+j(x)/v_F, \rho(x') + j(x')/v_F] \\
	&=& -i (\hbar/(2 v_F m)) \rho(x) \partial_x \delta(x - x') \\
	&=& -i (1/(2 \pi)) \partial_x \delta(x - x')  
\end{eqnarray*}
where in going to the last line we have substituted $\rho_0$ for  $\rho(x)$ on the right, since it is assumed to be much larger than $\rho_+$ and $\rho_-$. 
We also  used that for spinless one-dimensional fermions $\rho_0 = k_F/\pi = m v_F /(\pi \hbar)$.  
The chiral operators $\rho_+$ are what we called $\rho(x)$ in section \ref{ssec:edge}, and  Fourier transforming gives Eq.~\eqref{eq:KacMoody}. 

Note that just as in the ``infinite hotel'' derivation of the axial anomaly, we only  considered effects close to the Fermi surface. The connection between the 
two calculations follows if we couple the right moving charge to an electromagnetic potential by $H= e\int A_0 \rho_+ $ and use \eqref{dencurrcom} and partial integration to  get $\partial_t Q_+ = \int dxdy\, [\rho_+(x), eA_0(y) \rho_+(y)] = e\int dx\, \partial_x A_0(x) = (  1 / \pi) \int dx\, eE (x)$ which is the integrate form of 
the fermion current anomaly for spatially constant fields.

\bibliography{hierarchy}

\end{document}